\documentclass[useAMS,usenatbib]{mn2e}

\usepackage{aas_macros,graphicx,amsmath,amssymb,lscape,soul}

\title[Resolved spectroscopy of gravitationally lensed galaxies]{Resolved spectroscopy of gravitationally lensed galaxies: global dynamics and star-forming clumps on $\sim 100$\,pc scales at $1 < z < 4$}
\author[R.~C.~Livermore et~al.]{R.~C.~Livermore,$^{1,2}$ \thanks{E-mail:
r.c.livermore@astro.as.utexas.edu}
T.~A.~Jones,$^{3,4}$
J.~Richard,$^{5}$
R.~G.~Bower,$^{1}$
A.~M.~Swinbank,$^{1}$
\newauthor T.-T.~Yuan,$^{6}$
A.~C.~Edge,$^{1}$
R.~S.~Ellis,$^{3}$
L.~J.~Kewley,$^{6}$
Ian Smail,$^{1}$
K.~E.~K. Coppin$^{7}$
\newauthor and H.~Ebeling$^{8}$\\
$^{1}$Institute for Computational Cosmology, Durham University, South Road, Durham DH1 3LE, UK\\
$^{2}$Department of Astronomy, The University of Texas at Austin, 2515 Speedway Stop C1400, Austin, TX 78712, USA\\
$^{3}$Astronomy Department, California Institute of Technology, 249-17, Pasadena, CA 91125, USA\\
$^{4}$Department of Physics, University of California, Santa Barbara, CA 93106, USA\\
$^{5}$CRAL, Universit\'e Lyon 1, Observatoire de Lyon, 9 avenue Charles Andr\'e, 69561, Saint Genis Laval Cedex, France\\
$^{6}$Research School of Astronomy and Astrophysics, The Australian National University, Cotter Road, Weston Creek, ACT 2611, Australia\\
$^{7}$Centre for Astrophysics, Science \& Technology Research Institute, University of Hertfordshire, Hatfield AL10 9AB, UK\\
$^{8}$Institute for Astronomy, University of Hawaii, 2680 Woodlawn Drive, Honolulu, HI 96822, USA
}
\begin{document}

\date{}

\pagerange{\pageref{firstpage}--\pageref{lastpage}} \pubyear{}

\maketitle

\label{firstpage}

\begin{abstract}
We present adaptive optics-assisted integral field spectroscopy around the H$\alpha$ or H$\beta$ lines of 12 gravitationally lensed galaxies obtained with VLT/SINFONI, Keck/OSIRIS and Gemini/NIFS. We combine these data with previous observations and investigate the dynamics and star formation properties of 17 lensed galaxies at $1 < z < 4$. Thanks to gravitational magnification of $1.4-90\times$ by foreground clusters, effective spatial resolutions of $40 - 700$\,pc are achieved. The magnification also allows us to probe lower star formation rates and stellar masses than unlensed samples; our target galaxies feature dust-corrected SFRs derived from H$\alpha$ or H$\beta$ emission of $\sim 0.8 - 40$M$_{\odot}$yr$^{-1}$, and stellar masses $M_{\ast} \sim 4 \times 10^8 - 6 \times 10^{10}$M$_{\odot}$. All of the galaxies have velocity gradients, with 59\% consistent with being rotating discs and a likely merger fraction of 29\%, with the remaining 12\% classed as `undetermined.' We extract 50 star-forming clumps with sizes in the range 60\,pc - 1\,kpc from the H$\alpha$ (or H$\beta$) maps, and find that their surface brightnesses, $\Sigma_{\rm clump}$ and their characteristic luminosities, $L_0$, evolve to higher luminosities with redshift. We show that this evolution can be described by fragmentation on larger scales in gas-rich discs, and is likely to be driven by evolving gas fractions.
\end{abstract}


\section{Introduction}
\label{sec:intro}

Advances in observational facilities and instrumentation over recent decades have led to a rapid accumulation of data on the statistical properties of galaxy populations and their evolution with cosmic time. Most notably, it is established that the cosmic star formation rate density peaked around $z \sim 1$ and has since declined by an order of magnitude \citep{1996MNRAS.283.1388M,1996ApJ...460L...1L,2006ApJ...651..142H,2012MNRAS.420.1926S}, and that more than half of the present-day stellar mass had been formed by $z \sim 1$ \citep[e.g.][]{2003ApJ...587...25D,2013arXiv1304.2395P}.

The challenge now is to interpret these statistical trends and to uncover the underlying physical processes at work. How did these early galaxies form their stars, and what conditions have changed over time to result in the observed evolution of their integrated properties?

Star formation is a complex process driven by the interplay of competing effects: self-gravitating gas is supported against collapse by turbulent motion and by feedback in the form of radiation pressure from hot, young stars and active galactic nuclei. An understanding of the evolution of star-forming galaxies therefore requires an understanding of the evolving dynamics of galaxies at different epochs. A population of rapidly star-forming galaxies at high redshift implies the presence of large gas reservoirs. Since the dynamics of gas and stars are intrinsically different - stars comprise a collisionless system, while gas is collisional and can dissipate energy - it is to be expected that high-redshift galaxies will have systematically different dynamics than those found locally.

The advent of near-infrared Integral Field Units (IFUs) on 8-10m class telescopes presented the opportunity to study two-dimensional velocity fields. At $z \gtrsim 1$, the optical nebular emission lines such as H$\alpha$ are redshifted into the near-infrared, with [O{\sc ii}] visible out to $z \sim 5$, providing bright tracers of the underlying dynamics in ionised gas across a large redshift range.

IFU studies of the dynamics of high-$z$ galaxies quickly revealed that a significant fraction, around 1/3, are rotating systems in place as early as $z \sim 2$, but that they show higher velocity dispersions than local galaxies. They also found that $\sim 1/3$ are `dispersion dominated' non-rotators, with the remaining $\sim 1/3$ comprising merging systems \citep[see \citealp{2013arXiv1305.2469G} for a comprehensive review]{2006ApJ...645.1062F,2006Natur.442..786G,2007ApJ...656....1L,2009ApJ...697.2057L,2007ApJ...658...78W,2009ApJ...699..421W,2009A&A...504..789E,2012A&A...539A..92E}.

A common feature in H$\alpha$ maps of high-redshift galaxies from IFU studies is the irregular, clumpy morphology of the H$\alpha$ emission. These have also been observed in high-resolution imaging \citep{1995AJ....110.1576C,2004ApJ...603...74E,2005ApJ...627..632E,2009ApJ...692...12E} and termed `clump cluster' or `chain' galaxies. The prevailing view is that these clumps form from internal gravitational instabilities in gas-rich discs \citep{2007ApJ...658..763E,2009ApJ...692...12E,2008ApJ...687...59G,2010MNRAS.409.1088B}. Their ubiquity in galaxies with ordered rotation supports this view, as clumps forming through major mergers would be expected to disrupt the dynamics \citep[e.g.][]{2009ApJ...694L.158B,2011ApJ...730....4B,2009ApJ...703..785D}.

The study of dynamics and star formation morphologies of high-redshift galaxies is hampered by spatial resolution. Even with adaptive optics, the current generation of telescopes can achieve resolution of $\sim 1 - 1.5$\,kpc at $z \sim 2$. This limits the number of spatial resolution elements, as not only are galaxies intrinsically smaller at high redshift, but cosmological surface brightness dimming $\propto \left( 1 + z \right)^4$ limits observations to the bright central regions. Beam-smearing also affects measurements of velocity dispersion, which includes contributions from the underlying velocity gradient as well as local turbulent motion, and can smooth out rotation curves to give slowly rotating galaxies the appearance of being dispersion-dominated \citep[e.g.][]{2009ApJ...699..421W,2013ApJ...767..104N}.

\subsection{Gravitational magnification}

Enhanced spatial resolution can be achieved with current facilities by targeting galaxies that benefit from strong gravitational lensing by massive foreground clusters. As well as stretching the galaxy images and thus increasing the spatial resolution of observations, lensing conserves surface brightness, which means that the total flux is effectively magnified. This enables us to study galaxies whose intrinsic luminosities lie below the detection limits of current surveys, opening up more `normal' galaxies, where unlensed surveys have by necessity focused on the more extreme star-forming population.

By taking advantage of magnification due to strong gravitational lensing, high-resolution dynamics in high-redshift galaxies have been observed, revealing a high fraction to have ordered rotation \citep{2006ApJ...650..661N,2007ApJ...657..725N,2006MNRAS.368.1631S,2008Natur.455..775S,2009MNRAS.400.1121S,2010ApJ...725L.176J}. Star-forming clumps are also visible in these galaxies, and due to the high spatial resolution can be observed on $\sim 100$\,pc scales, enabling direct comparisons with the H{\sc ii} regions in which stars form in the local Universe \citep{2009MNRAS.400.1121S,2010ApJ...725L.176J}. \citet{2012MNRAS.427..688L} used H$\alpha$ narrowband imaging to study clumps in star-forming galaxies at $0 < z < 2$ and showed that both their surface brightnesses and luminosity functions evolve with redshift, with the higher-redshift galaxies having more massive, brighter clumps with higher surface brightnesses. The high-$z$ clumps can be explained by the same formation process as H{\sc ii} regions in local galaxies: a marginally stable disc fragments on scales related to the disc's Jeans mass. At high redshift, higher gas fractions cause collapse on larger scales, leading to clumps large enough to dominate the galaxy's morphology. This factor alone, however, overpredicts the luminosities of high-$z$ clumps. \citet{2012MNRAS.427..688L} therefore suggested a contribution from the epicyclic frequency, which is higher at high redshift and acts to stabilise the disc, causing collapse on smaller scales.

This model effectively explains the evolution in observed clump properties, but relies on the dynamics of the galaxies, which were derived from disc scaling relations and not measured in the data. In this paper, therefore, we use integral field spectroscopy of lensed galaxies to examine the evolution of galaxy dynamics in combination with the properties of star-forming clumps. The paper is organised as follows: we present the sample and describe the data reduction and derivation of galaxy properties in Section \ref{sec:sample}. In Section \ref{sec:results}, we analyse the results and discuss first the dynamics and then the star-forming clumps, and we discuss the formation and evolution of star-forming clumps in Section \ref{sec:disc}. Finally, we summarise our conclusions in Section \ref{sec:conc}. Throughout, we adopt a $\Lambda$CDM cosmology with $H_0 = 70$km\,s$^{-1}$Mpc$^{-1}$, $\Omega_{\Lambda} = 0.7$ and $\Omega_{\rm{m}} = 0.3$.

\section{Observations and Data Reduction}
\label{sec:sample}

\subsection{Integral Field Spectroscopy}

\begin{table*}
  \caption{Targets and observations.}
  \label{tab:sample}
  \begin{tabular}{l c c c c c c c}
    \hline
    Name & $\alpha_{2000}$ & $\delta_{2000}$ & $z$ & Instrument & Filter & Emission line & Exposure time (ks)\\
(a) & $(b)$ & $(b)$ & $(b)$ & & & & \\
    \hline
MACS0744-system3 & 07:44:50.90 & +39:27:34.5 & 1.28 & NIFS & H & H$\alpha$ & 19.2 \\
MACS1149-arcA1.1 & 11:49:35.30 & +22:23:45.8 & 1.49 & OSIRIS & Hn3 & H$\alpha$ & 17.1 \\
MACS0451-system7 & 04:51:57.22 & +00:06:21.3 & 2.01 & SINFONI & H & H$\beta$ & 19.2 \\
A1413-arc2.1a & 11:55:18.31 & +23:23:55.5 & 2.04 & SINFONI & K & H$\alpha$ & 21.6 \\
A1413-arc2.1b & 11:55:18.31 & +23:23:55.5 & 2.04 & SINFONI & K & H$\alpha$ & 21.6 \\
A1835-arc7.1 & 14:01:00.99 & +02:52:23.2 & 2.07 & SINFONI & K & H$\alpha$ & 19.2 \\
MS1621+26-system1 & 16:23:35.56 & +26:34:29.1 & 2.14 & SINFONI & K & H$\alpha$ & 21.6 \\
RXJ1720+26-arc1.1+1.2 & 17:20:10.26 & +26:37:27.4 & 2.22 & SINFONI & K & H$\alpha$ & 24.0 \\
A1689-arc2.1 & 13:11:26.52 & -01:19:55.5 & 2.54 & SINFONI & H+K & H$\beta$ & 19.2 \\
A1689-arc1.2 & 13:11:26.43 & -01:19:57.0 & 3.04 & SINFONI & H+K & H$\beta$ & 19.2 \\
A2895-arc1.1+1.2 & 01:18:11.15 & -26:58:03.9 & 3.40 & SINFONI & K & H$\beta$ & 19.2 \\
A2895-arc2.2 & 01:18:10.45 & -26:58:14.2 & 3.72 & SINFONI & K & H$\beta$ & 26.4 \\
\hline 
Cl0024-arc1.1 & 00:26:34.42 & +17:09:55.4 & 1.68 & OSIRIS & Hn5 & H$\alpha$ & 16.5 \\
Cl0949-arc1 & 09:52:49.78 & +51:52:43.7 & 2.39 & OSIRIS & Hn4 & H$\alpha$ & 19.2 \\
MACS0712-system1 & 07:12:17.51 & +59:32:16.3 & 2.65 & OSIRIS & Kc5 & H$\beta$ & 16.2 \\
MACSJ0744-arc1 & 07:44:47.82 & +39:27:25.7 & 2.21 & OSIRIS & Kn2 & H$\alpha$ & 14.4 \\
CosmicEye & 21:35:12.73 & -01:01:43.0 & 3.07 & OSIRIS & Kn1 & H$\beta$ & 21.6 \\
\hline \\
\multicolumn{8}{l}{\begin{minipage}{\textwidth}Notes: $(a)$ To identify the arcs, we use the cluster names and the arc ID from the published lens model or the system number where we observe multiple images of the same galaxy, except for the Cosmic Eye which is also known as LBG J213512.73-010143. MACS clusters are truncations of MACSJ0744.8+3927, MACSJ1149.5+2223, MACSJ0451.9+0006, MACSJ0712.3+5931 and MACSJ2135.2−0102 \citep{2001ApJ...553..668E,2007ApJ...661L..33E,2010MNRAS.407...83E,2012MNRAS.420.2120M}. $(b)$ Positions and redshifts are given for the target lensed arcs. \end{minipage}}
\end{tabular}
\end{table*}

Table \ref{tab:sample} lists the galaxies in our sample, which were selected from clusters with existing mass models, and to have spectroscopically confirmed redshifts that place their H$\alpha$ or H$\beta$ emission in windows of atmospheric transmission in the near-infrared and away from OH emission lines. In order to ensure detection in $\sim 5$\,hours, we targeted galaxies with integrated emission line fluxes of $>10^{-16}$erg\,s$^{-1}$\,cm$^{-2}$, known either from prior spectroscopy or from spectroscopic pre-imaging \citep[e.g.][]{2011MNRAS.413..643R}.

The new sample comprises twelve galaxies, of which ten were observed with the SINFONI Integral Field Unit (IFU) on the ESO/VLT \citep{2003SPIE.4841.1548E} between 2009 April 30 and 2012 July 16. The SINFONI targets were selected to be close to sufficiently bright stars for the use of NGS+AO, resulting in a median FWHM $= 0.2$''. As the lensed arcs are extended, usually over several arcseconds, we used the 8''$\times$8'' field of view with a spaxel size of 0.25''. Due to the elongated shape typical of lensed arcs, the targets were kept in one half of the IFU and nodded across in an ABBA sequence. We observed each target for six ABBA sequences in the $H$ or $K$-band filter according to the redshifted position of the target emission line (see Table \ref{tab:sample}). We also observe [N{\sc ii}] in the H$\alpha$ targets and [O{\sc iii}] in the H$\beta$ targets; the properties of these lines will be discussed in a future paper.

The data were reduced using the {\sc esorex} package, which performs flat-fielding and wavelength calibration and reforms the image into a data cube with two spatial and one spectral dimension. It also carries out sky subtraction by subtracting each B frame from its closest A frame, which with our observing strategy results in cubes that contain two images of the target, one positive and one negative. Standard stars were observed on the same nights and in the same filters as the science exposures, and were reduced in the same way. We used the standard stars to individually flux-calibrate each cube before combining them.

To combine the cubes, we first cut them in half to separate the two images of the target, and subtracted the negative image from the positive one. We then aligned the cubes by collapsing them into continuum images. As the targets all lie behind cluster lenses, they commonly have elliptical cluster members close to their line of sight. Where possible, a foreground elliptical galaxy was intentionally positioned within the SINFONI field of view to allow careful alignment between cubes. Where this was not possible (in the case of the two arcs lensed by the cluster Abell 2895), we aligned the cubes by centring on a bright feature within the lensed galaxy. Once aligned, the cubes were median-combined with the {\sc iraf} task \verb'imcombine' with the \verb'crreject' algorithm for cosmic ray rejection.

\begin{figure*}
\includegraphics[height=\textwidth, angle=90]{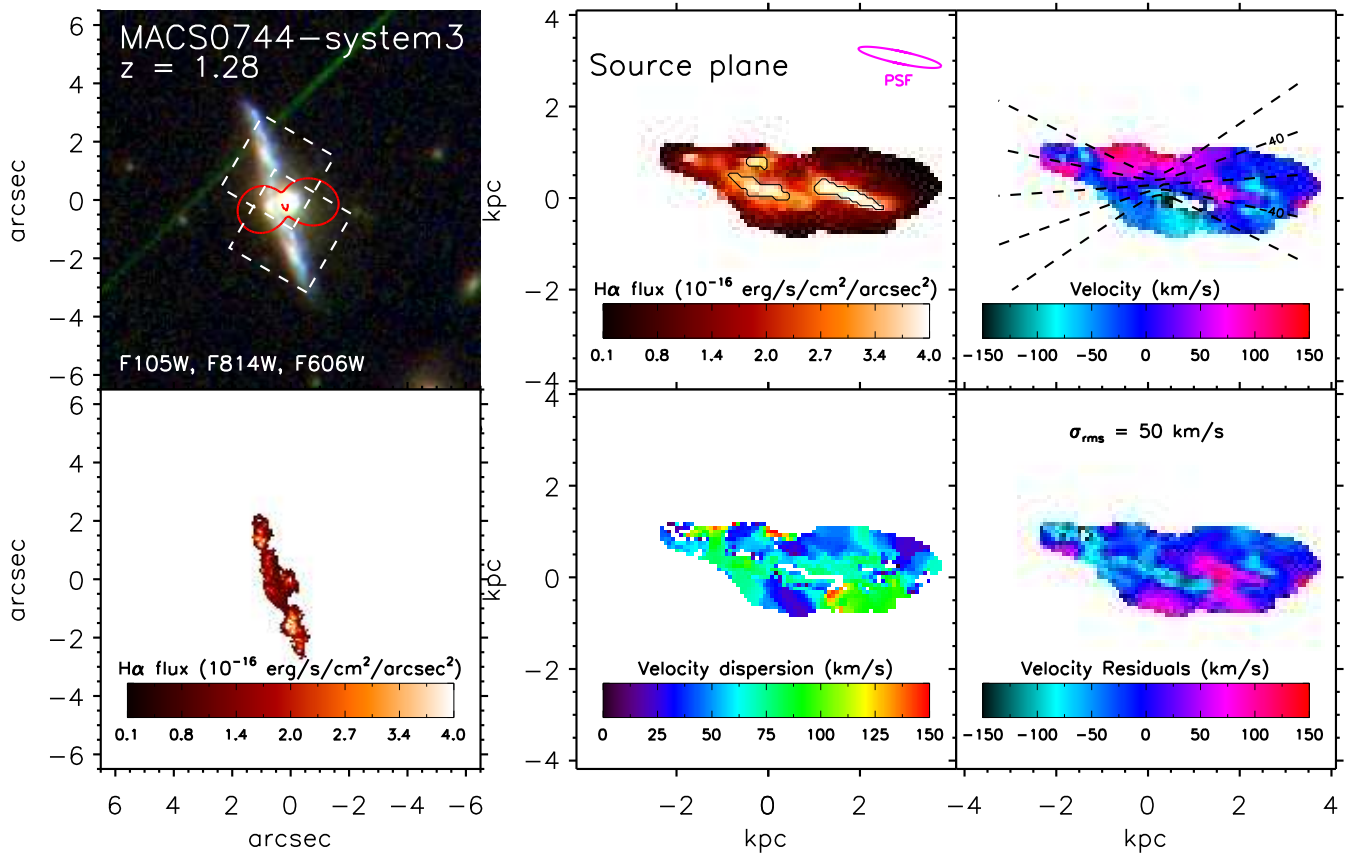}\\
\caption{Image plane colour images (where available) and reconstructed source plane maps of the target galaxies. The top-left image in each panel shows an \emph{HST} image of the arc constructed from archival \emph{ACS} and/or \emph{WFC3} data, in colour where there are multiple bands in the archive. The filters used are given in the lower-left corner of the image. The critical line at the target galaxy's redshift is overlaid in red. Where the IFU field of view cannot cover the entire arc, it is shown as a dashed white box. The lower-left image shows the H$\alpha$ or H$\beta$ emission line flux, aligned to the same astrometry as the \emph{HST} image for context. The right four images in each panel show the source plane reconstructions: from top left to bottom right, they show the H$\alpha$ or H$\beta$ intensity maps, velocity fields, line-of-sight velocity dispersion and velocity field residuals after subtracting the best-fit disc model. The star-forming clumps are contoured over the intensity maps. Contours overlaid on the velocity field show the best-fit disc model after smoothing by the effective source plane PSF, shown by a magenta ellipse. As surface brightness is conserved by lensing, the image and source plane intensity maps are displayed in terms of surface brightness, with the same scaling.}
\label{fig:im}
\end{figure*}

In addition to the ten targets observed with SINFONI, we observed a $z = 1.28$ galaxy lensed by the cluster MACS J0744+3927 with Gemini/ Near-Infrared Integral Field Spectrometer \citep[NIFS;][]{2003SPIE.4841.1581M}. The lensed target is denoted MACS0744-system3. As the target is too long to fit in the 3''$\times$3'' field of view (see first panel of Figure \ref{fig:im}), we observed in an ABC sequence, where two halves of the arc were positioned diagonally across the field in the A and C frames, and the B frame was a blank field used for sky subtraction. The A and C frames were chosen to overlap at the foreground elliptical in the middle of the arc to enable precise alignment of the A and C frames. We observed for 8 ABC sequences with the $H$-band filter. Data reduction was carried out using the \verb'gemini' package in {\sc iraf} of both the science frames and standard stars observed on the same nights and with the same setup as the science observations. Each cube was then individually flux-calibrated and combined as for the SINFONI observations described above.

One further target, the spiral galaxy at $z = 1.49$ lensed by the cluster MACS J1149.5+2223 (MACS1149-arcA1.1), was observed with Keck/OSIRIS. The observations and data reduction are described by \citet{2011ApJ...732L..14Y}, who use the data to show that the galaxy's metallicity gradient is strongly negative.

Throughout this paper we also make use of the sample of six lensed galaxies observed with Keck/OSIRIS from \citet{2010MNRAS.404.1247J}. They presented the resolved kinematics of these galaxies, which we adopt in the first half of this paper for comparison with our new data. They also demonstrated that individual star-forming clumps could be extracted from the data and discussed their properties in relation to local H{\sc ii} regions, deriving luminosity densities up to $100\times$ higher than local star-forming regions. So that they can be combined with the new data in a self-consistent manner, we undertake a new analysis of the clump properties in this sample in the latter half of this paper, but our results are entirely consistent with those presented by \citet{2010MNRAS.404.1247J}. One of these galaxies - MACS0451-system7, observed in H$\alpha$ by \citet{2010MNRAS.404.1247J} - is also included in the SINFONI sample, where we observed it in H$\beta$ and [O{\sc iii}]. The combined sample of lensed arcs thus comprises 17 galaxies at $1.28 < z < 3.72$.

\subsubsection{Unlensed samples}

For comparison purposes, we refer throughout this paper to three other IFU surveys of high-redshift galaxies, which are not lensed. For ease of reference, we summarise the main properties of these surveys as follows:

\begin{description}
\item \textbf{SINS+AO:} This is a subset of the Spectroscopic Imaging survey in the Near-infrared with SINFONI \citep[SINS;][]{2006Msngr.125...11F} sample which has been observed with adaptive optics \citep{2013ApJ...767..104N}. The sample comprises 38 star-forming galaxies selected via their UV or optical emission, and has $z \sim 2.2$. These galaxies sample the massive end of the star-forming `main sequence,' with stellar masses of $10^{9.2} - 10^{11.5}$\,M$_{\odot}$ and SFRs of $13 - 850$\,M$_{\odot}$yr$^{-1}$. The spatial resolution achieved with SINFONI+AO is $\sim 0\farcs 2$, equivalent to $1.7$\,kpc.
\item \textbf{SHiZELS:} A sample of nine galaxies at $0.84 < z < 2.23$ drawn from the High-$Z$ Emission Line Survey (HiZELS), observed with SINFONI+AO \citep{2012MNRAS.426..935S}. The spatial resolution of $0\farcs 1$ equates to $0.8$\,kpc at the sample's median redshift, $z = 1.47$. The sample covers a similar range of stellar masses $(M_{\ast} \sim 10^9 - 10^{11}$\,M$_{\odot})$ to the SINS sample, but with lower SFRs of $7 \pm 2$\,M$_{\odot}$yr$^{-1}$.
\item \textbf{WiggleZ:} This sample comprises 13 $z \sim 1.3$ galaxies selected from the WiggleZ Dark Energy Survey and observed with Keck/OSIRIS+LGSAO \citep{2011MNRAS.417.2601W,2012MNRAS.422.3339W}. The spatial resolution of these observations is $\sim 0\farcs 1$, or $0.8$\,kpc. The range of stellar masses covered is similar to the other two surveys at $10^{9.8}$\,M$_{\odot} < M_{\ast} < 10^{11.7}$\,M$_{\odot}$, and SFRs are in the range $30 - 200$\,M$_{\odot}$yr$^{-1}$.
\end{description}

\begin{figure*}
\includegraphics[height=\textwidth, angle=90]{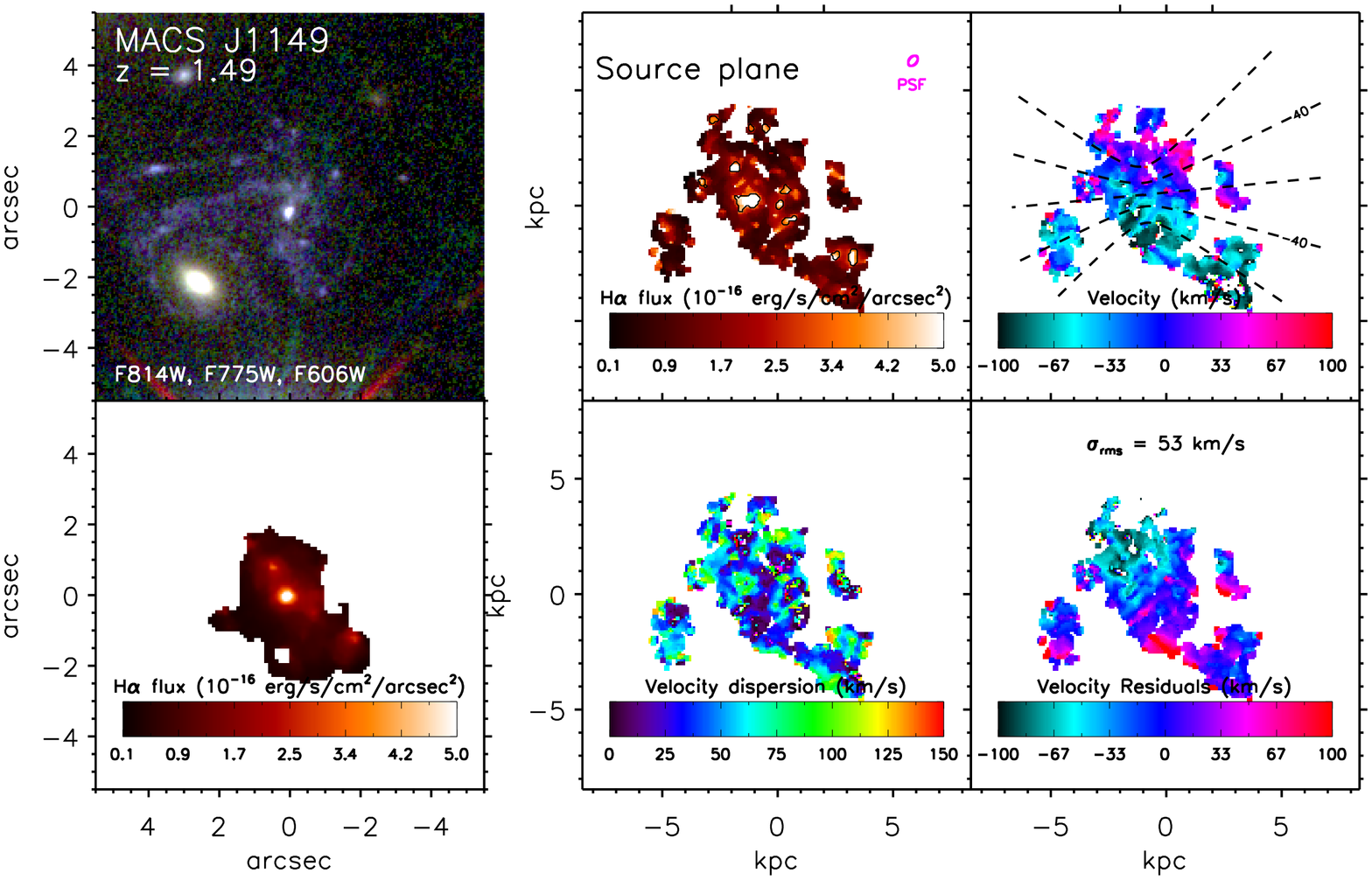}\\
\includegraphics[height=\textwidth, angle=90]{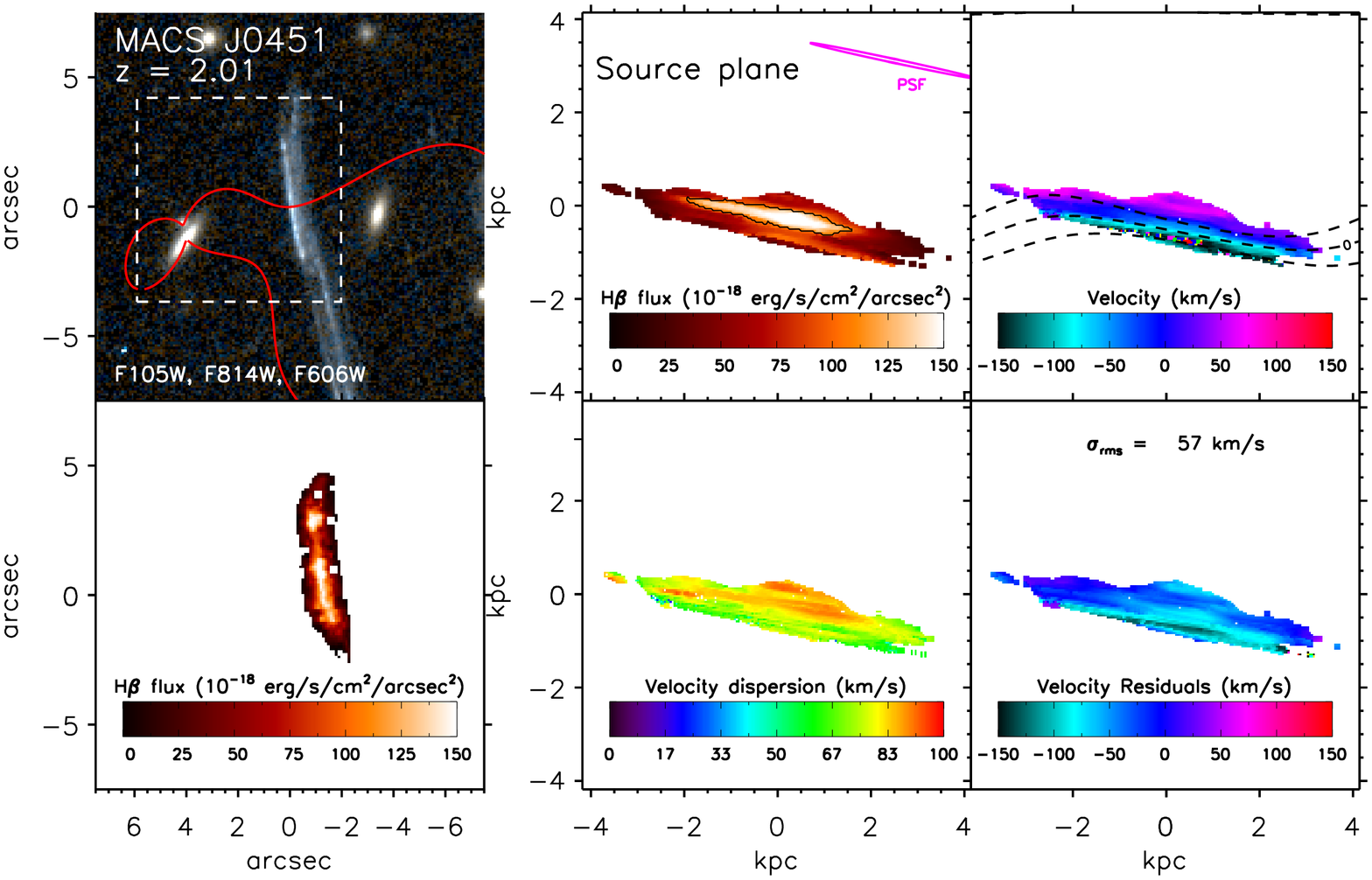}\\
\begin{center}
\textbf{Figure~\ref{fig:im}} (continued)
\end{center}
\end{figure*}
\begin{figure*}
\includegraphics[height=\textwidth, angle=90]{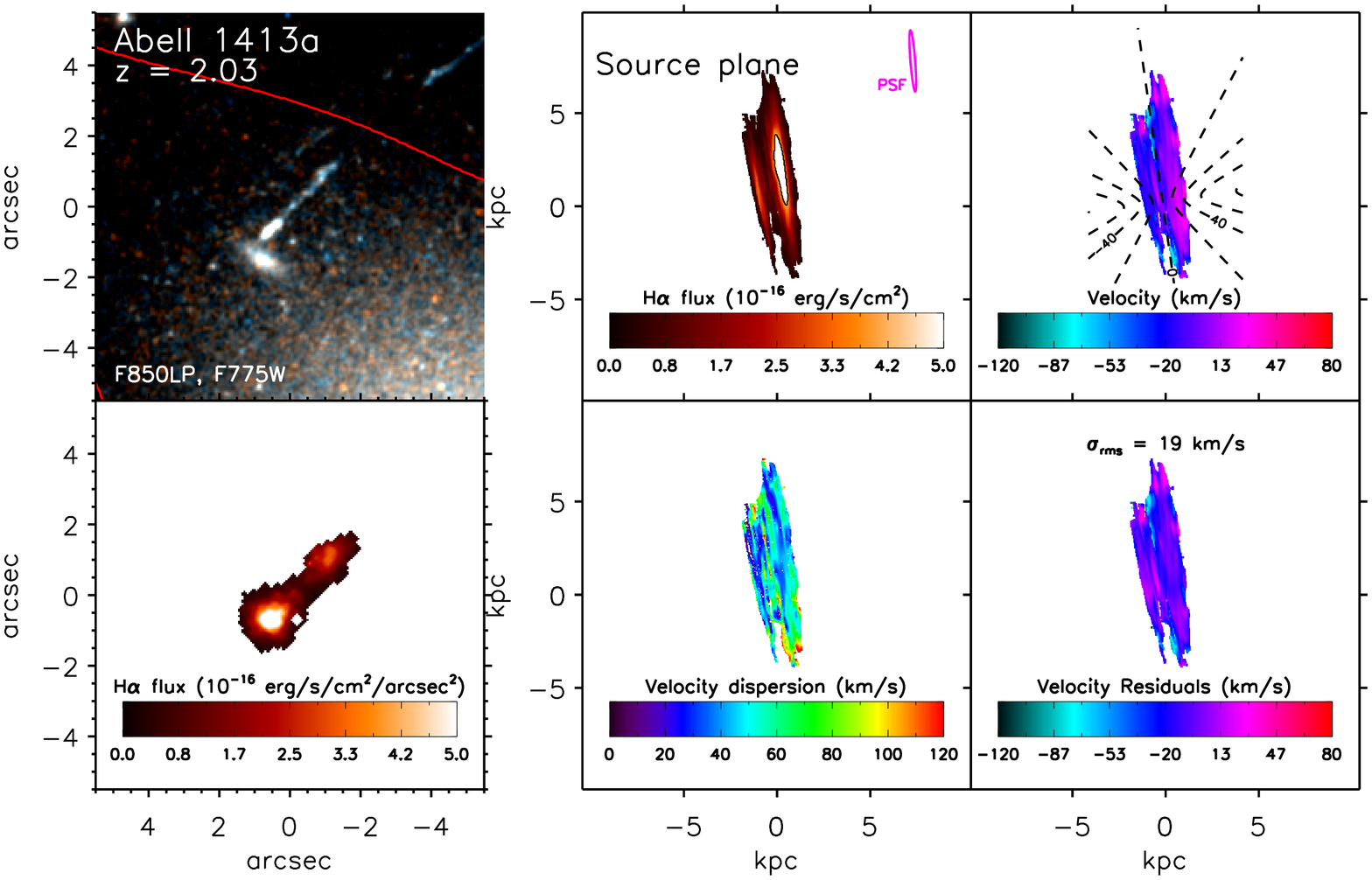}\\
\includegraphics[height=\textwidth, angle=90]{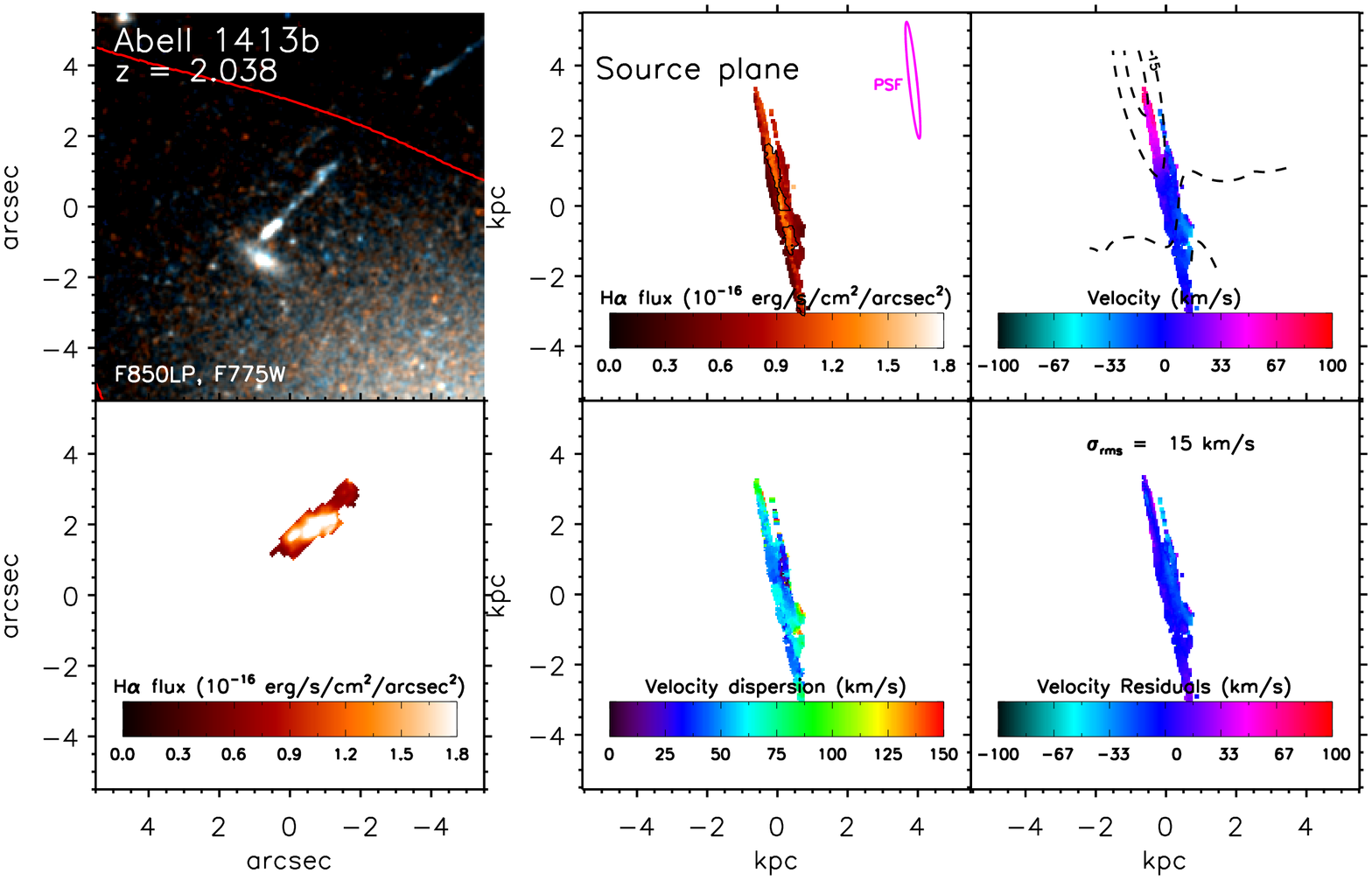}\\
\begin{center}
\textbf{Figure~\ref{fig:im}} (continued)
\end{center}
\end{figure*}
\begin{figure*}
\includegraphics[height=\textwidth, angle=90]{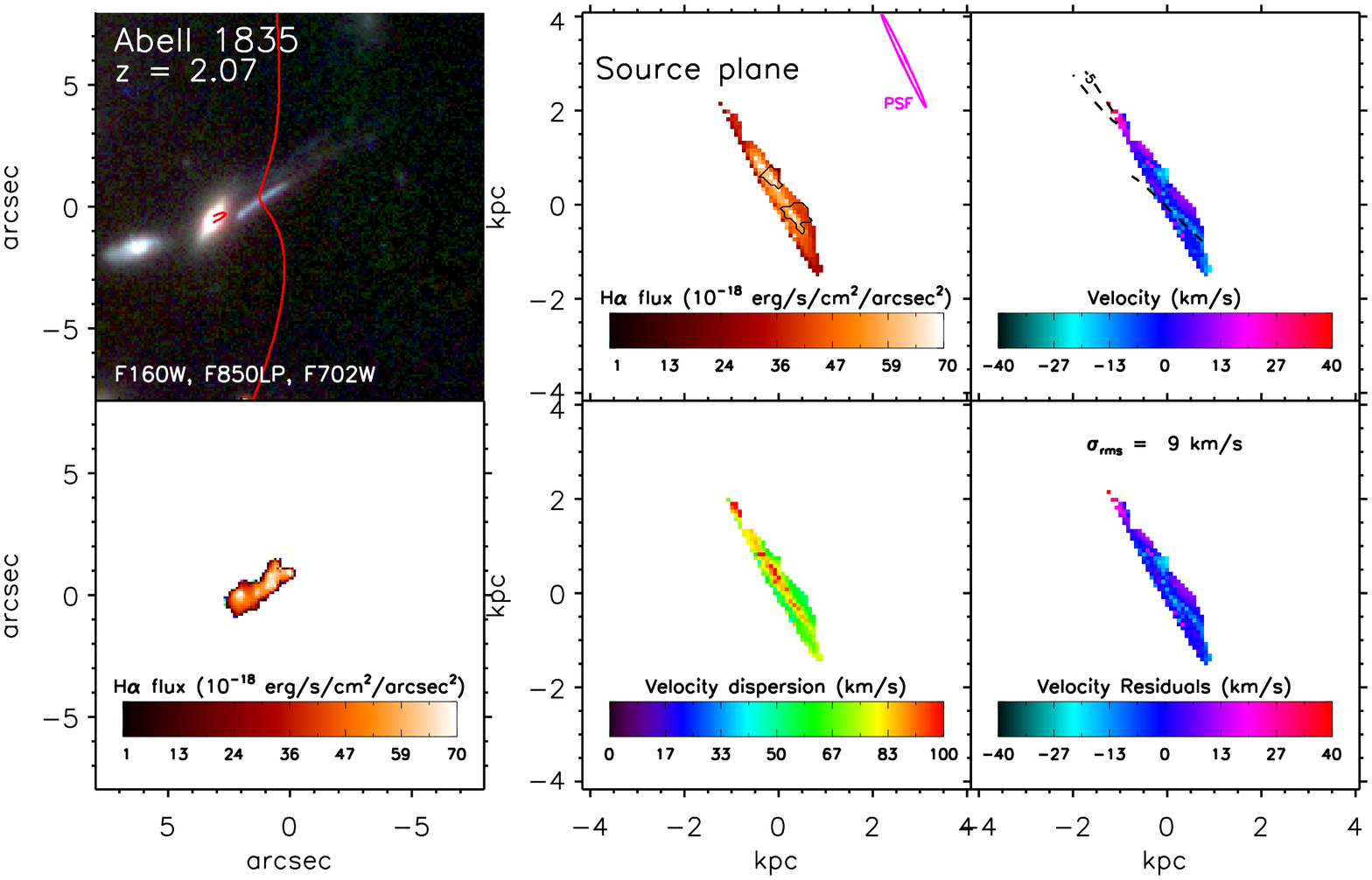}\\
\includegraphics[height=\textwidth, angle=90]{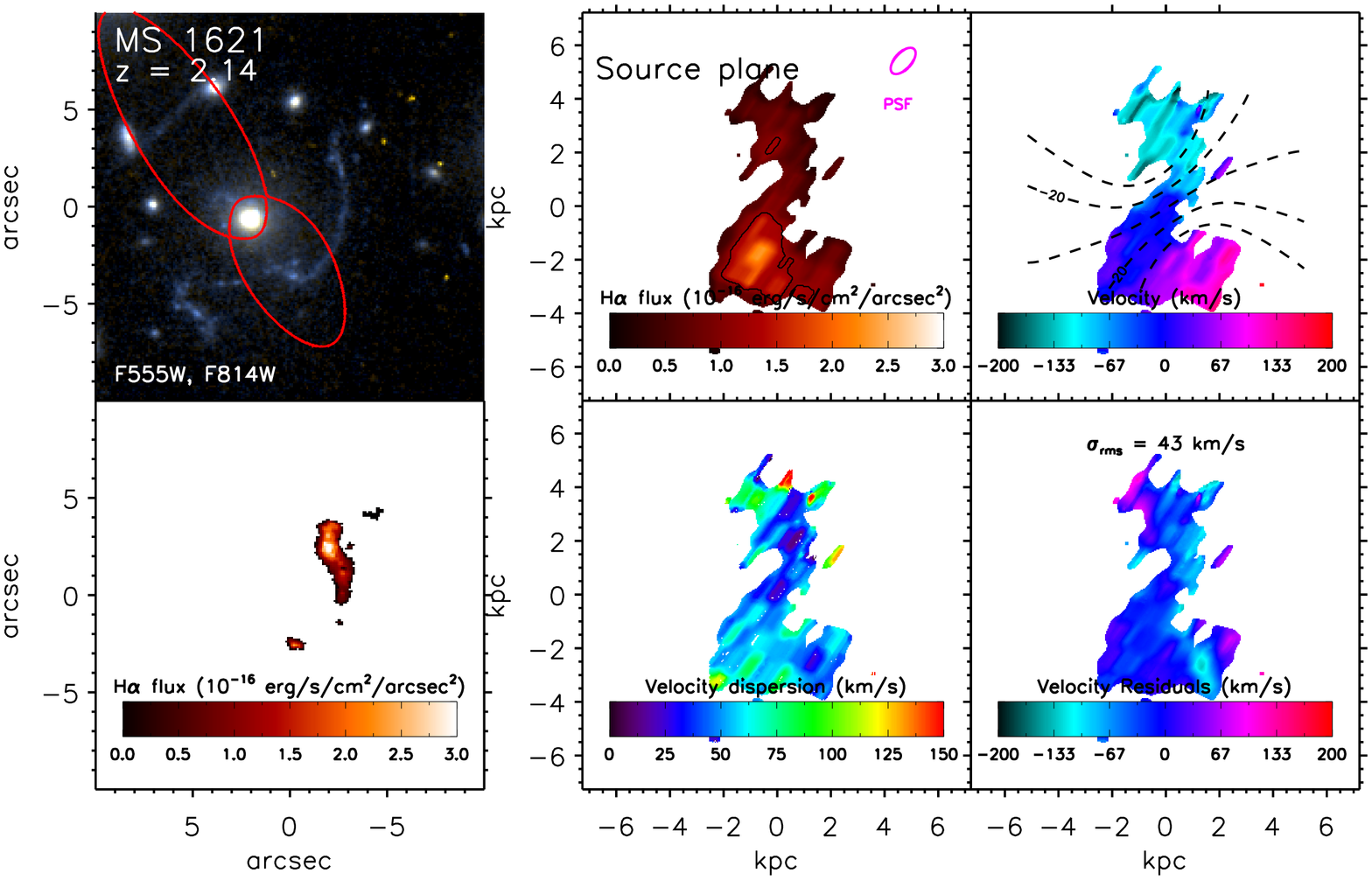}\\
\begin{center}
\textbf{Figure~\ref{fig:im}} (continued)
\end{center}
\end{figure*}
\begin{figure*}
\includegraphics[height=\textwidth, angle=90]{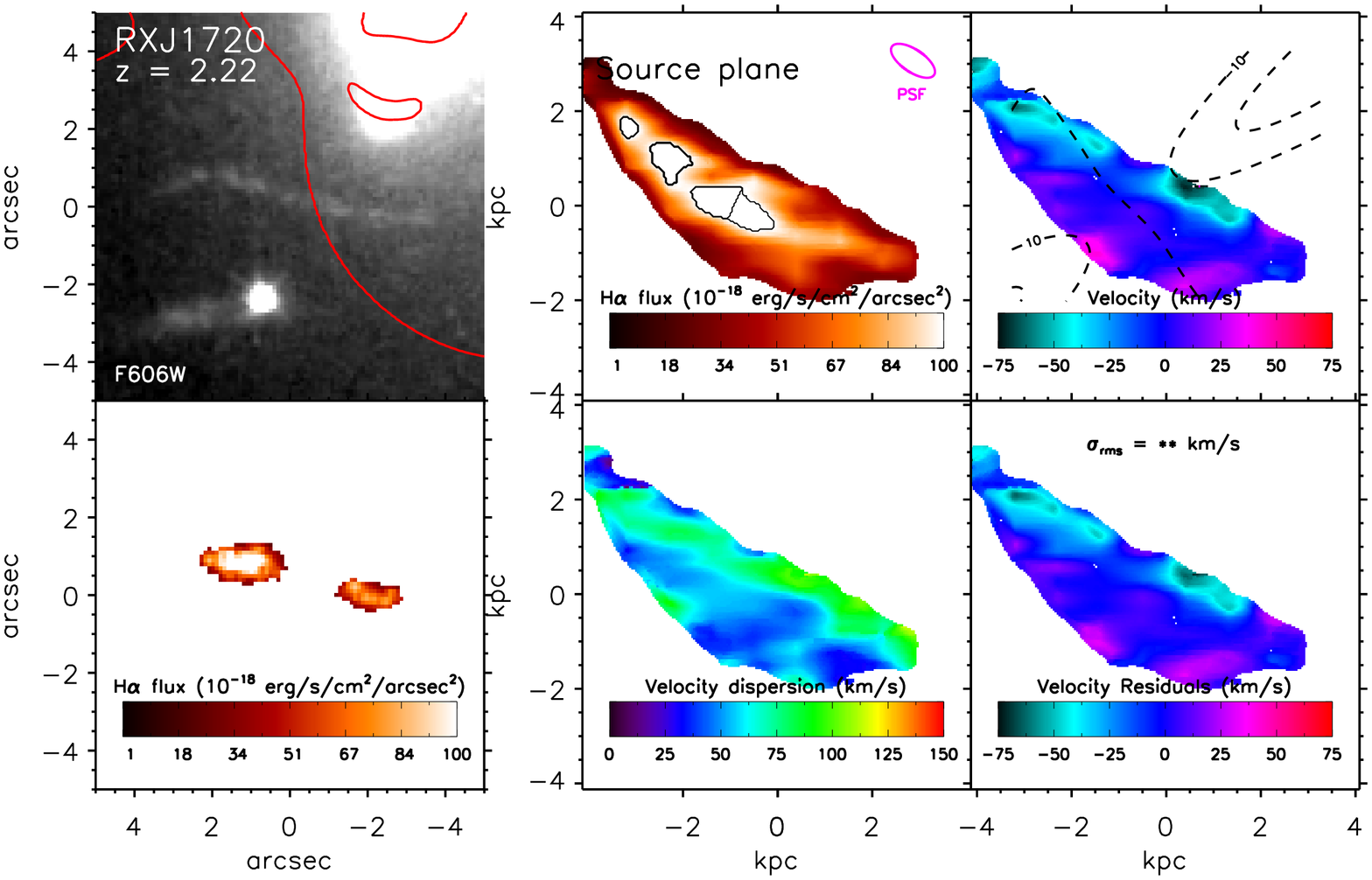}\\
\includegraphics[height=\textwidth, angle=90]{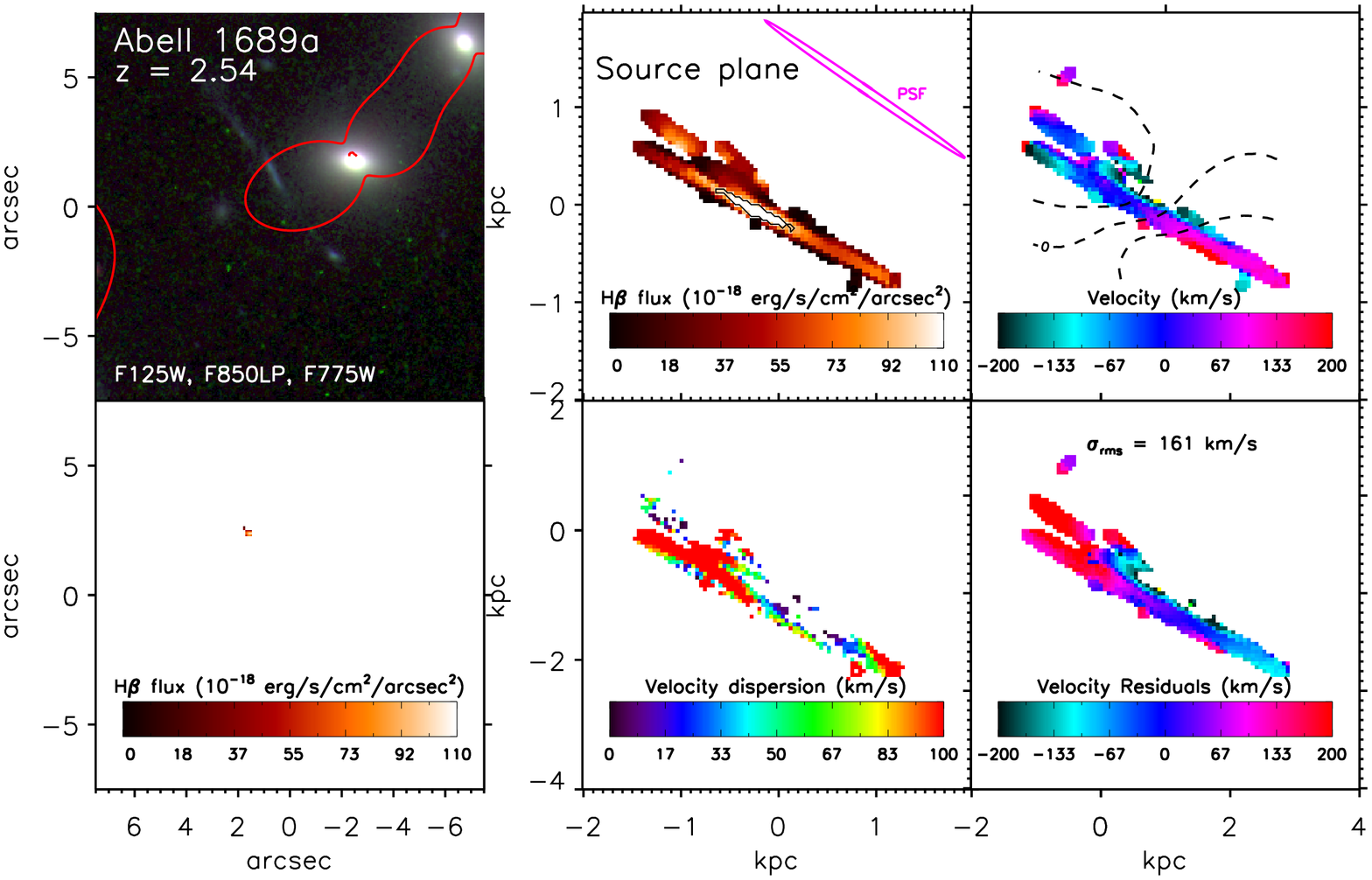}\\
\begin{center}
\textbf{Figure~\ref{fig:im}} (continued)
\end{center}
\end{figure*}
\begin{figure*}
\includegraphics[height=\textwidth, angle=90]{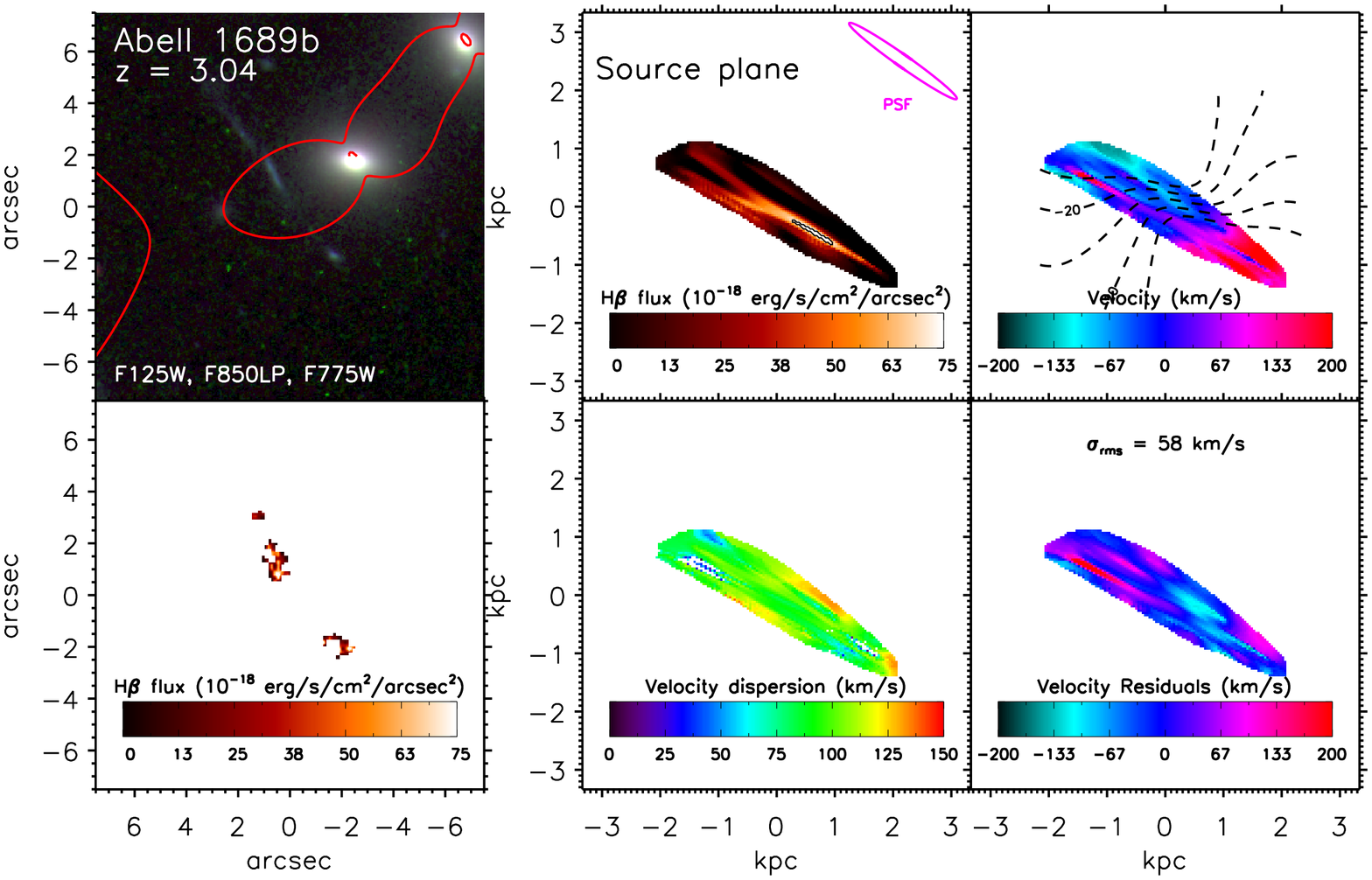}\\
\includegraphics[height=\textwidth, angle=90]{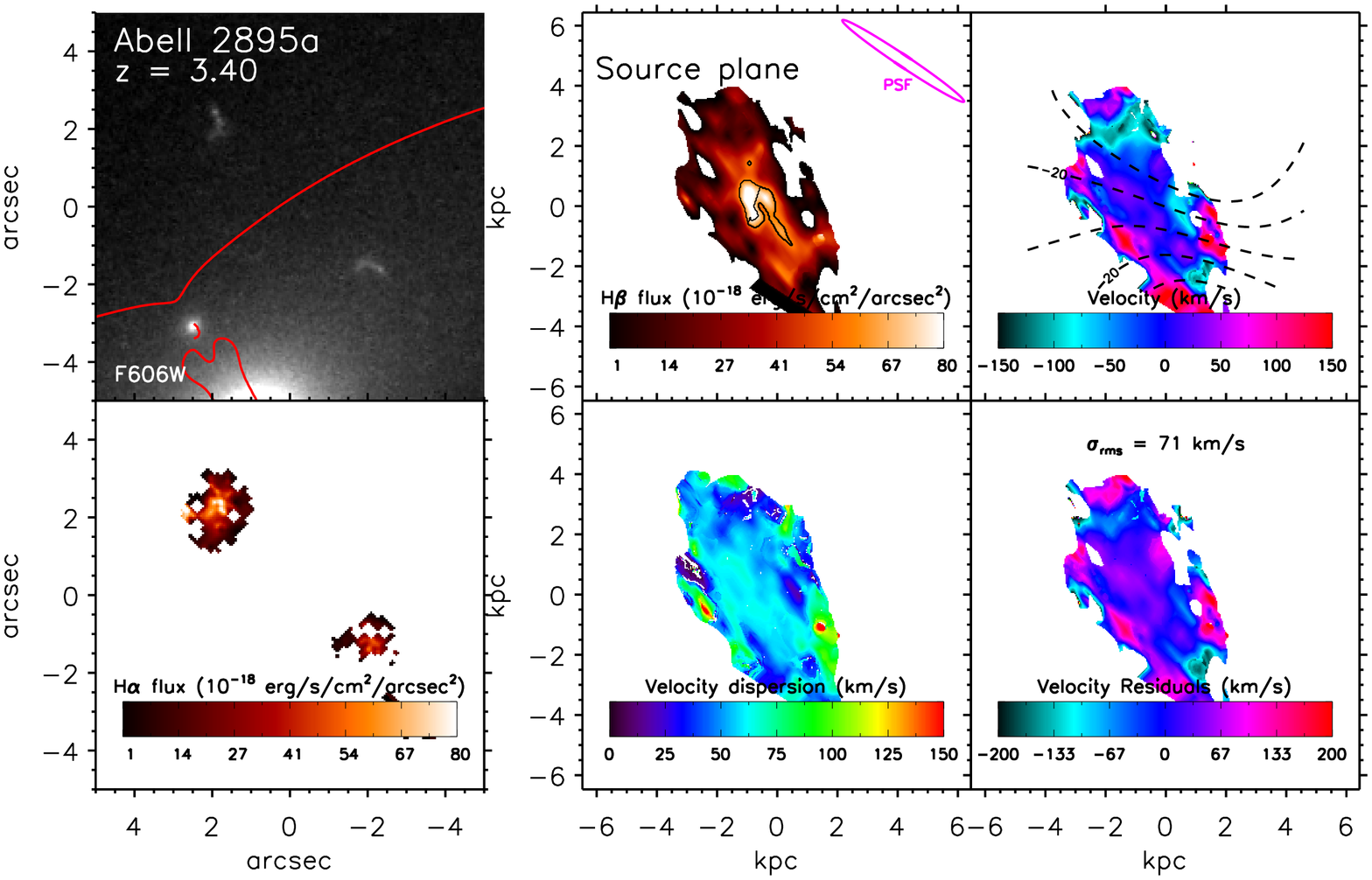}\\
\begin{center}
\textbf{Figure~\ref{fig:im}} (continued)
\end{center}
\end{figure*}
\begin{figure*}
\includegraphics[height=\textwidth, angle=90]{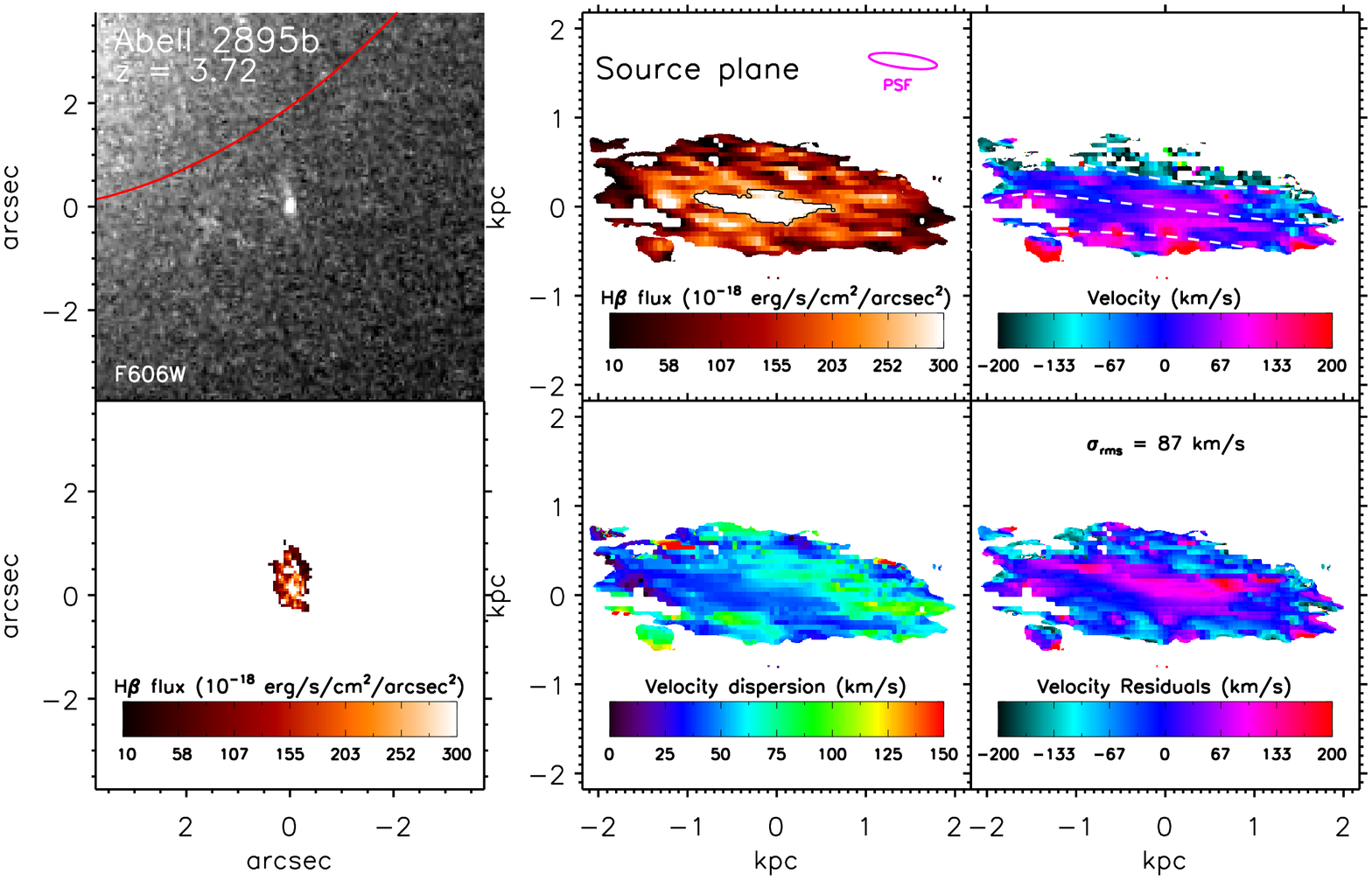}\\
\begin{center}
\textbf{Figure~\ref{fig:im}} (continued)
\end{center}
\end{figure*}

\subsection{\emph{HST} imaging and lens modelling}

\begin{table*}
  \caption{Gravitational lens properties of the sample.}
  \label{tab:lens}
  \begin{tabular}{l c c c l}
    \hline
    Name & $\mu$ & $\mu_{\rm{lin}}$ & Resolution (pc) & Lens model reference \\
 & $(a)$ & $(b)$ & $(c)$ & \\
    \hline
MACS0744-system3 & 63 $\pm$ 22 & 21 $\times$ 3 & 134 $\pm$ 47 &  \citet{2010MNRAS.404.1247J,2011ApJ...734...10K} \\
MACS1149-arcA1.1 & 16 $\pm$ 2 & 5 $\times$ 3 & 169 $\pm$ 21 & \citet{2009ApJ...707L.163S} \\
MACS0451-system7 & 90 $\pm$ 7 & 63 $\times$ 1 & 84 $\pm$ 7 & \citet{2010MNRAS.404.1247J} \\
A1413-arc2.1a & 24 $\pm$ 3 & 18 $\times$ 1 & 299 $\pm$ 37 & \citet{2010MNRAS.404..325R} \\
A1413-arc2.1b & 29 $\pm$ 2 & 21 $\times$ 1 & 265 $\pm$ 18 & \citet{2010MNRAS.404..325R} \\
A1835-arc7.1 & 63 $\pm$ 3 & 37 $\times$ 2 & 118 $\pm$ 6 & \citet{2010MNRAS.404..325R} \\
MS1621+26-system1 & 8 $\pm$ 2 & 4 $\times$ 2 & 722 $\pm$ 165 & Richard et~al. (in prep) \\
RXJ1720+26-arc1.1+1.2 & 86 $\pm$ 18 & 14 $\times$ 6 & 560 $\pm$ 115 & \citet{2010MNRAS.404..325R} \\
A1689-arc2.1 & 15 $\pm$ 3 & 21 $\times$ 1 & 96 $\pm$ 19 & \citet{2007ApJ...668..643L} \\
A1689-arc1.2 & 61 $\pm$ 11 & 25 $\times$ 2 & 249 $\pm$ 47 & \citet{2007ApJ...668..643L} \\
A2895-arc1.1+1.2 & 7 $\pm$ 1 & 10 $\times$ 1 & 454 $\pm$ 88 & May et~al. (in prep) \\
A2895-arc2.2 & 9 $\pm$ 2 & 7 $\times$ 1 & 167 $\pm$ 40 & May et~al. (in prep) \\
\hline 
Cl0024-arc1.1 & 1 $\pm$ 0 & 2 $\times$ 1 & 508 $\pm$ 55 & \citet{2010MNRAS.404.1247J} \\
Cl0949-arc1 & 7 $\pm$ 2 & 6 $\times$ 1 & 138 $\pm$ 38 & \citet{2010MNRAS.404.1247J} \\
MACS0712-system1 & 28 $\pm$ 8 & 19 $\times$ 1 & 40 $\pm$ 11 & \citet{2010MNRAS.404.1247J} \\
MACSJ0744-arc1 & 16 $\pm$ 3 & 9 $\times$ 2 & 83 $\pm$ 15 & \citet{2010MNRAS.404.1247J} \\
CosmicEye & 28 $\pm$ 3 & 8 $\times$ 3 & 96 $\pm$ 10 & \citet{2010MNRAS.404.1247J} \\
\hline \\
\multicolumn{5}{l}{\begin{minipage}{\textwidth}Notes: $(a)$ $\mu$ is the total magnification of the source, derived from the ratio of the emission line flux in the image plane to that in the source plane. $\mu_{\rm{lin}}$ gives the linear magnification factors in the direction of greatest magnification and perpendicular to that axis. $(c)$ Resolution is given in the direction of greatest magnification. \end{minipage}}
\end{tabular}
\end{table*}

\emph{HST} colour images (where available) of the target galaxies are shown in the top-left panels of Figure \ref{fig:im}. In order to determine the intrinsic properties of the target galaxies, it is necessary to account for the effects of lensing. To do this, we use the best-fit cluster mass models, the details of which are given in Table \ref{tab:lens}. Where the cluster's critical line at the redshift of the lensed arc crosses the field, it is overlaid in red. The critical line denotes regions of theoretically infinite magnification, and arcs which cross this line are multiply imaged within the IFU field of view. To reconstruct the source-plane images of the galaxies, we use {\sc lenstool} \citep{1993PhDT.......189K,2007NJPh....9..447J} to ray-trace each pixel to its source-plane origin. After using a central pixel to obtain the position of the galaxy in the source plane, we construct a regular grid in the source plane, ray-trace each pixel to the image plane and obtain the spectrum in that position by interpolating between pixels in the IFU data cube. In the cases of A1835-arc7.1 and A2895-arc2.2, the high magnification gradient in the vicinity of the critical line causes this method to omit data from the image plane. For these galaxies, we therefore carry out the process in reverse by ray-tracing each pixel from the IFU data cube to the source plane. This results in an irregular grid in the source plane, which is gridded into square pixels using a delauney tesselation. Once the source plane cubes have been constructed, we apply conservation of surface brightness to each pixel to obtain the instrinsic source plane flux. The magnification factor $\mu$ for the galaxy given in Table \ref{tab:sample} is then the ratio of image- to source-plane flux measured from the H$\alpha$ or H$\beta$ emission lines.

With both methods, the source plane pixel scale is chosen so that each pixel in the IFU data cube is represented in the source plane, and we impose a lower limit of 0.001'' to ensure manageable file sizes. The pixel scale is therefore dictated by the direction or region of highest magnification. Gravitational lensing usually acts preferentially in one direction, and galaxies lying close to the critical line will experience strong magnification gradients. Hence, the source plane cubes are oversampled both in the direction of lower magnification and in regions of the galaxy that lie furthest from the critical line. To estimate the actual resolution achieved, we apply the same reconstruction as described above to the standard star observations and measure the FWHM in the image and source plane. The result is an ellipse, shown in magenta in Figure \ref{fig:im}. We remind the reader that while lensing brings a major improvement in the spatial resolution that can be achieved, this gain is not without its problems. For example, in many systems the magnification is much higher in one direction than the other, leading to a distorted image of the galaxy. While we find that reconstruction of the original image is not problematic for the lenses considered here, this limitation should be born in mind when assessing the overall morphology of the source. The effective PSFs shown in Figure \ref{fig:im} illustrate the extent of anisotropy in each system, and we give values for the magnification in each direction in Table \ref{tab:lens}.

In order to estimate the errors on the magnification factors for each galaxy, we use the family of 100 best-fit lens models. We use {\sc lenstool} to reconstruct the H$\alpha$ or H$\beta$ intensity maps for each galaxy for each one of the possible lens models, and then measure the ratio of image- to source-plane flux in each. The 1$\sigma$ deviation in the derived magnifications are given in Table \ref{tab:sample} as the error on the total magnification. As we measure the magnification for the purpose of this paper from the H$\alpha$ (or H$\beta$) line intensity (effectively a weighted mean), the largest errors are found where there is a large magnification gradient across the arc without strong constraints on the mass distribution of the primary lensing source. The largest fractional error, 35\%, is found in MACS0744-system3, due to the strong magnification gradient across the image arising from lensing by the foreground cluster galaxy positioned along the line of sight to the arc. The smallest error, 5\%, is found in A1835-arc7.1, which also has a strong magnification gradient, but in this case the two images straddling the critical line provide good constraints on the precise position of the line. We note that changing the lens model tends to result in the galaxy image being stretched differently, so that the effective resolution described above would change, but the underlying morphology cannot be materially altered. Crucially for the discussion in the latter half of this paper, new features such as clumps cannot be created during the lensing recontruction.

\subsection{Dynamical maps}

With both the image- and source-plane cubes, we fit the intensity, velocity and velocity dispersion of the emission lines in each pixel using a $\chi^2$ minimisation technique. In each pixel, we simultaneously fit the H$\alpha$ and [N{\sc ii}]$\lambda$6583 or H$\beta$ and [O{\sc iii}]$\lambda\lambda$4959,5007 lines. To reduce the number of parameters in the fit, we require all lines to be of the same velocity and velocity dispersion, and we impose a ratio of [O{\sc iii}]$\lambda\,5007/\lambda\,4959 = 3$ \citep{2000MNRAS.312..813S}. The fit is accepted if it results in a $\Delta \chi^2 > 25$ compared to a straight-line fit, equivalent to a signal-to-noise of $S/N > 5$. If no fit is obtained, we adaptively bin up to 3$\times$3 pixels in the image plane, or the equivalent area of 3$\times$3 PSF areas in the source plane. We deconvolve the velocity dispersion for spectral resolution by subtracting in quadrature the median width of sky lines measured from blank exposures taken with the standard stars. The high spatial resolution of the source-plane cube means that most of the contribution to the velocity dispersion of the velocity gradient across the galaxy is removed. To remove the remainder, we subtract in quadrature the velocity gradient across each pixel, measured over the PSF. For each parameter, we also estimate the errors in each pixel. Taking the best fit to the emission lines, we vary each parameter in turn, while allowing the others to find their new minima, until we obtain a $\Delta \chi^2 \geq 1$.

\begin{table*}
  \caption{Dynamical properties of the sample}
  \label{tab:dyn}
  \begin{tabular}{l r r r r r r c}
    \hline
    Name & Inclination & PA & $v_{2.2}$ & $\sigma$ & $v/r$ & $v/\sigma$ & Disc/ Merger/ \\
         & $\theta$ & & (km\,s$^{-1}$) & (km\,s$^{-1}$) & (km\,s$^{-1}$kpc$^{-1}$) & & Undetermined\\
    \hline
MACS0744-system3 & 61 & 266 & 80 $\pm$ 10 & 60 $\pm$ 20 & 230 $\pm$ 30 & 1.4 $\pm$ 0.8 & Disc \\
MACS1149-arcA1.1 & 45 & 86 & 59 $\pm$ 3 & 50 $\pm$ 10 & 88 $\pm$ 4 & 2 $\pm$ 1 & Disc \\
MACS0451-system7 & 55 & 100 & 100 $\pm$ 10 & 77 $\pm$ 9 & 220 $\pm$ 30 & 1.6 $\pm$ 0.6 & Disc \\
A1413-arc2.1a & 50 & 172 & 17 $\pm$ 4 & 50 $\pm$ 10 & 24 $\pm$ 5 & 0.5 $\pm$ 0.3 & Disc \\
A1413-arc2.1b & 66 & 249 & 20 $\pm$ 3 & 60 $\pm$ 20 & 8 $\pm$ 1 & 0.6 $\pm$ 0.3 & Merger \\
A1835-arc7.1 & 89 & 236 & 40 $\pm$ 10 & 70 $\pm$ 10 & 4 $\pm$ 1 & 0.2 $\pm$ 0.2 & Undetermined \\
MS1621-system1 & 84 & 247 & 126 $\pm$ 4 & 60 $\pm$ 20 & 47 $\pm$ 1 & 3 $\pm$ 1 & Merger \\
RXJ1720-arc1.1+1.2 & 70 & 144 & 63.2 $\pm$ 0.5 & 60 $\pm$ 20 & 27.1 $\pm$ 0.2 & 0.4 $\pm$ 0.2 & Merger \\
A1689-arc2.1 & 81 & 38 & 64 $\pm$ 7 & 55 $\pm$ 1 & 180 $\pm$ 20 & 1 $\pm$ 1 & Disc \\
A1689-arc1.2 & 80 & 234 & 78 $\pm$ 4 & 90 $\pm$ 10 & 53 $\pm$ 4 & 1.3 $\pm$ 0.4 & Merger \\
A2895-arc1.1+1.2 & 85 & 260 & 60 $\pm$ 10 & 60 $\pm$ 20 & 62 $\pm$ 3 & 1.9 $\pm$ 0.7 & Undetermined \\
A2895-arc2.2 & 47 & 276 & 200 $\pm$ 10 & 60 $\pm$ 20 & 530 $\pm$ 30 & 4 $\pm$ 2 & Disc \\
\hline
Cl0024-arc1.1 & 50 & 122 & 91 $\pm$ 8 & 63 $\pm$ 6 & 15 $\pm$ 3 & 2.6 $\pm$ 0.3 & Disc \\
Cl0949-arc1 & \ldots & \ldots & \ldots & 68 $\pm$ 7 & 16 $\pm$ 3 & \ldots & Merger \\
MACS0712-system1 & 40 & 351 & 33 $\pm$ 3 & 75 $\pm$ 7 & 25 $\pm$ 3 & 0.81 $\pm$ 0.08 & Disc \\
MACSJ0744-arc1 & 45 & 124 & 118 $\pm$ 9 & 110 $\pm$ 10 & 210 $\pm$ 20 & 2.4 $\pm$ 0.2 & Disc \\
CosmicEye & 55 & 113 & 51 $\pm$ 4 & 44 $\pm$ 4 & 55 $\pm$ 5 & 1.9 $\pm$ 0.2 & Disc \\
\hline
\multicolumn{8}{l}{\begin{minipage}{\textwidth}Notes: Inclination and PA are derived from the best-fit disc models. $v_{2.2}$ is the velocity at $2.2r_{1/2}$, where $r_{1/2}$ is the half-light radius. Velocity dispersion, $\sigma$, is the luminosity-weighted mean local value, after deconvolving for local velocity gradient. The velocity gradient $v/r$ is measured at $r_{1/2}$. Targets below the line are from \citet{2010MNRAS.404.1247J} with values reported in their Table 2, except for $v_{2.2}$, $v/r$ and $v/\sigma$, which we measure for consistency with the new data. No values for $\theta$, PA or $v_{2.2}$ are given for Cl0949-arc1 as this galaxy cannot be fit by a disc model.\end{minipage}}
\end{tabular}
\end{table*}

We show the resulting maps in Figure \ref{fig:im}. In the bottom left of each panel is the H$\alpha$ or H$\beta$ emission line intensity from the image-plane IFU data, mapped to the same astrometry as the \emph{HST} image. The source plane reconstructions are in the right-four images of each panel, showing the H$\alpha$ or H$\beta$ emission line intensity, the velocity and velocity dispersion maps and the velocity residuals after subtracting the best-fit disc models (contoured over the velocity map and described in Section \ref{sec:dynamics}). We note that as the velocity and velocity dispersion are fixed between multiple emission lines, their values are dominated by the line with the highest signal-to-noise. In the case of the H$\beta$/[O{\sc iii}] observations, the [O{\sc iii}]$\lambda\,5007$ line is far brighter than the H$\beta$ line; in some cases we thus obtain velocity and velocity dispersion values in pixels where no H$\beta$ is measured. We note that the method above is the same as that used by \citet{2010MNRAS.404.1247J}, so the properties derived from the two samples are self-consistent.

\subsection{Disc modelling}
\label{sec:discmodel}

\begin{figure*}
\includegraphics[height=84mm, angle=90]{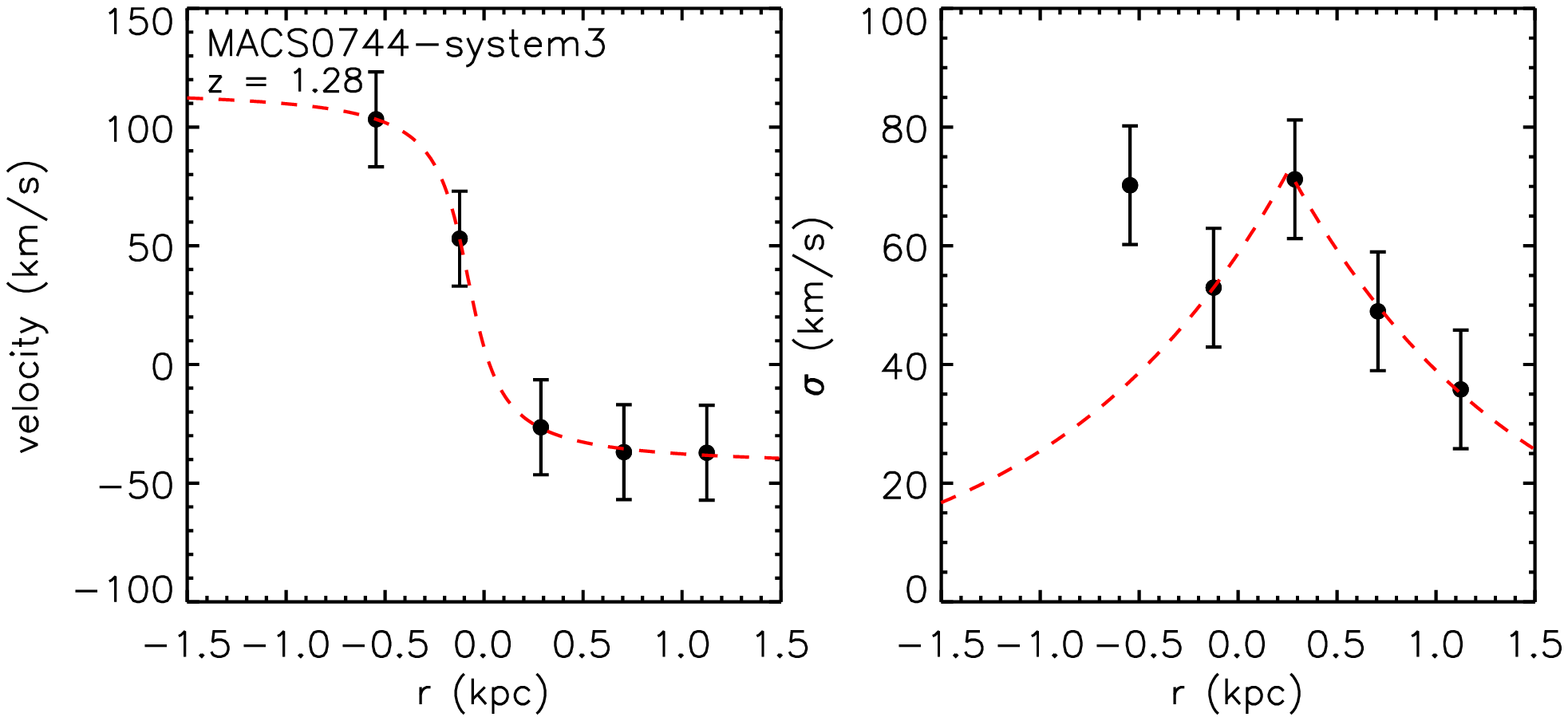} \includegraphics[height=84mm, angle=90]{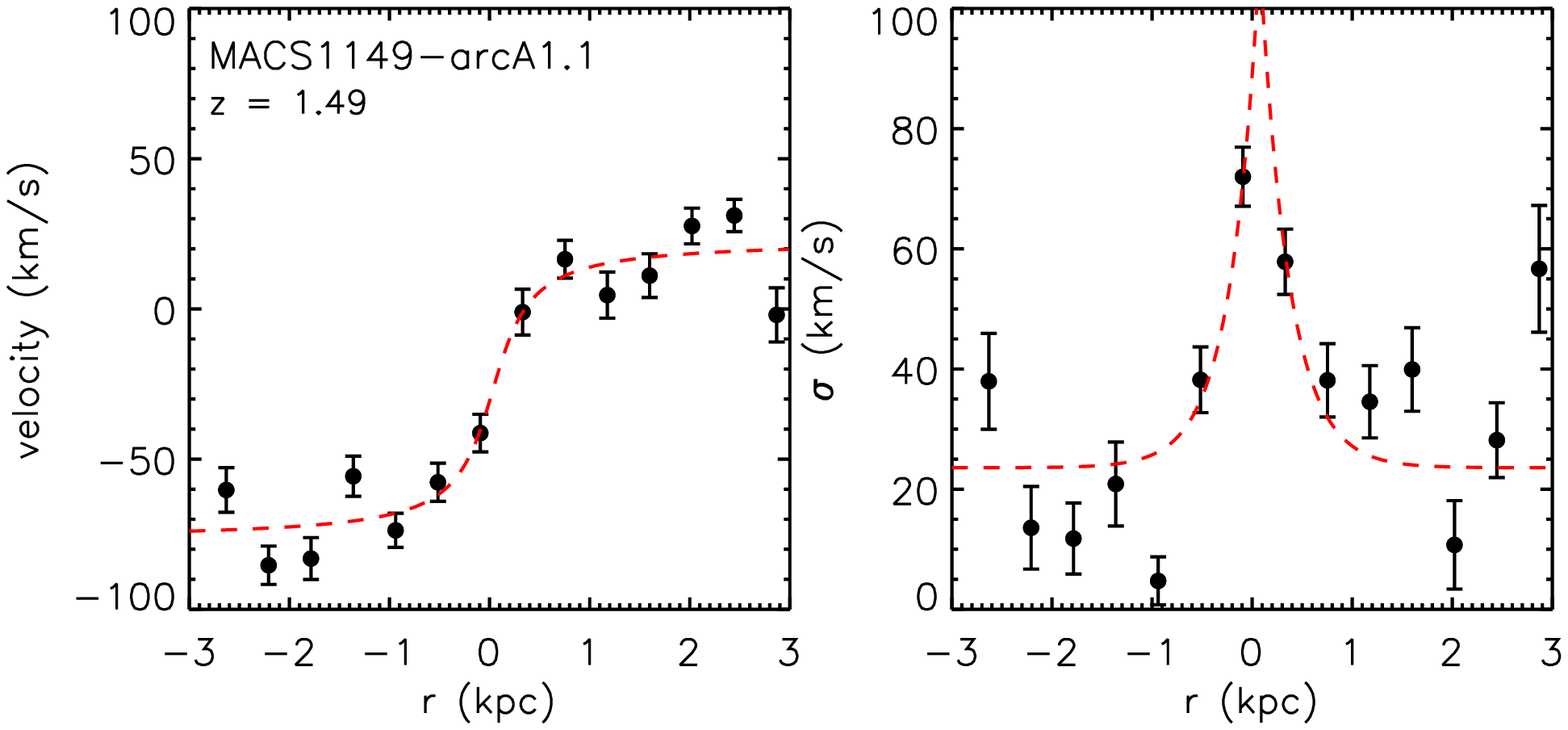}\\
\includegraphics[height=84mm, angle=90]{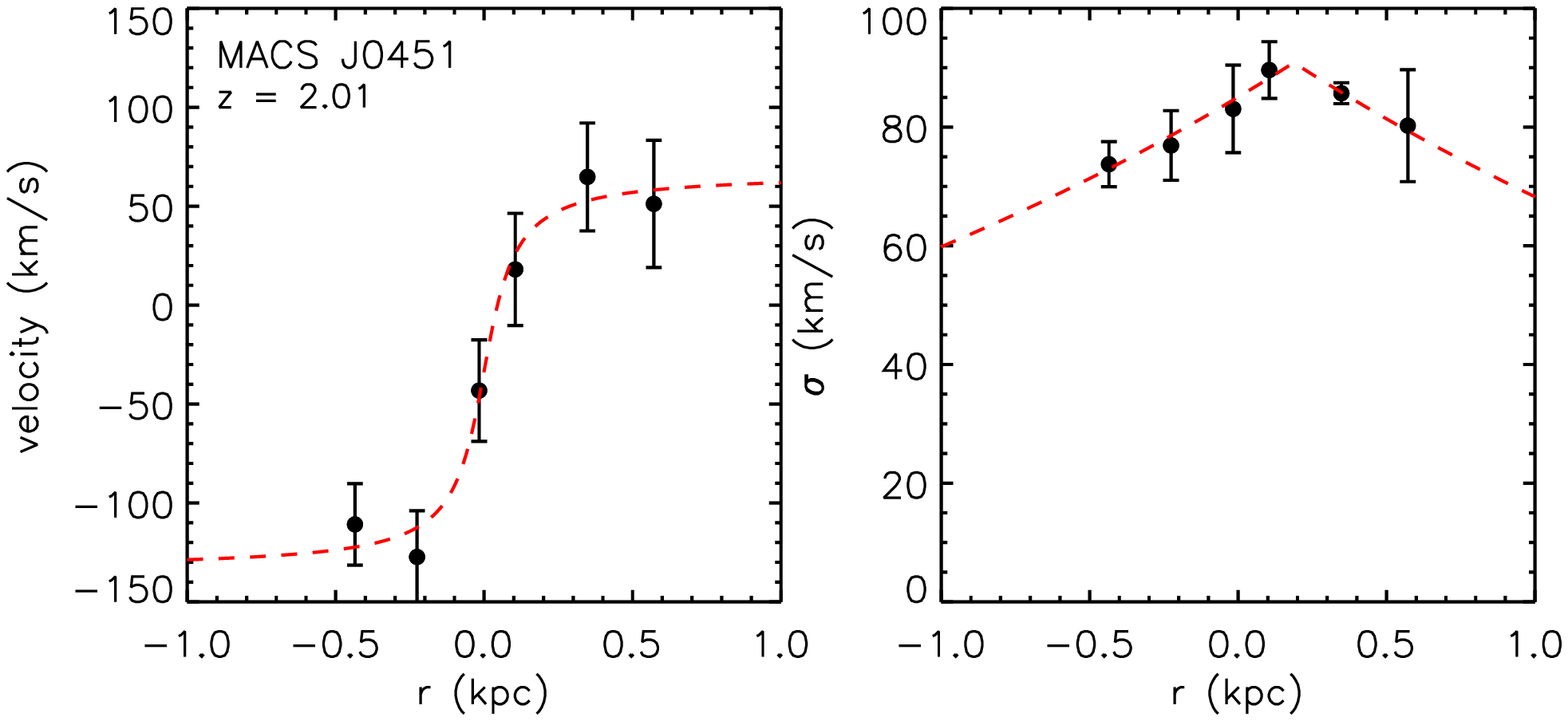} \includegraphics[height=84mm, angle=90]{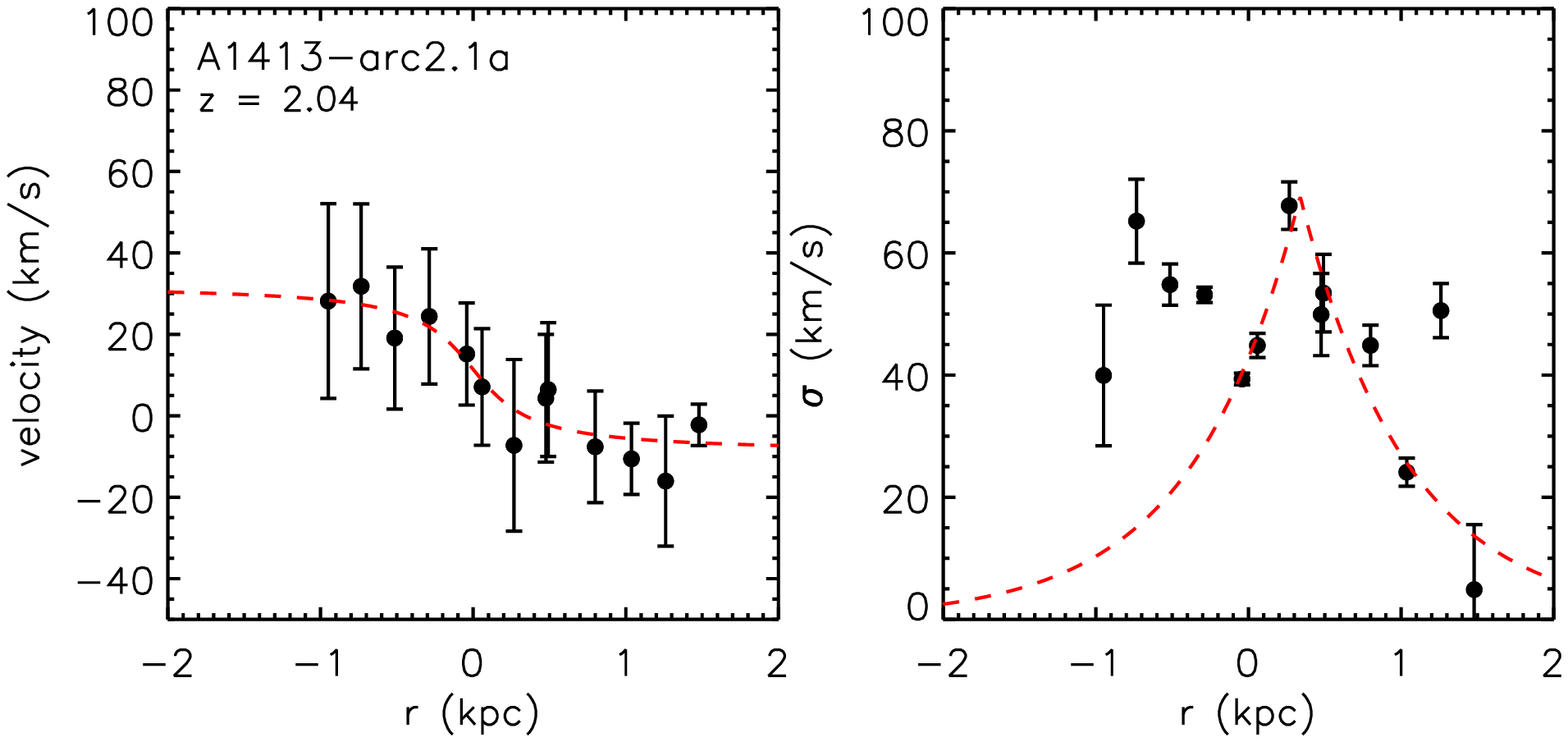}\\
\includegraphics[height=84mm, angle=90]{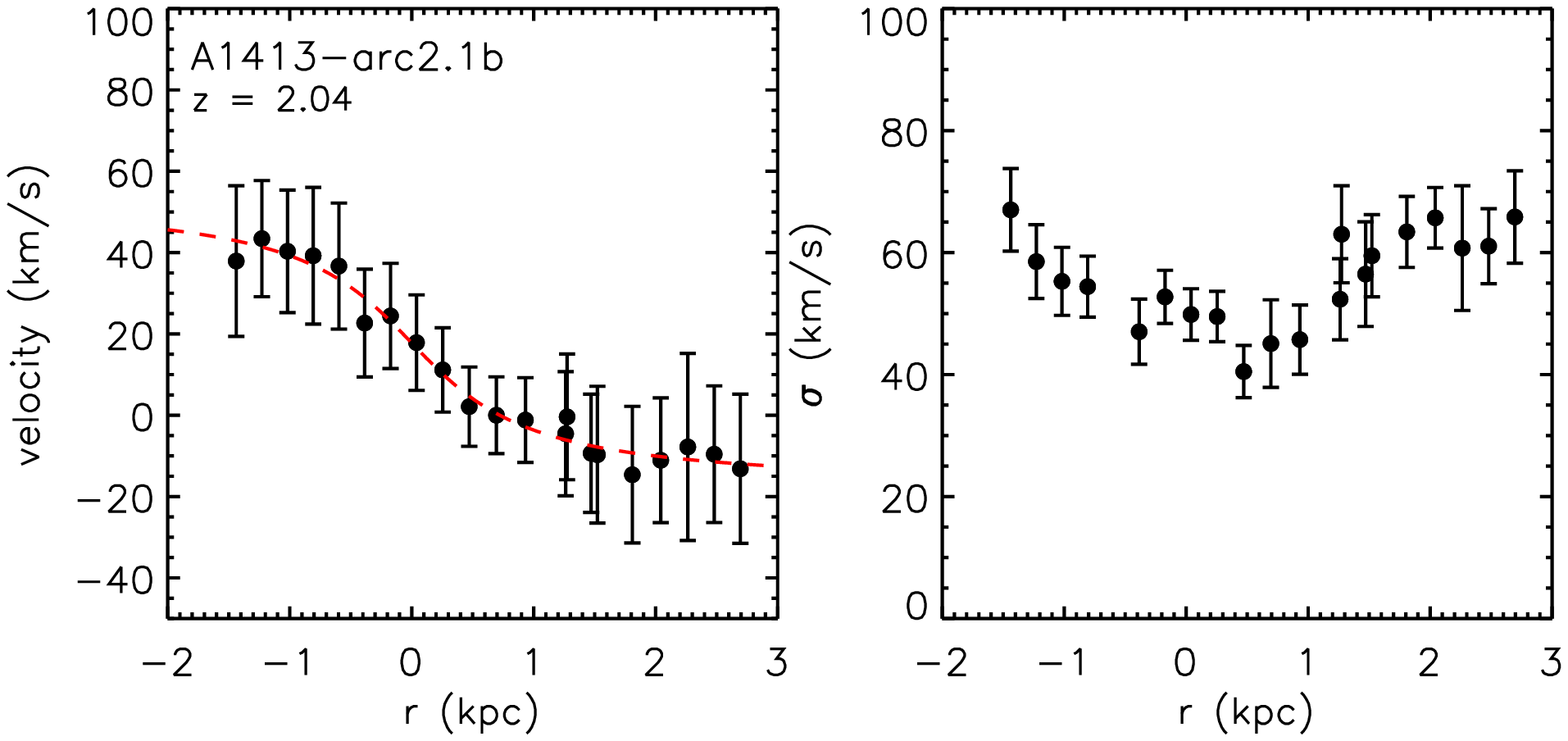} \includegraphics[height=84mm, angle=90]{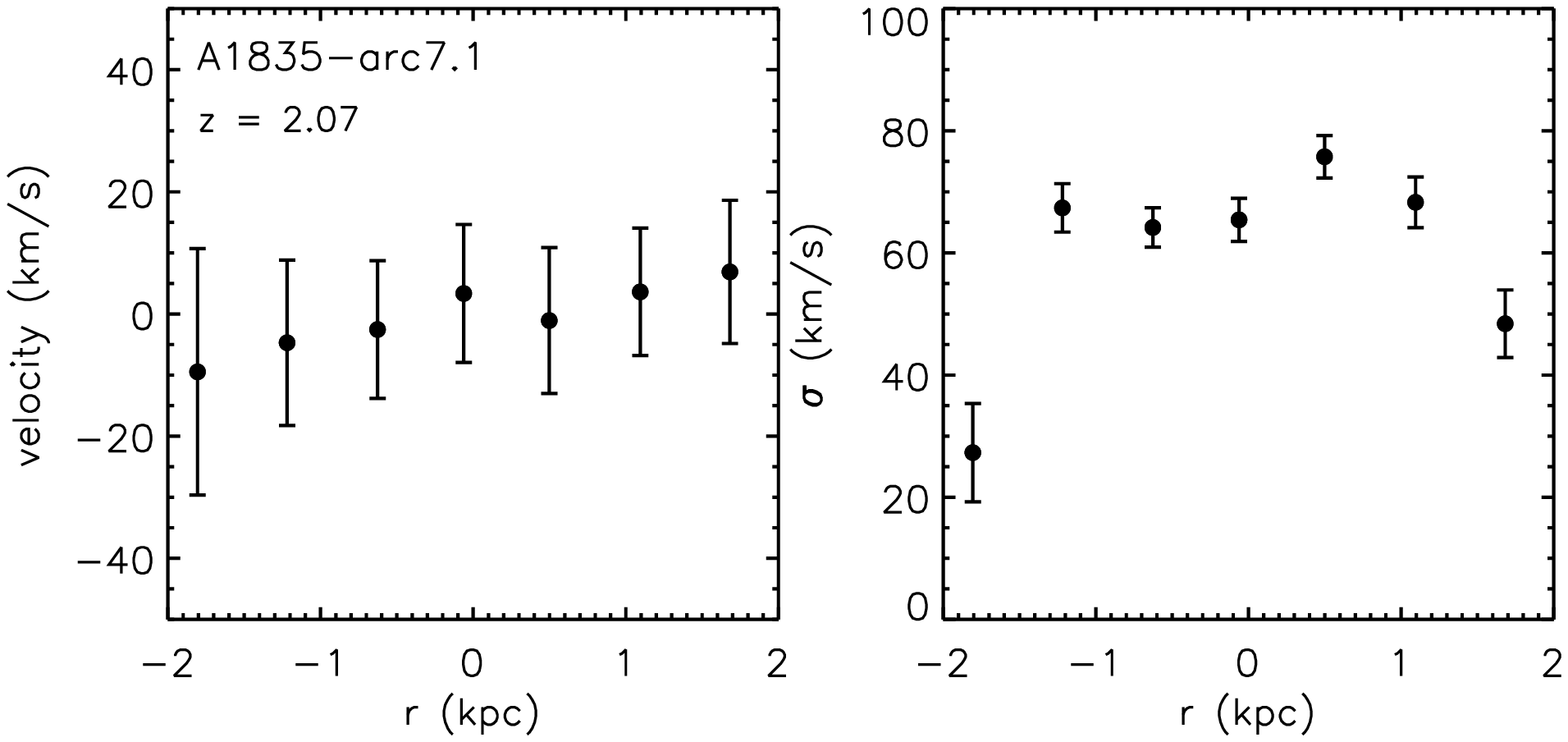}\\
\includegraphics[height=84mm, angle=90]{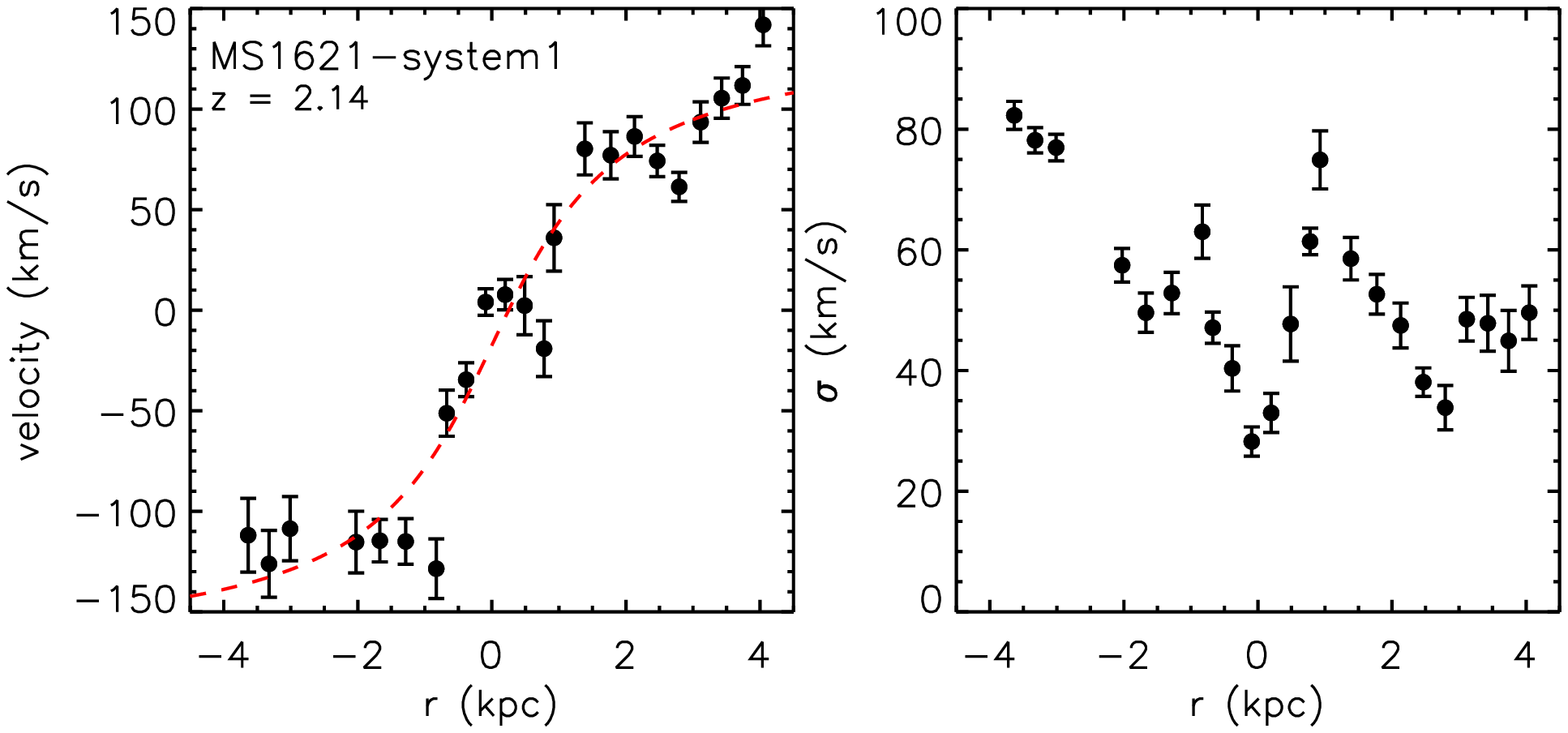} \includegraphics[height=84mm, angle=90]{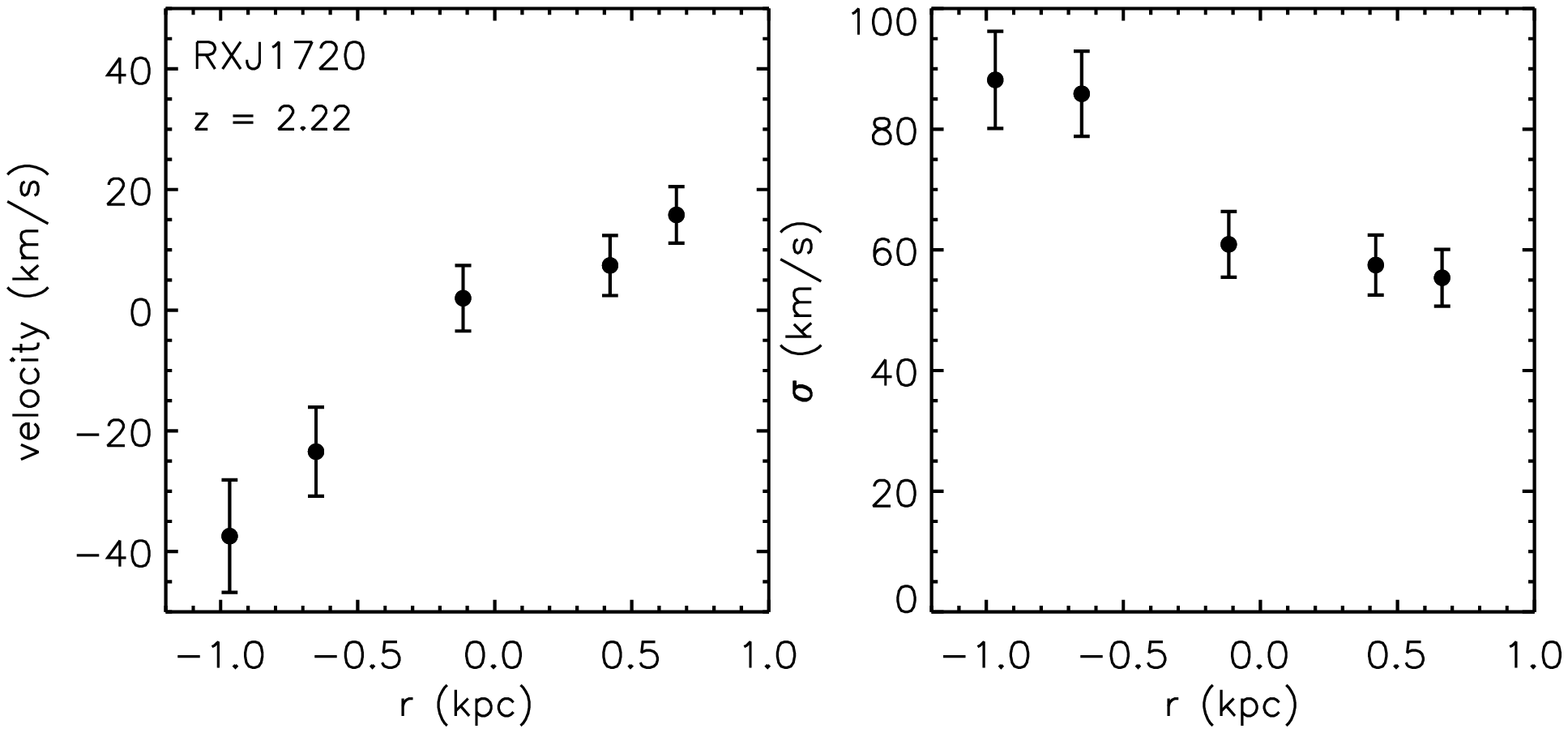}\\
\includegraphics[height=84mm, angle=90]{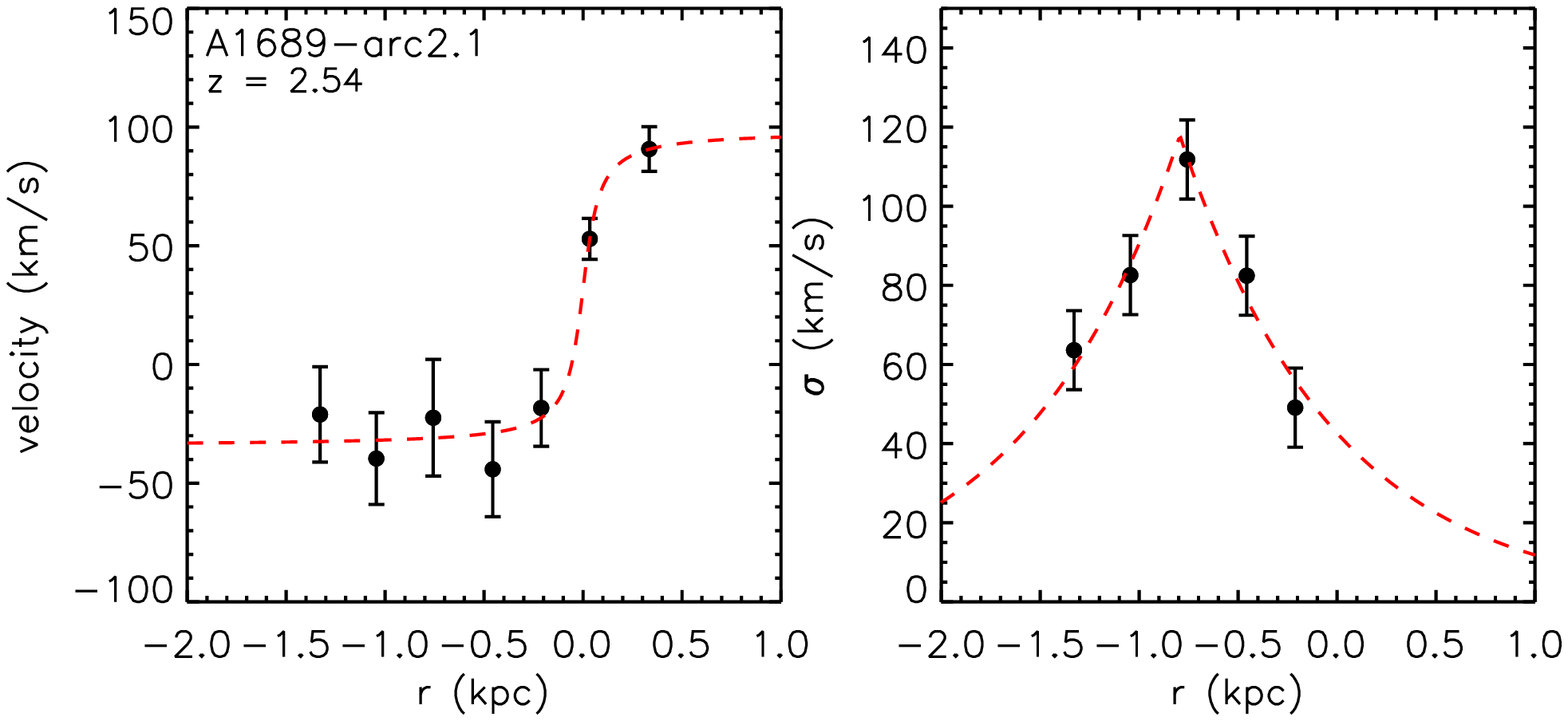} \includegraphics[height=84mm, angle=90]{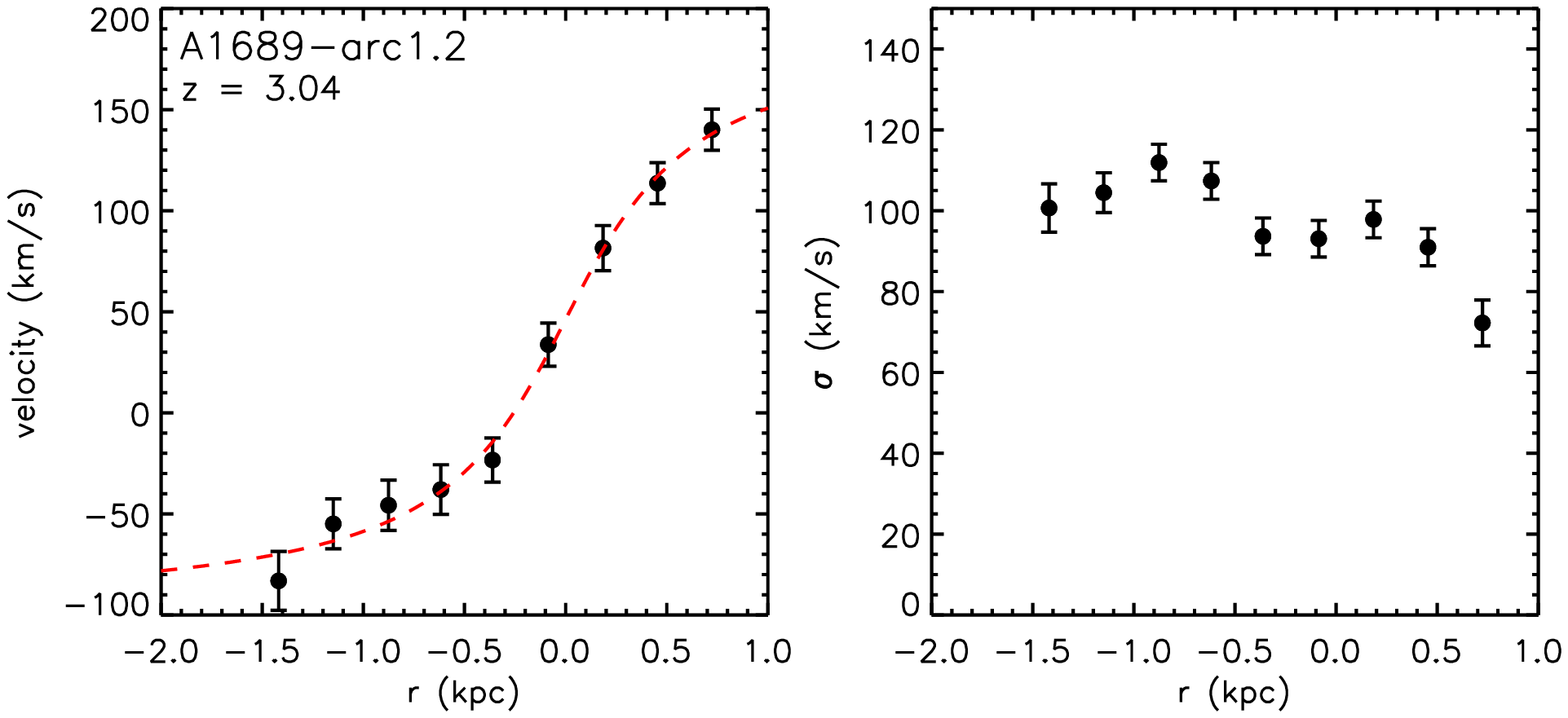}\\
\includegraphics[height=84mm, angle=90]{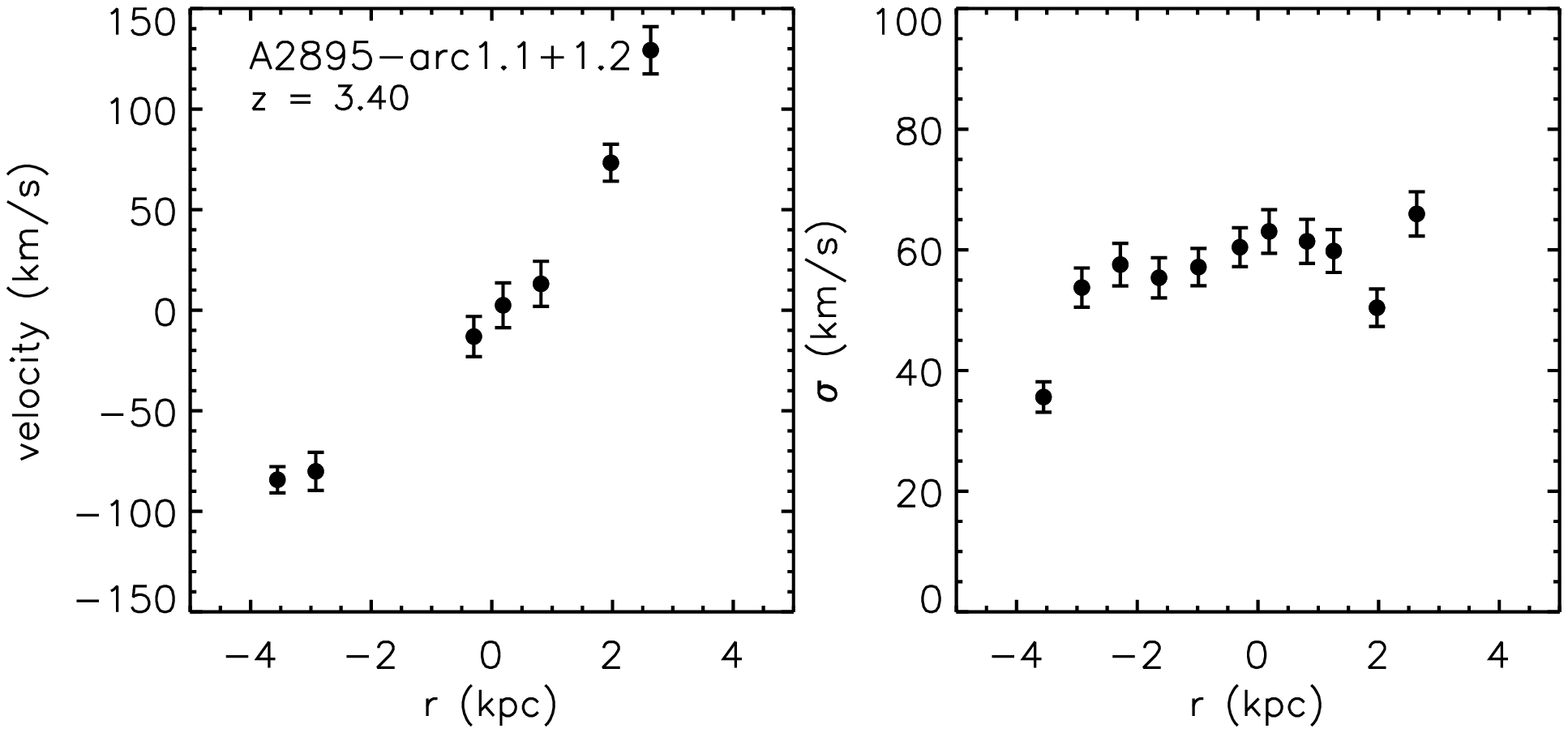} \includegraphics[height=84mm, angle=90]{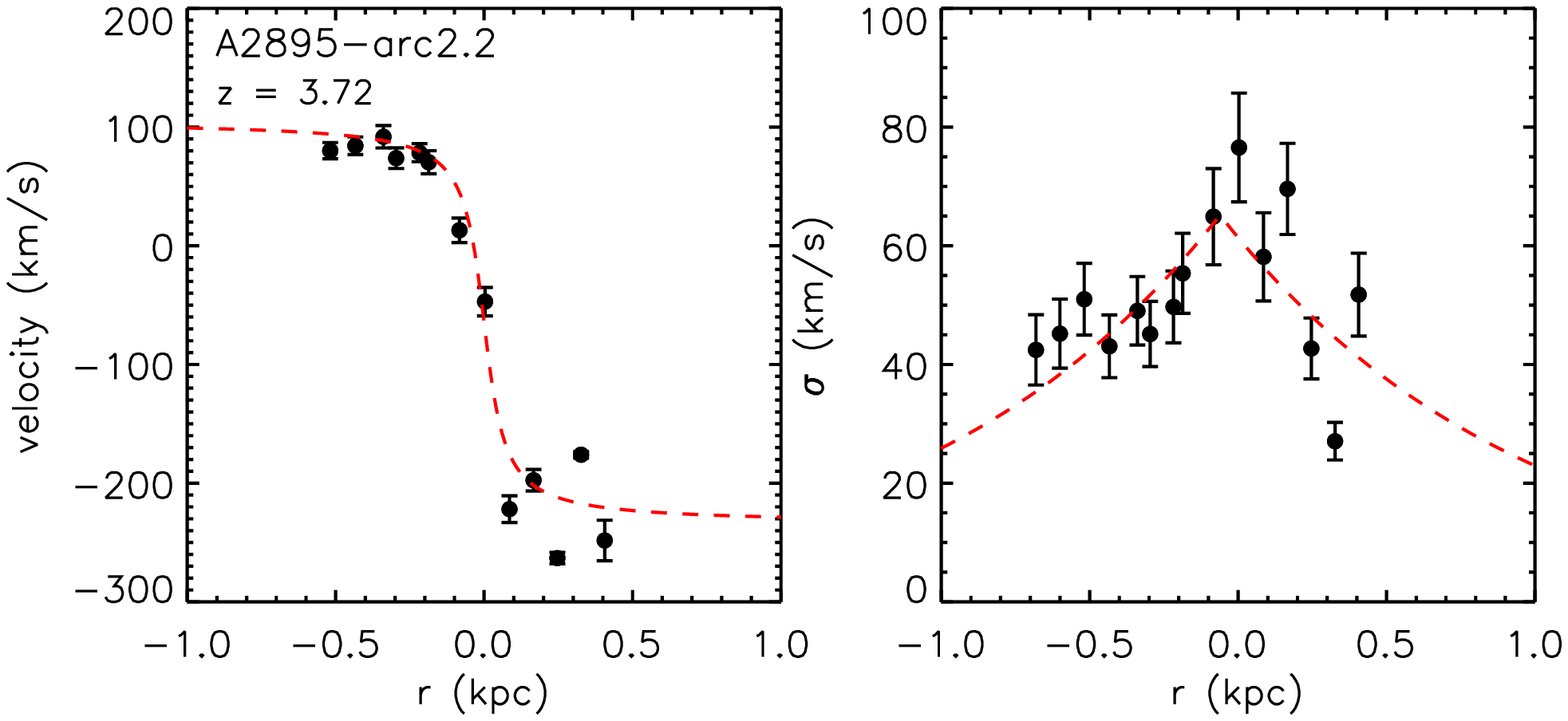}\\
\caption{One-dimensional rotation curves (left) and velocity dispersion profiles (right), extracted from a slit aligned along the axis of the best-fit disc model. For the galaxies whose dynamics are well fit by a rotating disc model, we overlay the best-fit rotation curve and velocity dispersion profile.}
\label{fig:profiles}
\end{figure*}

To each of the velocity fields, we follow \citet{2010MNRAS.404.1247J} and attempt to fit a rotating disc model and hence infer the true disc rotation speed and inclination. The model is described by six parameters: the disc centre $x$ and $y$, the asymptotic rotational velocity $v_{\rm{asym}}$, the turnover radius $r_t$, the position angle $\phi$ and inclination $\theta$. The values of these parameters are constrained so that $x$ and $y$ are within the range of the data, and $r_t < 50$kpc. Rotation is described by an arctan function \citep{1997AJ....114.2402C} of the form

\begin{equation}
v(r) = \left( \frac{2}{\pi}\right) v_{\rm{asym}}\arctan \left( \frac{r}{r_t} \right).
\end{equation}

The fit is carried out with an iterative procedure whereby $10,000$ random sets of parameters are generated, and the resulting 2D velocity models are smoothed by the effective source plane PSF of the galaxy in order to account for the effect of asymmetric magnification. The $\chi^2$ is then calculated for each one. The parameter space is contracted by discarding up to 10\% of the worst fitting models in each iteration. This process is repeated up to 100 times, and the fit is deemed to have converged once all models in a generation have $\Delta\chi^2 < 1$. The smoothed best-fit disc models are contoured on the observed velocity fields in Figure \ref{fig:im}, and the velocity residuals after subtracting the best-fit disc model are shown in the bottom-right image of each panel. The dynamical properties of the sample are given in Table \ref{tab:dyn}.

While it is always possible to fit a rotating disc model to the data, this is insufficient to determine that the galaxy is a rotating disc. A rotating disc should have symmetric velocity and velocity dispersion fields, the latter peaking in the centre of the galaxy. The traditional requirement that the velocity field form a `spider' pattern is rarely found in lensed galaxies due to the asymmetric magnification (the contours overlaid in Figure \ref{fig:im} demonstrate how much smoothing by the elliptical source plane PSFs distorts even an idealised rotation field).

\citet{2010MNRAS.404.1247J} illustrate the symmetry of the kinematics in their sample by extracting one-dimensional profiles along the kinematic axis, concluding that 5/6 of the galaxies have velocity profiles indicative of rotation, with one likely merger (Cl0949-arc1). In order to examine the symmetry in our new data, we extract a one-dimensional profile of the velocity and velocity dispersion from each galaxy. We do this by using the best-fit disc model to identify the dynamical centre and major kinematic axis of the galaxy. The velocity and velocity dispersion are extracted along a slit five pixels wide, and binned into independent resolution elements along the long axis according to the galaxies' effective source plane PSF. The resulting dynamical profiles for the 12 galaxies in our sample are shown in Figure \ref{fig:profiles}. The best fit one-dimensional rotation curves and exponential velocity dispersion profiles (where applicable) are overlaid in red.

All of the galaxies in the sample exhibit velocity gradients, and all can be fit by an arctan function. However, we only see indications of a turnover in the rotation curve in 9/12 of the sample. Furthermore, the velocity dispersion profiles are less ordered. In Figure \ref{fig:profiles} we attempt to fit an exponential profile to the velocity dispersions. In 50\% of cases the exponential function is a good fit to the data. In two cases, A1413-arc2.1b and RXJ1720-arc1.1+1.2, there is no central peak. A1413-arc2.1b is probably interacting with the larger A1413-arc2.1a, from which it is offset by only $\sim 1000$km\,s$^{-1}$, and the morphology of RXJ1720-arc1.1+1.2 with its extended tail suggests that it also may be an interacting system. In the remaining four galaxies - MACS0744-system3, A1413-arc2.1a, MS1621-system1 and A1689-arc1.2 - there are suggestions of a double peak in the velocity dispersion profile, which may also be indicative of a late-stage merger. Alternatively, the irregular and asymmetric profiles could be due to turbulence within the discs. 

The kinematic properties of the galaxies are similar to those in the \citet{2010MNRAS.404.1247J} sample, which also exhibited rotation-like velocity fields with irregular velocity dispersion profiles. The galaxies in the new sample tend to be smaller than those from \citet{2010MNRAS.404.1247J}, with median $r_{1/2} \sim 0.5\,$kpc and $1.2$\,kpc respectively, but both lensed samples probe systematically smaller galaxies than unlensed studies such as SINS+AO, SHiZELS and WiggleZ, which have median $r_{1/2} \sim 4.3$\,kpc, 2.3\,kpc and 2.5\,kpc respectively \citep{2013ApJ...767..104N,2012MNRAS.426..935S,2011MNRAS.417.2601W}. The two lensed samples are closely matched in rotational velocity with median rotational velocity measured at 2.2\,$r_{1/2}$ of $v_{2.2} \sim 64$\,km\,s$^{-1}$ and $51$\,km\,s$^{-1}$ for the new data and \citet{2010MNRAS.404.1247J} sample respectively. The median rotational velocity in the WiggleZ survey is marginally higher at 74\,km\,s$^{-1}$, with $\sim 110$\,km\,s$^{-1}$ in SHiZELS and $\sim 150$\,km\,s$^{-1}$ in the SINS AO sample. There are two factors that bias lensing surveys towards more slowly rotating systems: firstly, that they are sensitive to smaller galaxies, which tend to be slower rotators \citep[e.g.][see also Section \ref{sec:dynprops}]{2013ApJ...767..104N}, and secondly, the high spatial resolution enables us to measure small velocity gradients that would be flattened by beam smearing in unlensed data and thus categorised as non-rotating.

As the properties of the two lensed samples are similar, and differ systematically from the unlensed data, we discuss them in the remainder of this paper as a combined sample. In the interest of providing a comparison to other studies, we give our best estimate of the nature of the galaxies in Table \ref{tab:dyn}, but caution that these are by no means certain.

\subsection{SED fitting and stellar mass estimates}

In order to estimate the extinction due to dust and the stellar masses of our sample, we model their spectral energy distributions (SEDs) from available archival \emph{HST} and \emph{Spitzer/IRAC} imaging. We first subtract any foreground cluster members lying close to the line of sight to the target galaxy, using the {\sc iraf} tasks \verb'ellipse' and \verb'bmodel'. We then degrade all of the imaging for each galaxy to the poorest resolution, that of the longest-wavelength \emph{IRAC} band. We then extract the photometry using an elliptical aperture large enough to encompass the galaxy, with a 2'' annulus for sky subtraction. We apply aperture corrections to the \emph{IRAC} fluxes based on the area of the aperture used. 

As the continuum light does not necessarily follow the same distribution as the nebular emission, we do not assume the same magnification factor from gravitational lensing; instead, the measured fluxes are corrected for lensing using the flux-weighted mean magnification within the aperture, from magnification maps created with {\sc lenstool}. The resulting magnification factors are within 1$\sigma$ of the values given for the nebular emission in Table \ref{tab:sample}. The complete photometry, corrected for lensing, is given in Tables \ref{tab:phot1}, \ref{tab:phot2} and \ref{tab:phot3} in the appendix. When calculating stellar masses, we omit the arcs in three clusters that do not have imaging in at least two bands either side of the $4000\AA'$ break (MS1621, RXJ1720 and Abell 2895), as well as the galaxy A1413-arc2.1b which is undetected in the \emph{HST} imaging.

We perform the SED fitting using the {\sc cigale} code \citep{2009A&A...507.1793N}. We use the stellar population models of \citet{2005MNRAS.362..799M}, allow either exponentially decreasing or continuous star formation histories and constrain the oldest stellar populations to be younger than the age of the Universe at the target redshift. The {\sc cigale} code creates SEDs from the far-ultraviolet to the infrared based on a dust-attenuated stellar population, infrared dust emission and spectral line templates. The best-fit galaxy parameters are derived with a Bayesian-like analysis from the distribution of probability-weighted best-fit models.

The stellar masses and dust extinction, $A_V$, obtained for each galaxy from the SED fit is given in Table \ref{tab:props}. We find stellar masses of $M_{\ast} = 6 \times 10^8 - 2 \times 10^{10}$M$_{\odot}$; thus, lensing allows us to probe systematically smaller galaxies than unlensed samples such as SINS (median $M_{\ast} \sim 3 \times 10^{10}$M$_{\odot}$; \citealp{2009ApJ...706.1364F}) or SHiZELS (median $M_{\ast} \sim 2 \times 10^{10}$M$_{\odot}$; \citealp{2012MNRAS.426..935S}). The dust extinctions we derive are in the range $A_V = 0.4 - 1.1$ with a median $A_V \sim 0.9$, similar to those of SINS ($A_V \sim 0.8$; \citealp{2009ApJ...706.1364F}) and SHiZELS ($A_V \sim 0.9$; \citealp{2012MNRAS.426..935S}).

The stellar masses and dust extinctions of the \citet{2010MNRAS.404.1247J} sample and A1835-arc7.1 of the new observations are derived by \citet{2011MNRAS.413..643R} and given in their Table 4. We note that the method used is similar to the one we employ, so the results can be directly compared.

\subsection{Integrated galaxy properties}
\begin{table*}
  \caption{Integrated properties of the sample}
  \label{tab:props}
  \begin{tabular}{l r r r r r r}
    \hline
    Name & Intrinsic f$_{\rm{H}\alpha}$ & Intrinsic f$_{\rm{H}\beta}$ & SFR & $r_{1/2}$ & log\,M$_{\ast}$ & A$_V$ \\
         & \multicolumn{2}{c}{($10^{-18}$erg\,s$^{-1}$cm$^{-2}$)} & (M$_{\odot}$yr$^{-1}$) & (kpc) & (M$_{\odot}$) &  \\
    \hline
MACS0744-system3 & 18 $\pm$ 2 && 1.7 $\pm$ 0.2 & 0.3 $\pm$ 0.1 & 9.2 $\pm$ 0.4 & 0.9 $\pm$ 0.4 \\
MACS1149-arcA1.1 & 12 $\pm$ 2 && 1.2 $\pm$ 0.2 & 0.6 $\pm$ 0.2 & 9.3 $\pm$ 0.4 & 0.5 $\pm$ 0.3 \\
MACS0451-system7 & & 5 $\pm$ 2 & 5 $\pm$ 2 & 0.18 $\pm$ 0.03 & 8.8 $\pm$ 0.3 & 0.7 $\pm$ 0.3 \\
A1413-arc2.1a & 34 $\pm$ 4 && 6.6 $\pm$ 0.7 & 0.46 $\pm$ 0.08 & 8.9 $\pm$ 0.4 & 0.4 $\pm$ 0.3 \\
A1413-arc2.1b & 10 $\pm$ 1 && 3 $\pm$ 0.3 & \ldots & \ldots \\
A1835-arc7.1 & 2.6 $\pm$ 0.3 && 0.8 $\pm$ 0.2 & 0.6 $\pm$ 0.2 & 8.8 $\pm$ 0.2$^a$ & 1.1 $\pm$ 0.3$^a$ \\
MS1621-system1 & 32 $\pm$ 4 && 11 $\pm$ 1 & 2 $\pm$ 1 & \ldots & \ldots \\
RXJ1720-arc1.1+1.2 & 3.7 $\pm$ 0.5 && 1.3 $\pm$ 0.2 & 0.7 $\pm$ 0.3 & \ldots & \ldots \\
A1689-arc2.1 & & 0.7 $\pm$ 0.5 & 1.1 $\pm$ 0.2 & 0.3 $\pm$ 0.1 & 9.5 $\pm$ 0.1 & 0.9 $\pm$ 0.1 \\
A1689-arc1.2 & & 1 $\pm$ 0.5 & 1.5 $\pm$ 0.3 & 0.23 $\pm$ 0.08 & 9.2 $\pm$ 0.2 & 0.9 $\pm$ 0.2 \\
A2895-arc1.1+1.2 & & 10.4 $\pm$ 0.1 & 22 $\pm$ 4 & 0.5 $\pm$ 0.3 & \ldots & \ldots \\
A2895-arc2.2 & & 9 $\pm$ 3 & 27 $\pm$ 5 & 0.17 $\pm$ 0.09 & \ldots & \ldots \\
\hline
Cl0024-arc1.1 & 130 $\pm$ 30 && 27 $\pm$ 6 & 1.8 $\pm$ 0.2 & 10.4 $\pm$ 0.1$^a$ & 1.1 $\pm$ 0.2$^a$ \\
Cl0949-arc1 & 50 $\pm$ 10 && 20 $\pm$ 6 & 3.5 $\pm$ 0.9 & 10.2 $\pm$ 0.5$^a$ & 1.0 $\pm$ 0.0$^a$ \\
MACS0712-system1 & & 2.8 $\pm$ 0.6 & 5 $\pm$ 0.1 & 0.5 $\pm$ 0.2 & 10.5 $\pm$ 0.4$^a$ & 0.8 $\pm$ 0.2$^a$ \\
MACSJ0744-arc1 & 8 $\pm$ 2 && 2.4 $\pm$ 0.5 & 1.2 $\pm$ 0.2 & 10.0 $\pm$ 0.1$^a$ & 0.8 $\pm$ 0.2$^a$ \\
CosmicEye & & 18 $\pm$ 2 & 40 $\pm$ 5 & 0.8 $\pm$ 0.2 & 10.8 $\pm$ 0.1$^a$ & 0.7 $\pm$ 0.1$^a$ \\
\hline
\end{tabular}
\begin{flushleft}\footnotesize{Notes: All quantities are corrected for lensing magnification. SFR is calculated from the H$\alpha$ or H$\beta$ luminosity corrected for dust extinction based on A$_V$. The half-light radius $r_1/2$ is based on the H$\alpha$ or H$\beta$ morphology as described in the text. A$_v$ and $M_{\ast}$ are estimated from SED fitting of broadband photometry as described in the text, except for those marked $(a)$, which are from \citet{2011MNRAS.413..643R}. Targets below the line are from \citet{2010MNRAS.404.1247J}, with SFRs given in their Table 2 and half-light radii reported by \citet{2011MNRAS.413..643R}.}\end{flushleft}
\end{table*}

\begin{figure}
\includegraphics[height=84mm,angle=90]{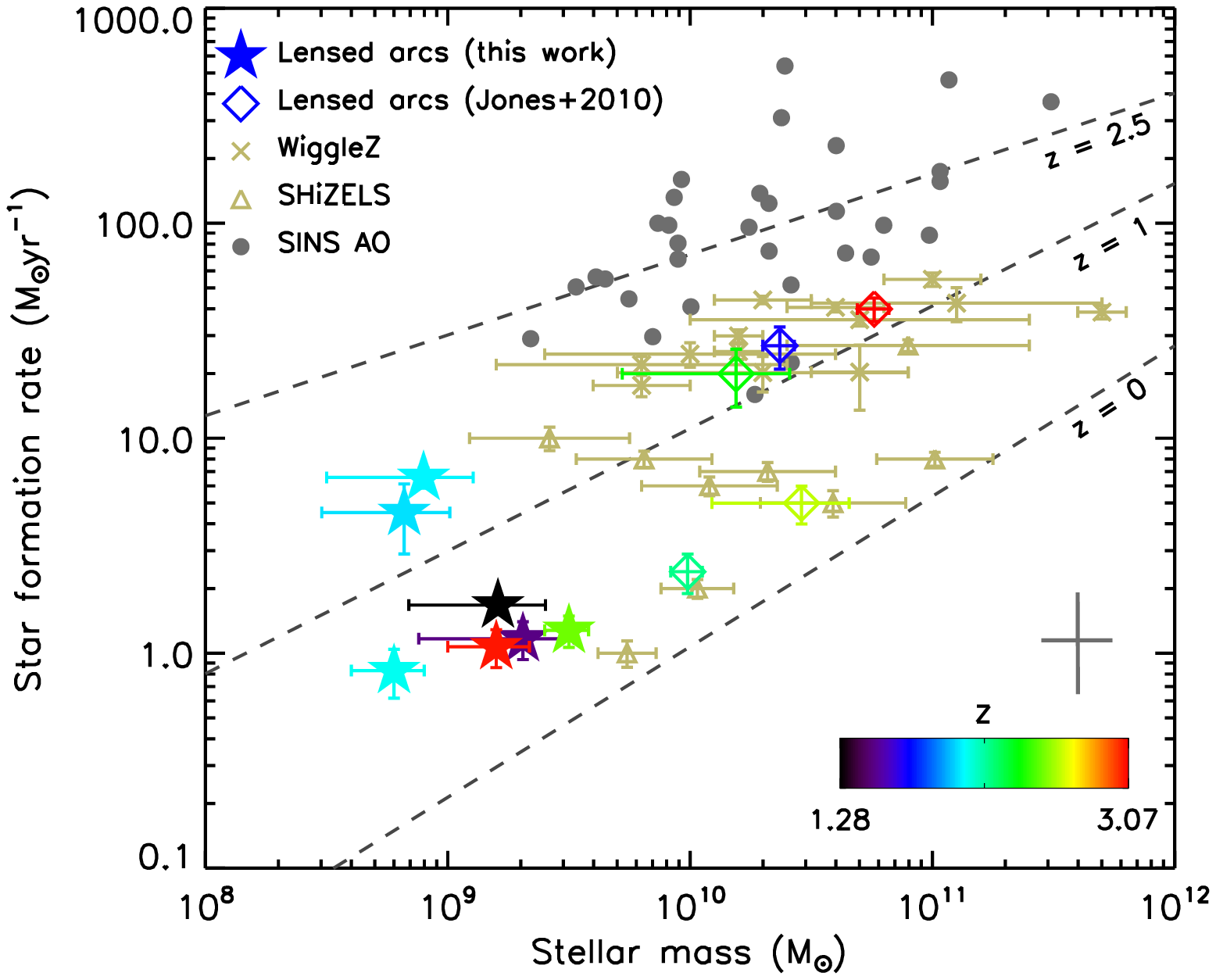}
\caption{Total galaxy SFR against stellar mass for the lensed arcs for which we are able to derive stellar masses, in comparison with other high-redshift kinematic studies. The lensed arcs are colour-coded by redshift. The dashed lines indicate the star-forming galaxy `main sequence' at $z = 0$, $1$ and $2.5$ \citep{2012ApJ...754L..29W}. Comparison data are from SHiZELS \citep{2012MNRAS.426..935S}, WiggleZ \citep{2011MNRAS.417.2601W} and the SINS AO sample \citep{2013ApJ...767..104N}, and the grey cross indicates the typical error on the SINS points. The lensed arcs, as well as the WiggleZ and SHiZELS samples, are similar to $z \sim 1$ main sequence star-forming galaxies, though the lensed arcs probe the lower-mass end of the relation. The SINS AO sample covers higher specific star formation rates, similar to the $z \sim 2.5$ star-forming main sequence.}
\label{fig:sfrmass}
\end{figure}

We now derive some integrated properties of the sample, so that we can relate the lensed galaxies to the population as a whole and explore how their internal structures evolve as a function of their global properties.

A summary of the integrated properties of the sample is given in Table \ref{tab:props}. Total fluxes of the H$\alpha$ or H$\beta$ emission lines are calculated by summing every pixel in the source plane cubes with signal-to-noise $>5$. We measure the total flux by fitting a Gaussian profile to the emission lines, and the ratio of the total flux in the image plane to that in the source plane gives the total magnification given in Table \ref{tab:sample}.

Intrinsic star formation rates (SFRs) in Table \ref{tab:props} are calculated from the H$\alpha$ (or H$\beta$, assuming case B recombination) flux, corrected for dust according to the extinction $A_V$ derived from SED fitting, by applying the \citet{1998ARA&A..36..189K} prescription. We correct this conversion to a Chabrier initial mass function (IMF), which includes a turnover at the low-mass end and produces more realistic mass-to-light ratios. This correction has the effect of reducing the SFR by a factor $1.7 \times$.

To find the half-light radius $r_{1/2}$, we first construct an array in which each pixel value is the distance from the dynamical centre, $[i_c,j_c]$, based on an estimated position angle, $\phi$, and inclination ($\theta$; the derivation of the dynamical centre, position angle and inclination is discussed in Section \ref{sec:dynamics}). The result is a series of concentric ellipses with the galactocentric radius in pixel $i,j$ given by

\begin{equation}
r_{i,j} = \sqrt{ \left( x\cos\phi + y\sin\phi \right)^2 + \left( \frac{x\sin\phi - y\cos\phi}{\cos\theta}\right)^2},
\end{equation}

where $x = |i - i_c|$ and $y = |j - j_c|$. We then convolve this array with the flux-weighted source plane PSF, to account for the fact that the preferential direction of magnification causes the source plane images to appear smeared in one direction.

Starting from the dynamical centre, we then sum the flux within contours of constant $r$, incrementing $r$ until half the total flux is enclosed. This then gives the half-light radius $r_{1/2}$ given in Table \ref{tab:props}. The half-light sizes we find are in the range $r_{1/2} \sim 0.2 - 2.1$\,kpc. Where we have rest-frame optical imaging of a galaxy we carry out the same procedure and find that the results are consistent within the quoted errors. The half-light radii of the \citet{2010MNRAS.404.1247J} sample are $r_{1/2} = 0.5 - 3.5$\,kpc. As noted in Section \ref{sec:discmodel}, we find galaxies that are systematically smaller than those observed in unlensed samples due to the lensing magnification.

\section{Results and Analysis}
\label{sec:results}

\subsection{Dynamics}
\label{sec:dynamics}

\begin{figure*}
\includegraphics[height=0.49\textwidth, angle=90]{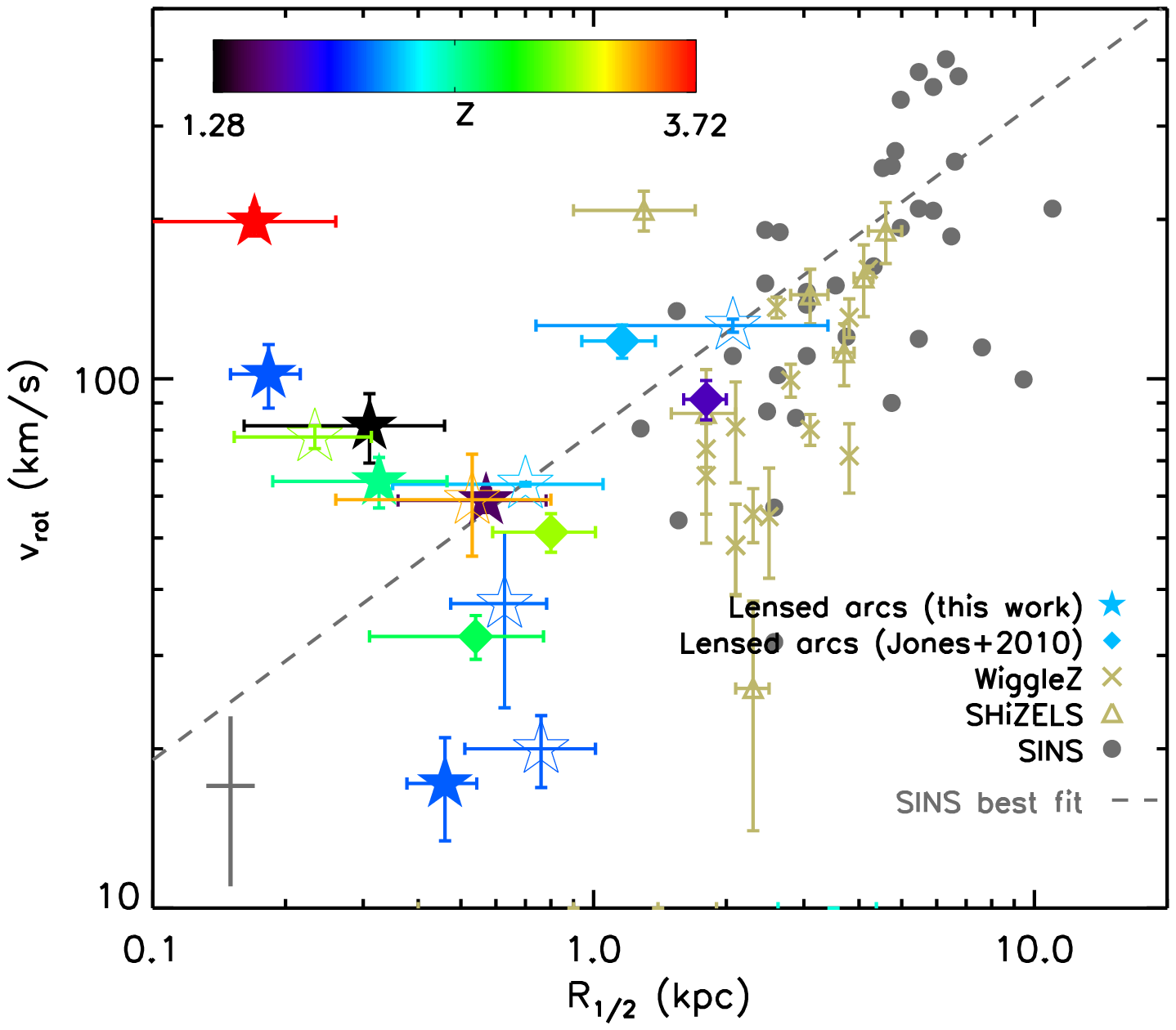} \includegraphics[height=0.49\textwidth, angle=90]{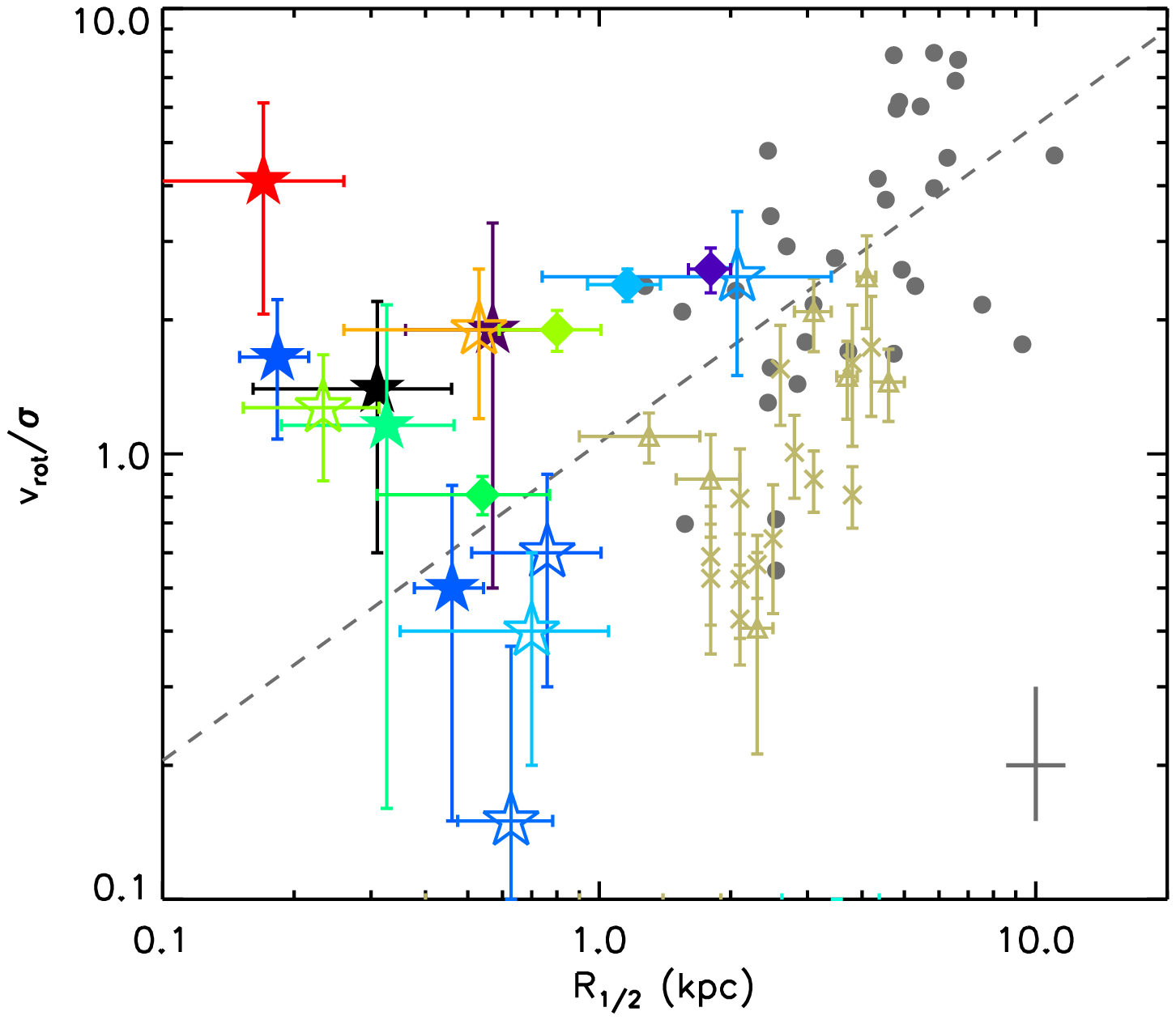} 
\caption{\emph{Left:} Relation between rotational velocity and half-light radius. \emph{Right:} $v_{\rm{rot}}/\sigma$ against half-light radius. The lensed arcs are colour-coded by redshift on the same scale in both plots, with filled symbols representing those galaxies identified as discs, and unfilled symbols mergers or undetermined systems. Comparison samples are included from the high-$z$ lensed sources of \citet{2010MNRAS.404.1247J}, and the unlensed high-$z$ IFU surveys SHiZELS \citep{2012MNRAS.426..935S}, WiggleZ \citep{2011MNRAS.417.2601W} and the SINS AO sample \citep{2013ApJ...767..104N}. The grey crosses indicate the average error on the SINS points. The dashed line is the fit to the SINS data from \citet{2013ApJ...767..104N}. The intrinsically smaller lensed galaxies extend the non-lensed samples to smaller sizes, but the correlation is weaker.}
\label{fig:vsigr}
\end{figure*}

\subsubsection{Dynamical properties of the sample}
\label{sec:dynprops}

From our best estimates of kinematic classification given in Table \ref{tab:dyn}, we estimate that $10/17 \left( 59\% \right)$ of the combined lensed sample are rotating discs, $5/17 \left( 29\% \right)$ are probable mergers, and the remaining two are undetermined. The merger fraction in our sample is very close to those in unlensed surveys, typically $1/3$, but we find a higher fraction of rotating systems, which typically make up 33-44\% of other samples \citep{2009ApJ...706.1364F,2009ApJ...697.2057L,2009ApJ...699..421W,2012A&A...539A..92E}. While the merger classification adopted is relatively insensitive to spatial resolution, it is likely that many of the galaxies classed as `dispersion-dominated' in unlensed studies would present with velocity gradients given higher spatial resolution. For example, \citet{2013ApJ...767..104N} found that when the same 34 galaxies observed in seeing-limited conditions were re-observed with adaptive optics, the fraction classified as dispersion-dominated fell from $41\%$ to $6 - 9\%$. The two galaxies we classify as undetermined are A1835-arc7.1 and MS1621-system1. The former lies on the critical line so it is possible that we have not observed the whole velocity gradient in this source. Alternatively, the turnover in the rotation curves of these galaxies might be missed due to the presence of a low surface brightness disc.

The quantity $v_{\rm{rot}}/\sigma$ is a measure of the rotational support in a system. Galaxies exist on a continuum in $v_{\rm{rot}}/\sigma$; they all have some degree of rotation, but those with low $v_{\rm{rot}}/\sigma$ values have a larger fraction of their kinematic support from random motion rather than ordered rotation. Generally, a simple cut is applied, where galaxies with $v_{\rm{rot}}/\sigma > 0.4$ are considered rotating systems \citep{2009ApJ...706.1364F}. Of our combined sample, only A1835-arc7.1 falls beneath this cut, with $v_{\rm{rot}}/\sigma = 0.2 \pm 0.2$. However, as discussed above, we may not have observed the full velocity gradient in this galaxy. Three of the galaxies that appear to be interactions or mergers - A1413-arc2.1a, A1413-arc2.1b and RXJ1720-arc1.1+1.2 - all have marginal $v_{\rm{rot}}/\sigma$ within 1$\sigma$ of the boundary between rotation-dominated and dispersion-dominated systems. The remaining possible merger - MS1621 - has high $v_{\rm{rot}}/\sigma = 3 \pm 1$, which could be due to a velocity difference between two merging components.

In non-lensed galaxies observed with adaptive optics, \citet{2013ApJ...767..104N} found a correlation between $v_{\rm{rot}}/\sigma$ and the half-light radius $r_{1/2}$. This was suggested to be due to a relationship between $v_{\rm{rot}}$ and $r_{1/2}$, with larger galaxies having higher rotation velocities. We add our sample to these relationships in Figure \ref{fig:vsigr}. As gravitational lensing allows us to observe intrinsically smaller galaxies, we are able to extend these relations to smaller sizes, and we find much more scatter than in the unlensed systems, with a Spearman rank correlation coefficient of $\rho = 0.61$ (where $\rho = 0$ indicates no correlation, and $\rho = 1$ implies perfect correlation).

Comparing $v_{\rm{rot}}/\sigma$ to $r_{1/2}$ in Figure \ref{fig:vsigr} gives similar results; we find higher scatter at low $r_{1/2}$, but there remains a weak positive correlation (Spearman rank correlation coefficient $\rho = 0.52$) in the combined lensed and unlensed data.

In both cases, there is no systematic difference in this scatter between the rotation-dominated galaxies and those identified as non-disc systems. The lack of correlation seen in the lensed sample with respect to the unlensed galaxies therefore implies a fundamental difference in the kinematic properties. As discussed below, this is likely due to the systematically smaller sizes and lower masses of the lensed galaxies.

\begin{figure}
\includegraphics[height=84mm, angle=90]{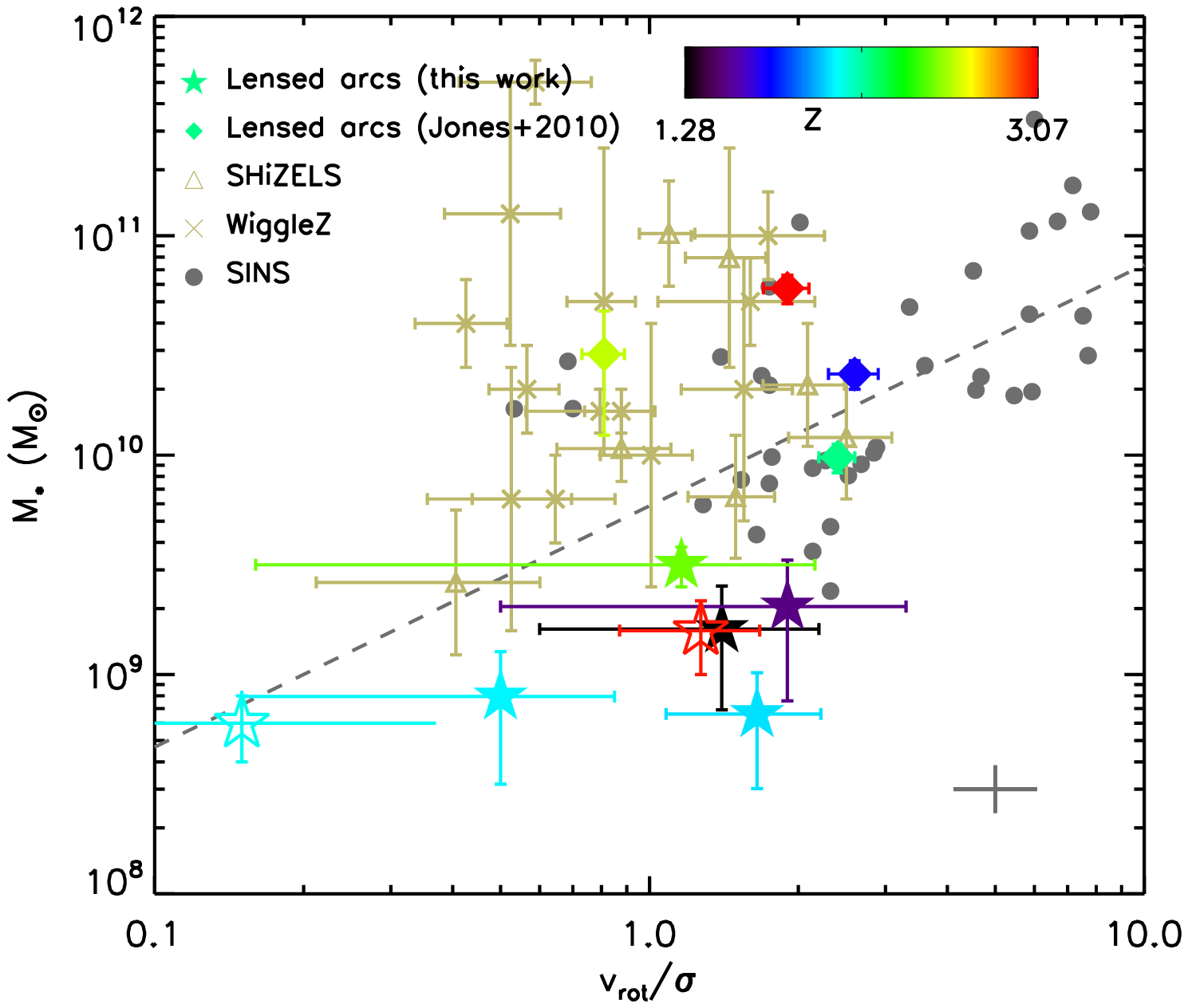}
\caption{Relation between stellar mass, $M_{\ast}$, and $v_{\rm{rot}}/\sigma$. The rotational velocity $v_{\rm{rot}}$ is corrected for inclination, and $\sigma$ is the luminosity-weighted mean velocity dispersion, corrected for beam smearing. The SINS AO sample \citep{2013ApJ...767..104N}, SHiZELS \citep{2012MNRAS.426..935S} and WiggleZ \citep{2011MNRAS.417.2601W} are also shown for comparison with higher stellar mass systems at high redshift. The grey cross indicates the typical error on the SINS data, and the dashed line is the best fit to the SINS + lensed samples. We find that lensed galaxies with high $v_{\rm{rot}}/\sigma$ tend to have high $M_{\ast}$, indicating that galaxies become more dynamically `settled' as they build up more mass.}
\label{fig:mstarvsig}
\end{figure}

In Figure \ref{fig:mstarvsig}, we plot stellar mass, $M_{\ast}$, as a function of $v_{\rm{rot}}/\sigma$, finding that our combined sample continues the relation observed in the non-lensed SINS+AO sample to lower stellar masses, although with significant scatter. However, there is little correlation with the unlensed SHiZELS and WiggleZ data. This could be because they are observed with similar resolution to the SINS data, but have systematically smaller sizes, and thus a larger contribution from beam smearing. The best fit to the combined lensed and SINS+AO unlensed data is $\log \left(M_{\ast}\right) = 1.2\log\left( v_{\rm{rot}}/\sigma \right) + 9.7$.

If we interpret $v_{\rm{rot}}/\sigma$ as a measure of the `order' in a system, where $v_{\rm{rot}}$ measures the ordered rotation and $\sigma$ constitutes turbulent or disordered motion, this relationship suggests that galaxies become more ordered as they build up higher stellar masses, a process described as `kinematic settling' \citep{2012ApJ...758..106K}. Therefore, in the lensed sample, which probes systematically smaller intrinsic sizes and lower masses than the unlensed surveys, we see less contribution from ordered rotation as shown in Figure \ref{fig:vsigr}, and hence do not see the correlations exhibited by the larger, more massive unlensed galaxies.

\begin{figure}
\includegraphics[height=84mm, angle=90]{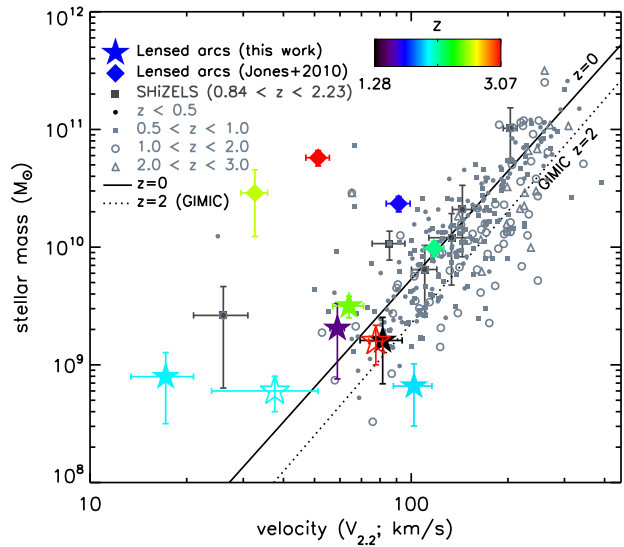}
\caption{The stellar mass Tully-Fisher relation. The stellar mass, $M_{\ast}$, is estimated from SED fitting to the available \emph{HST} and \emph{Spitzer} imaging. The rotation velocity $v_{2.2}$ is measured from the model rotation curves interpolated at a radius of $r = 2.2\,r_{1/2}$, where $r_{1/2}$ is the half-light radius. The $z = 0$ data are from \citet{2005ApJ...633..844P}, and the solid line is the best fit to these points. High-redshift comparison data are from the $z \sim 2$ lensed arcs of \citet{2010MNRAS.404.1247J}, the $z=1$ lensed arcs of \citet{2006MNRAS.368.1631S}, the DEIMOS $z \sim 0.6$ and $z \sim 1.3$ samples of \citet{2011ApJ...741..115M,2012ApJ...753...74M}, the $z \sim 2 - 3.5$ SINS and AMAZE surveys \citep{2009ApJ...697..115C,2011A&A...528A..88G} and the SHiZELS $0.84 < z < 2.3$ sample of \citet{2012MNRAS.426..935S}. The dotted line indicates the predicted evolution of the Tully-Fisher relation at $z = 2$ from simulations \citep{2009MNRAS.399.1773C}. We colour-code our lensed arcs by redshift, and show rotating disc-like systems with filled symbols, and mergers and dispersion-dominated systems with open symbols. We find that most of our sample is largely consistent with the $z=0$ relation, but with significant scatter discussed further in the text.}
\label{fig:tf}
\end{figure}

\subsubsection{The stellar mass Tully-Fisher relation}

The Tully-Fisher relation relates the stellar content of galaxies to their rotational velocity \citep{1977A&A....54..661T}. In a model in which gas cools from a dark matter halo into a rotating disc, maintaining the angular momentum of its parent halo, the Tully-Fisher relation is interpreted as a relationship between the baryonic content of galaxies and the angular momentum of their dark matter halos. As such, it is a key parameter that models of galaxy evolution must reproduce.

Attempts to place observational constraints on the redshift evolution of the Tully-Fisher relation have had inconclusive results. Observations of the $B$-band Tully-Fisher relation at high redshift demonstrate modest evolution, but are affected by recent star formation and so evolution of the B-band mass-to-light ratio \citep{1996ApJ...465L..15V,1997ApJ...479L.121V}. We therefore concentrate on the more physical stellar mass Tully-Fisher relation, but this too has produced mixed results, with no coherent evidence of evolution out to $z \sim 2$ \citep{2005ApJ...628..160C,2006A&A...455..107F,2006A&A...458..385C,2007ApJ...660L..35K,2008A&A...484..173P,2012A&A...546A.118V,2011ApJ...741..115M,2012ApJ...753...74M,2012MNRAS.426..935S}. \citet{2013ApJ...762L..11M} found that bulgeless galaxies at high-$z$ are the ones that show the greatest offset from the local relation, which they interpret as evidence that these galaxes have yet to `mature' onto the Tully-Fisher relation, perhaps related to higher gas fractions in their discs. Similarly, \citet{2011A&A...528A..88G} investigated the stellar mass Tully-Fisher relation at $z \sim 3$ and found marginal evolution but with a large amount of scatter, concluding that the relation may not be in place at this epoch.

To compare different galaxies, it is important to measure $v_{\rm{rot}}$ in a consistent manner. Measurements of the rotation curve are naturally sensitive to the surface brightness of the galaxy and the depth of the observations. \citet{1997AJ....114.2402C} compared various methods of measuring $v_{\rm{rot}}$ and found that $v_{2.2}$ (the velocity interpolated from the model rotation curve at a radius $r = 2.2\,h$, where $h$ is the disc scale length) performed best in terms of minimal internal scatter and residuals from the Tully-Fisher relation and provided the best match to radio (21cm) results. In high-redshift galaxies with clumpy and disturbed morphologies, the light does not follow a pure exponential disc profile, so we cannot strictly use this definition with our data. We therefore adopt the half-light radius, $r_{1/2}$, as a proxy for the disc scale length, and measure $v_{2.2}$ from the model rotation curve fits, corrected for inclination, at $r = 2.2\,r_{1/2}$. For consistency across the sample, we re-analyse the velocity fields of the \citet{2010MNRAS.404.1247J} sample in order to extract $v_{2.2}$ from these galaxies. Our derived $v_{2.2}$ values are slightly lower than the inclination-corrected $v_{\rm max}$ that they derive by an average of 20\%.

\begin{figure}
\includegraphics[height=84mm, angle=90]{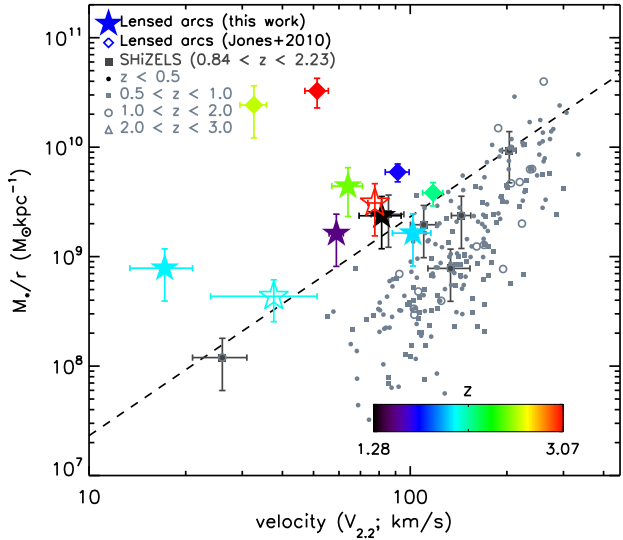}
\caption{The ratio between stellar mass and galaxy radius as a function of rotational velocity, where radius is defined as 2.2 disc scale lengths and the velocity is corrected for inclination. Comparison data is from the same sources as in Figure \ref{fig:tf}. The dashed line indicates $v^2 = GM_{\ast}/r$. The lensed galaxies and some data from the other samples, in particular from SHiZELS, lie along this line, indicating that their dynamics are dominated by baryons in the discs. There is no apparent systematic difference between disc-like (filled symbols) and mergers/ dispersion-dominated systems (open symbols) in the lensed samples. Offsets from this line are likely related to the gas fractions of the galaxies.}
\label{fig:mrv}
\end{figure}

We plot the stellar mass Tully-Fisher relation in Figure \ref{fig:tf}. We find that most of our lensed arcs are consistent with the Tully-Fisher relation observed in other samples, but with considerable scatter. Three galaxies lie well above the local relation, with low $v_{2.2}$ for their stellar masses. Of these, two are from the \citet{2010MNRAS.404.1247J} sample; these are Cl0949-arc1 and the Cosmic Eye. The former was identified as a merger by \citet{2010MNRAS.404.1247J}, though the latter appears to be a rotationally-supported disc \citep[see][]{2008Natur.455..775S}. The remaining galaxy lying above the local relation A1413-arc2.1a, which appears to be interacting with A1413-arc2.1b.

Simulations predict that the zero-point of the Tully-Fisher relation should decrease at high redshift \citep[e.g.][indicated by a dotted line in Figure \ref{fig:tf}]{2012MNRAS.427..379M,2009MNRAS.399.1773C}. Semi-analytic models in which the disc dynamics are driven by the dark matter halo make a similar prediction \citep{2011MNRAS.410.1660D}. However, \citet{2012NewA...17..175B} predicts a modest \emph{positive} increase in the zero-point at high redshift. As \citet{2012ApJ...753...74M} point out, the primary difference is that the \citeauthor{2012NewA...17..175B} model has a larger contribution from baryons. High-redshift galaxies are intrinsically smaller than their present-day counterparts - especially in the lensed sample that probes the smaller, fainter population - and so we might expect the dynamics within $2.2r_{1/2}$ to be dominated by baryons \citep[see also][]{2011ApJ...741..115M}. In the simplest case of orbital circular motion, we should see $v^2 = GM/r$ for total enclosed mass $M$. To test whether this is a better description of our data, we divide the y-axis of Figure \ref{fig:tf} by $2.2r_{1/2}$ and show the result in Figure \ref{fig:mrv}. The scatter is significantly reduced in comparison to Figure \ref{fig:tf}, and the line $v^2 = GM_{\ast}/r$ provides a good fit to the lensed and SHiZELS data, although the same three lensed galaxies remain far outside this relation. The varying gas fractions in the galaxies will contribute significantly to the scatter in this relation. The other comparison samples have a systematically higher velocities at a given $M_{\ast}/r$, perhaps suggesting that their dynamics do have contributions from the dark matter halo, whereas the dynamics of the lensed galaxies within the central few kpc are dominated by baryons.

\subsection{Star-forming clumps}
\label{sec:clumps}

\begin{table}
  \caption{Properties of the star-forming clumps}
  \label{tab:clumps}
  \begin{tabular}{l r r r r r r}
    \hline
    Name & radius & SFR & $\sigma$ \\
         & pc & M$_{\odot}$yr$^{-1}$ & km\,s$^{-1}$ \\
Notes: & $(a)$ & $(b)$ & $(c)$ \\
    \hline
MACS0744-system3-1 & 300 $\pm$ 100 & 0.26 $\pm$ 0.02 & 70 $\pm$ 7 \\
MACS0744-system3-2 & 300 $\pm$ 100 & 0.18 $\pm$ 0.01 & 84 $\pm$ 8 \\
MACS0744-system3-3 & 160 $\pm$ 30 & 0.04 $\pm$ 0.01 & 51 $\pm$ 8 \\
MACS1149-arcA1.1-1 & 450 $\pm$ 70 & 0.51 $\pm$ 0.05 & 74 $\pm$ 10 \\
MACS1149-arcA1.1-2 & 150 $\pm$ 50 & 0.15 $\pm$ 0.02 & 120 $\pm$ 19 \\
MACS1149-arcA1.1-3 & 160 $\pm$ 30 & 0.06 $\pm$ 0.01 & 134 $\pm$ 31 \\
MACS1149-arcA1.1-4 & 210 $\pm$ 50 & 0.12 $\pm$ 0.02 & 80 $\pm$ 17 \\
MACS1149-arcA1.1-5 & 180 $\pm$ 70 & 0.07 $\pm$ 0.01 & 95 $\pm$ 39 \\
MACS1149-arcA1.1-6 & 150 $\pm$ 20 & 0.05 $\pm$ 0.01 & 71 $\pm$ 17 \\
MACS1149-arcA1.1-7 & 160 $\pm$ 40 & 0.04 $\pm$ 0.01 & 123 $\pm$ 33 \\
MACS1149-arcA1.1-8 & 140 $\pm$ 50 & 0.03 $\pm$ 0.01 & 115 $\pm$ 30 \\
MACS1149-arcA1.1-9 & 140 $\pm$ 20 & 0.02 $\pm$ 0.01 & 122 $\pm$ 61 \\
MACS0451-system7-1 & 500 $\pm$ 200 & 1.2 $\pm$ 0.4 & 87 $\pm$ 6 \\
MACS0451-system7-2 & 300 $\pm$ 200 & 0.5 $\pm$ 0.2 & 84 $\pm$ 6 \\
A1413-arc2.1a-1 & 700 $\pm$ 300 & 2.3 $\pm$ 0.2 & 53 $\pm$ 8 \\
A1413-arc2.1b-1 & 200 $\pm$ 100 & 0.18 $\pm$ 0.02 & 57 $\pm$ 9 \\
A1413-arc2.1b-2 & 260 $\pm$ 60 & 0.15 $\pm$ 0.02 & 58 $\pm$ 10 \\
A1413-arc2.1b-3 & 300 $\pm$ 100 & 0.26 $\pm$ 0.03 & 54 $\pm$ 9 \\
A1413-arc2.1b-4 & 140 $\pm$ 40 & 0.02 $\pm$ 0.01 & 84 $\pm$ 32 \\
A1835-arc7.1-1 & 90 $\pm$ 60 & 0.10 $\pm$ 0.01 & 77 $\pm$ 14 \\
A1835-arc7.1-2 & 200 $\pm$ 100 & 0.12 $\pm$ 0.02 & 67 $\pm$ 13 \\
MS1621-system1-1 & 900 $\pm$ 400 & 3.1 $\pm$ 0.3 & 63 $\pm$ 9 \\
MS1621-system1-2 & 600 $\pm$ 200 & 0.8 $\pm$ 0.1 & 66 $\pm$ 11 \\
MS1621-system1-3 & 340 $\pm$ 70 & 0.18 $\pm$ 0.03 & 55 $\pm$ 12 \\
MS1621-system1-4 & 200 $\pm$ 100 & 0.08 $\pm$ 0.01 & 58 $\pm$ 12 \\
RXJ1720-arc1.1+1.2-1 & 700 $\pm$ 200 & 0.77 $\pm$ 0.08 & 61 $\pm$ 8 \\
RXJ1720-arc1.1+1.2-2 & 600 $\pm$ 100 & 0.59 $\pm$ 0.06 & 67 $\pm$ 10 \\
RXJ1720-arc1.1+1.2-3 & 400 $\pm$ 100 & 0.28 $\pm$ 0.03 & 76 $\pm$ 11 \\
RXJ1720-arc1.1+1.2-4 & 600 $\pm$ 100 & 0.65 $\pm$ 0.06 & 65 $\pm$ 9 \\
RXJ1720-arc1.1+1.2-5 & 300 $\pm$ 100 & 0.23 $\pm$ 0.03 & 67 $\pm$ 12 \\
A1689-arc2.1-1 & 180 $\pm$ 60 & 0.2 $\pm$ 0.1 & 116 $\pm$ 45 \\
A1689-arc1.2-1 & 150 $\pm$ 80 & 0.13 $\pm$ 0.04 & 103 $\pm$ 16 \\
A2895-arc1.1+1.2-1 & 400 $\pm$ 100 & 1.8 $\pm$ 0.7 & 63 $\pm$ 4 \\
A2895-arc1.1+1.2-2 & 500 $\pm$ 300 & 2.3 $\pm$ 0.8 & 61 $\pm$ 4 \\
A2895-arc1.1+1.2-3 & 60 $\pm$ 10 & 0.03 $\pm$ 0.01 & 57 $\pm$ 5 \\
A2895-arc2.2-1 & 290 $\pm$ 90 & 5. $\pm$ 2. & 51 $\pm$ 3 \\
Cl0024-arc1.1-1 & 420 $\pm$ 70 & 5. $\pm$ 1. & 65 $\pm$ 11 \\
Cl0024-arc1.1-2 & 600 $\pm$ 200 & 9. $\pm$ 2. & 60 $\pm$ 9 \\
Cl0024-arc1.1-3 & 370 $\pm$ 50 & 4. $\pm$ 1. & 63 $\pm$ 14 \\
Cl0024-arc1.1-4 & 400 $\pm$ 100 & 3.4 $\pm$ 0.8 & 62 $\pm$ 13 \\
Cl0949-arc1-1 & 1000 $\pm$ 100 & 16. $\pm$ 4. & 72 $\pm$ 13 \\
Cl0949-arc1-2 & 800 $\pm$ 300 & 10. $\pm$ 2. & 63 $\pm$ 8 \\
Cl0949-arc1-3 & 800 $\pm$ 300 & 11. $\pm$ 3. & 56 $\pm$ 6 \\
Cl0949-arc1-4 & 700 $\pm$ 100 & 8. $\pm$ 2. & 76 $\pm$ 14 \\
MACS0712-system1-1 & 220 $\pm$ 70 & 3.3 $\pm$ 0.8 & 74 $\pm$ 11 \\
MACSJ0744-arc1-1 & 300 $\pm$ 100 & 0.7 $\pm$ 0.2 & 93 $\pm$ 26 \\
MACSJ0744-arc1-2 & 300 $\pm$ 100 & 0.4 $\pm$ 0.1 & 136 $\pm$ 45 \\
MACSJ0744-arc1-3 & 300 $\pm$ 100 & 0.5 $\pm$ 0.1 & 118 $\pm$ 14 \\
CosmicEye-1 & 600 $\pm$ 100 & 25. $\pm$ 6. & 41 $\pm$ 15 \\
CosmicEye-2 & 500 $\pm$ 100 & 12. $\pm$ 3. & 40 $\pm$ 14 \\
\hline
\multicolumn{4}{l}{\begin{minipage}{\columnwidth}Notes: All quantities are corrected for lensing where appropriate. $(a)$: The clump radius is calculated as described in the text, using the area enclosed by an isophote and assuming circular symmetry. The error bar encompasses an alternative definition using the FWHM of the clump emission line profile. $(b)$: SFR is calculated from the total H$\alpha$ or H$\beta$ flux enclosed in the isophote and is corrected for extinction. $(c)$: The velocity dispersion, $\sigma$, is the mean local value within the clump and is deconvolved for the local velocity gradient. \end{minipage}} \\
\end{tabular}
\end{table}

\begin{figure}
\includegraphics[height=84mm, angle=90]{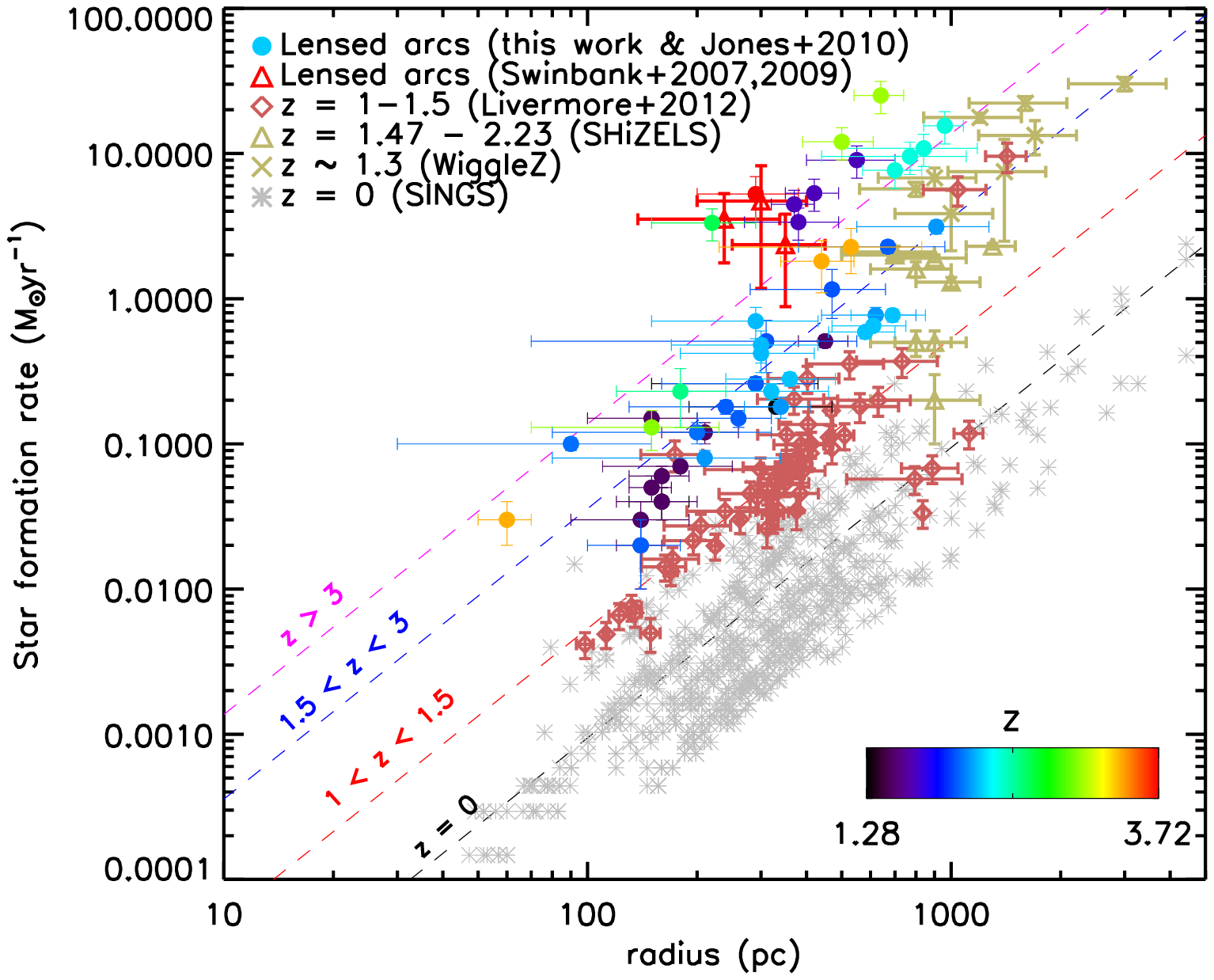}
\caption{Relation between Star formation rate (SFR) in clumps and their sizes. The $z=0$ sample from the SINGS survey \citep{2003PASP..115..928K} and the $z \sim 1 - 1.5$ lensed arcs from \citet{2012MNRAS.427..688L} are extracted from H$\alpha$ narrowband imaging. We also show the $z \sim 5$ lensed arcs of \citet{2007MNRAS.376..479S} and \citet{2009MNRAS.400.1121S} and the unlensed high-$z$ samples from SHiZELS \citep{2012ApJ...760..130S} and WiggleZ \citep{2012MNRAS.422.3339W} for comparison. The clumps from our sample of lensed arcs are colour-coded by redshift. The dashed lines are contours of constant surface brightness with the zero-points fit in four redshift bins. We find evolution in this zero-point such that high-redshift galaxies host clumps of higher surface brightnesses.}
\label{fig:csizelum}
\end{figure}

With the high spatial resolution afforded by gravitational lensing, it is possible to extract star-forming clumps from lensed galaxies on scales comparable to giant H{\sc ii} regions in local galaxies. It has long been noted that galaxies in the high-redshift Universe tend to be `clumpier' then their local counterparts, with clumps often dominating the host galaxies' morphologies and giving rise to `clump cluster' and `chain' galaxies \citep[e.g.][]{1995AJ....110.1576C,2004ApJ...603...74E,2005ApJ...627..632E}. Numerous high-redshift galaxy surveys have isolated clumps and analysed their properties on $\sim$\,kpc scales \citep{2011ApJ...739...45F,2011ApJ...733..101G,2012MNRAS.422.3339W,2012ApJ...760..130S}, but to compare them directly to star formation in the local Universe, which takes place in H{\sc ii} regions on scales of $\sim 100$\,pc, the spatial resolution afforded by gravitational lensing is required \citep{2011ApJ...742...11S,2009MNRAS.400.1121S,2010MNRAS.404.1247J,2012MNRAS.427..688L}.

Different studies of star-forming clumps have used a variety of methods to detect and isolate them from their host galaxies. Discussions of the merits of various techniques can be found in \citet{2012MNRAS.422.3339W} and \citet{2012MNRAS.427..688L}. We adopt the same method as in \citet{2012MNRAS.427..688L} and use the 2D version of {\sc clumpfind} \citep{1994ApJ...428..693W}. This routine uses multiple isophotes, starting by defining clumps in the brightest regions and then moving down through the isophote levels. Any isolated contours are defined as new clumps, and any which enclose an existing peak are allocated to that clump. A contour which encloses two or more existing peaks has its pixels divided between them using a `friends-of-friends' algorithm. We define the lowest contour at 3$\sigma$, where $\sigma$ is the standard deviation of pixels in the underlying galaxy. We then add additional contours in 1$\sigma$ intervals up to the peak in the image.

For IFU data, \citet{2011ApJ...733..101G} advocate defining clumps in slices in velocity. We test this method with our data and find that it produces similar results to using the integrated intensity maps, and where it does produce clumps that are not evident in the integrated maps, they are marginal detections found only in the lowest ($3 - 4\sigma$) contour levels. We therefore use the integrated H$\alpha$ (or H$\beta$) maps to define clumps. This has the added advantage of being the method most easily comparable to the narrowband imaging in \citet{2012MNRAS.427..688L}, where we used maps of H$\alpha$ integrated over a wide velocity range.

To ensure self-consistency across the entire sample of lensed arcs, we reanalyse the intensity maps of the \citet{2010MNRAS.404.1247J} sample to extract clumps with {\sc clumpfind}. \citet{2010MNRAS.404.1247J} used a single isophote per galaxy to extract their clump properties, whereas the algorithm developed by \citet{2012MNRAS.427..688L} used the noise properties of the image to robustly define a series of isophote levels in a way that does not require any fine tuning. The variation in methods means that the clumps extracted are not identical; a lower isophote will increase both the size and luminosity of the extracted clump, and vice versa. Generally, we use lower isophote levels than \citet{2010MNRAS.404.1247J}, leading to larger sizes (but within $\sim 1\sigma$ of the published values) with correspondingly higher luminosities. The primary difference is that we also extract some smaller, lower-luminosity clumps that are missed when adopting a single isophote.

Once the clumps are defined, we sum the spectra in their component pixels and fit a Gaussian emission line (a single line for H$\alpha$ or a triplet for H$\beta$ and [O{\sc iii}]) to obtain their intensity and velocity dispersion. For those galaxies observed in H$\beta$, we estimate the equivalent H$\alpha$ intensity by assuming case B recombination and a \citet{2000ApJ...533..682C} extinction curve based on the attenuation $A_V$ estimated from the SED fitting. No additional extinction is assumed in the nebular lines. Thus, the estimated H$\alpha$ luminosity, $L_{\rm{H}\alpha}$ is given by

\begin{equation}
  L_{\rm{H}\alpha} = 2.86 \times 10^{\frac{A_V}{7.96}} L_{\rm{H}\beta}.
\end{equation}

From $L_{\rm{H}\alpha}$, we calculate the SFR using the prescription of \citet{1998ARA&A..36..189K} adapted to a \citet{2003PASP..115..763C} initial mass function (IMF) and extinction $A_{\rm{H}\alpha} = 0.82\,A_V$.

The effective radius, $r$, of the clump is estimated by summing the area of all pixels contained in the clump, $A$. We then subtract the area of the effective source plane PSF in quadrature and define an effective radius

\begin{equation}
r = \sqrt{\frac{A}{\pi}}.
\end{equation}

We identify 50 clumps in our sample, with between 1 and 9 clumps per galaxy, and list the derived properties in Table \ref{tab:clumps}.

\citet{2010MNRAS.404.1247J} found that the clumps in high-redshift lensed galaxies had surface brightnesses that are higher than those of local H{\sc ii} regions by up to $100\times$. Such intense star-forming regions are found locally in merging systems such as the Antennae \citep{2006A&A...445..471B}, but appeared to be ubiquitious in isolated discs at high-$z$. Similar results were found at $z \sim 5$ \citep{2009MNRAS.400.1121S} and in kpc-scale clumps from unlensed sources \citep{2012MNRAS.tmp.2831W,2012ApJ...760..130S}. To investigate the origin of these high-surface brightness clumps, \citet{2012MNRAS.427..688L} observed 57 clumps in 8 galaxies with H$\alpha$ narrowband imaging, and demonstrated that their surface brightnesses evolve with redshift. In Figure \ref{fig:csizelum}, we expand on this relation adding our new clumps. We note that one galaxy - MACS1149-arcA1.1 - is included in both the current IFU sample and the narrowband imaging sample. Only one clump (the galaxy bulge) was observed in the narrowband sample, and we confirm that its luminosity is consistent with the previous observation within the measurement errors.

With the new data, the evolution in clump surface brightnesses is confirmed and extended out to $z > 3$. We show the mean surface brightness, $\Sigma_{\rm clump}$, in four redshift bins in Figure \ref{fig:csizelum}. The best fit to the redshift evolution of the clump surface brightness is

\begin{equation}
\log\left(\frac{\Sigma_{\rm clump}}{{\rm M}_{\odot}\,{\rm yr}^{-1}\,{\rm kpc}^{-2}}\right) = \left( 3.5 \pm 0.5 \right) \log\left( 1 + z\right) - \left( 1.7 \pm 0.2 \right).
\end{equation}

Clearly there is a selection effect involved, as we would not observe low-surface-brightness clumps in the high-$z$ galaxies. However, there appears to be evolution in the properties of the brightest clumps, as these are not seen in isolated local galaxies but seem to be ubiquitous at high redshift. We will discuss this further in the context of the clump luminosity function along with implications of this evolution in Section \ref{sec:clumpdisc}.

\begin{figure*}
\includegraphics[height=\textwidth, angle=90]{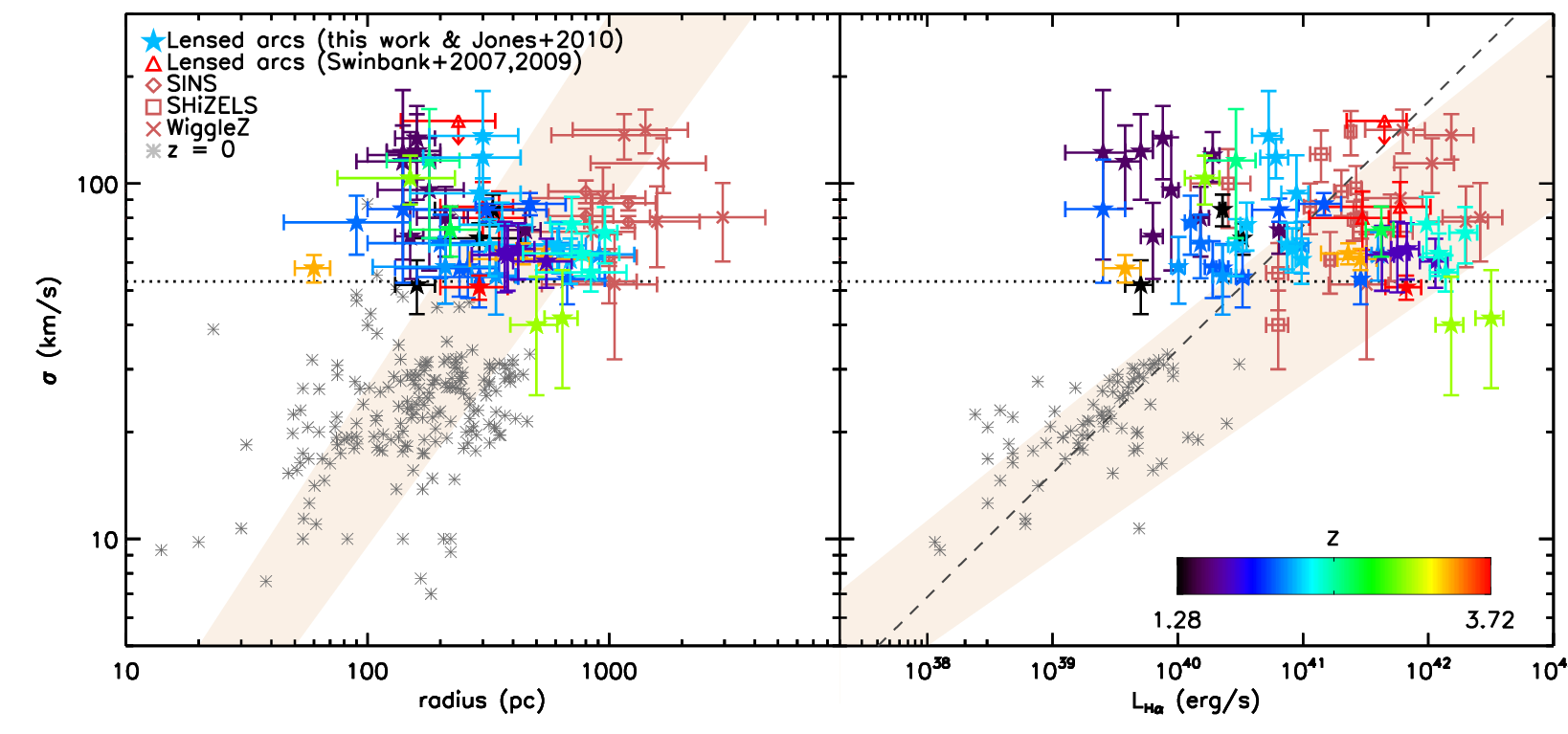}
\caption{Velocity dispersion, $\sigma$, of star-forming clumps, as a function of clump size (\emph{left}) and H$\alpha$ luminosity (\emph{right}). Clumps from the local Universe are taken from \citet{1981MNRAS.195..839T}, \citet{1990A&A...234...23A}, \citet{2011ApJ...735...52B}, \citet{2000AJ....120..752F} and \citet{2006A&A...455..539R}. The high-redshift comparison samples are from the $z \sim 5$ lensed arcs of \citet{2007MNRAS.376..479S} and \citet{2009MNRAS.400.1121S} (with $L_{{\rm H}\alpha}$ estimated from the [O{\sc ii}]-derived SFRs), and the unlensed SINS \citep{2011ApJ...733..101G}, WiggleZ \citep{2012MNRAS.422.3339W} and SHiZELS \citep{2012ApJ...760..130S} surveys. The shaded region shows the best fits from \citet{2012ApJ...760..130S}, the dashed line is the best-fit $L_{{\rm H}\alpha}-\sigma$ relation from \citet{2006A&A...455..539R}, and the dotted line indicates the estimated resolution limit (determined as described in the text). The addition of lensed galaxies extends the high-$z$ samples to smaller sizes and lower luminosities while $\sigma$ is similar to the unlensed samples. Although partially an effect of the spectral resolution limit, we find an excess of clumps with high resolved $\sigma$ for their sizes and luminosities.}
\label{fig:csigma}
\end{figure*}

The IFU data of our current sample allows us to add the extra dimension of velocity dispersion, $\sigma$, to this analysis. If the clumps are gravitationally bound, they should follow a relation of the form $L \propto \sigma^4$ \citep{1981MNRAS.195..839T}. However, not all studies of local star-forming regions have found this relation; \citet{1990A&A...234...23A} found that it applied only to the brightest, `first-ranked' regions in galaxies, with no $L-\sigma$ or $r-\sigma$ relation applying to their entire sample. Others have found a shallower $L-\sigma$ relation indicative of density-bounded regions, where only a fraction of the ionising photons are able to escape \citep{2001Ap&SS.276..413R,2006A&A...455..539R}. It is not clear that clumps should be expected to be virialised; indeed, Giant Molecular Clouds (GMCs) locally are not necessarily gravitationally bound \citep{2011MNRAS.413.2935D}.

In Figure \ref{fig:csigma}, we plot $\sigma$ as a function of both the clump size and luminosity. As a guide, we overlay the relations found by \citet{2012ApJ...760..130S}:

\begin{equation}
  \log \left( \frac{r}{\rm{kpc}} \right) = \left( 1.01 \pm 0.08 \right) \log \left( \frac{\sigma}{\rm{km\,s}^{-1}} \right) + \left( 0.8 \pm 0.1 \right) 
\end{equation}
\begin{equation}
  \log \left( \frac{L}{\rm{erg\,s}^{-1}} \right) = \left(3.81 \pm 0.29 \right) \log \left( \frac{\sigma}{\rm{km\,s}^{-1}}\right) + \left( 34.7 \pm 0.4 \right).
\end{equation}

Although our data scatter around these relations, we do not find any correlation between $\sigma$ in our clumps and either $r$ or $L$. 

We first consider whether this is an effect of spectral resolution. We deconvolve $\sigma$ for the instrumental resolution as measured from sky emission lines, but there is some lower limit at which we will not be able to measure the broadening of the line. To test where this limit is, we construct a set of 1000 Gaussian emission lines with varying widths of $< 100$\,km\,s$^{-1}$ and add noise so that the final signal-to-noise, S/N$ = 5$ (this gives us a conservative estimate, as S/N$>5$ is the constraint we set to detect an emission line). We then convolve the resulting spectrum with the instrumental resolution, and apply the same emission line fitting routine used on the data to recover the line properties. We find that we consistently recover the input line width to within $20\%$ at $\sigma > 53$\,km\,s$^{-1}$. We show this line on Figure \ref{fig:csigma}.

The clumps that lie close to this line may be affected by the resolution limit; however, there remains an excess of clumps at high-$\sigma$ that are well-resolved and lie above the expected relations. It therefore appears that the clumps are not necessarily virialised, and the observed $\sigma$ may contain contributions from a range of processes including gravitational instabilities and star formation feedback. A further possible contribution to the high $\sigma$ values could come from the superposition of multiple clumps along the line of sight. This is an additional source of uncertainty in all values derived from clumps, and motivates studies with higher spectral resolution so that any such multiple clumps can be kinematically separated.

\section{Discussion}
\label{sec:disc}

\subsection{Clump formation}
\label{sec:clumpdisc}

\begin{figure}
\includegraphics[height=84mm, angle=90]{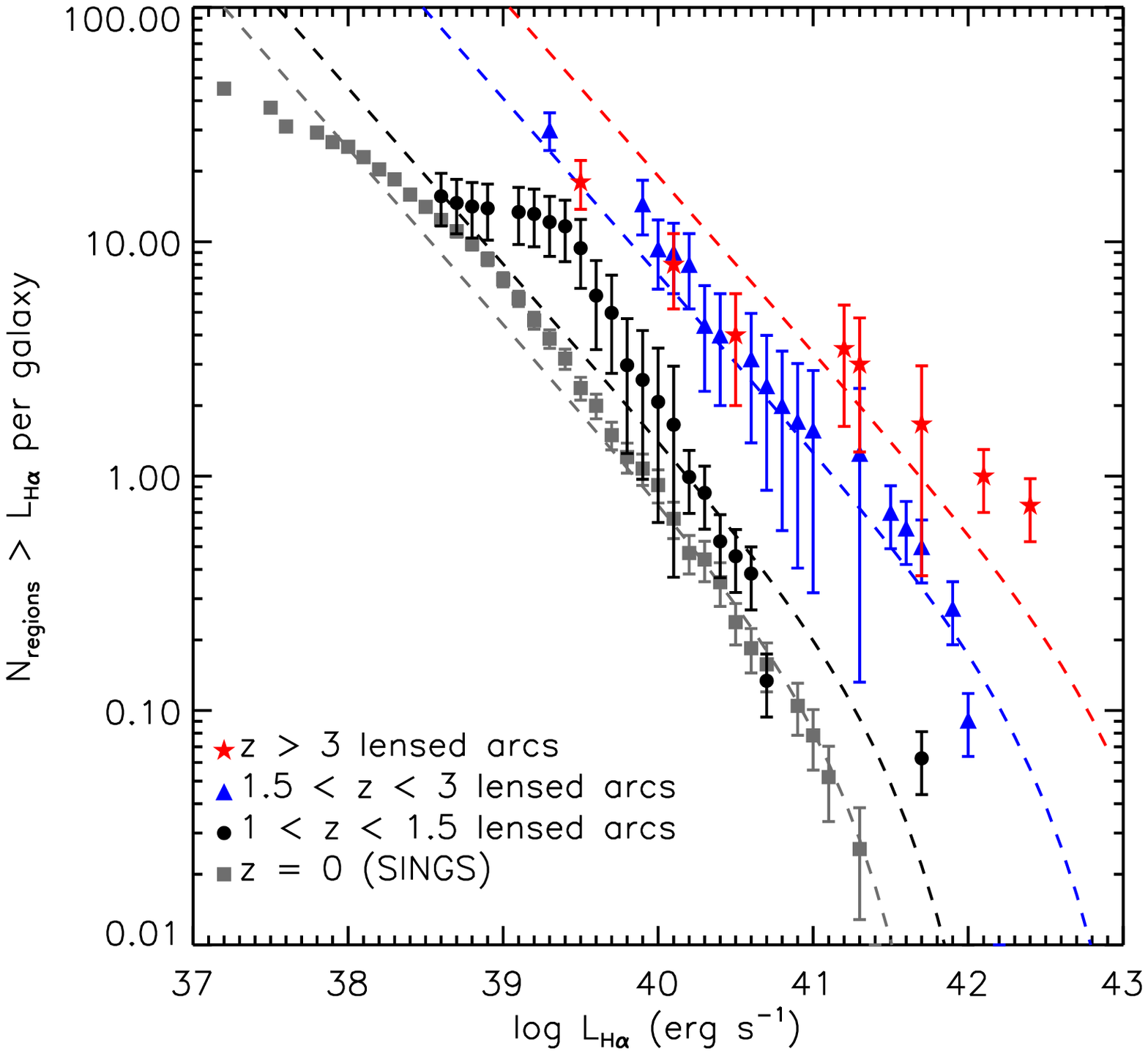}
\caption{Cumulative clump luminosity functions. The $z = 0$ data are from the SINGS survey \citep{2003PASP..115..928K}, with clumps extracted by \citet{2012MNRAS.427..688L}. The $z < 1.5$ bin includes the narrowband imaging sample of \citet{2012MNRAS.427..688L} combined with the two $z < 1.5$ galaxies from this work. Each luminosity bin is normalised by the number of galaxies contributing to that bin, accounting for variations in the depth of observations as described in the text. The dashed lines are Schechter function fits, where the $z = 0$ fit is from the simulations of \citet{2012MNRAS.421.3488H}, and we keep the same normalisation but allow the cutoff to vary to fit the high-redshift data. We find that this cut-off evolves with redshift.}
\label{fig:lumfunc}
\end{figure}

\citet{2012MNRAS.427..688L} demonstrated that an effective means of quantifying the `clumpiness' of a galaxy is the clump luminosity function. The luminosity function can be described by a Schechter function with a cut-off at some maximum luminosity $L_0$, which describes the brightest clumps. The presence of more bright clumps in a galaxy causes it to appear as `clumpy.' \citet{2012MNRAS.427..688L} showed that $L_0$ increases with redshift, giving rise to the appearance of more clumpy galaxies at high-redshift.

We show the luminosity functions in Figure \ref{fig:lumfunc} with the new data added. We account for the varying surface brightness limits of the observations by normalising each bin by the number of galaxies in which we should be able to identify clumps of that luminosity, and the error bars represent the Poisson error from counting clumps. The $z < 1.5$ bin combines the two $z < 1.5$ galaxies from the current sample with the seven galaxies from the \citet{2012MNRAS.427..688L} narrowband imaging sample (MACS1149-arcA1.1 is removed from the latter as it is duplicated in the current sample), and the $z > 3$ bin includes the two lensed $z \sim 5$ galaxies of \citet{2007MNRAS.376..479S} and \citet{2009MNRAS.400.1121S}.

To each redshift bin, we fit a Schechter function of the form

\begin{equation}
N \left( >L \right) = N_0 \left( \frac{L}{L_0} \right)^{\alpha} \exp \left( \frac{-L}{L_0} \right),
\end{equation}

where the power-law slope is fixed to $\alpha = -0.75$ from \citet{2012MNRAS.421.3488H}. We do not attempt to fit the faint-end slope as the sharp turnovers evident in Figure \ref{fig:lumfunc} are likely due to incompleteness and small number statistics in the faintest clumps. The normalisation $N_0$ is arbitrary, so we fit $N_0$ to the $z = 0$ data. We then keep $N_0$ and $\alpha$ fixed while allowing the cut-off $L_0$ to vary, and we find the best-fit $L_0$ for each bin with a $\chi^2$ minimisation procedure. The best fits are

\[
\frac{L_0}{\rm{erg\,s}^{-1}} = 
\begin{cases}
2.5^{+1.8}_{-1.2} \times 10^{41}, &z = 0 \\
4.0^{+0.7}_{-0.7} \times 10^{41}, &1 < z < 1.5 \\
4^{+3}_{-3} \times 10^{42}, &1.5 < z < 3 \\
1.3^{+0.9}_{-0.8} \times 10^{43}, &z > 3
\end{cases}
\]

where the errors are estimated using a bootstrap method.

We therefore find that the cut-off of the clump luminosity function evolves with redshift, such that high-redshift galaxies have more high-luminosity clumps. The best fit to the redshift evolution of $L_0$ is

\begin{equation}
\log\left(\frac{L_0}{{\rm erg\,s}^{-1}}\right) = \left( 2.0 \pm 0.7 \right) \log\left( 1 + z \right) + \left( 41.0 \pm 0.2 \right).
\end{equation}

In combination with the evolution in clump surface brightness discussed in the previous section, it follows that high-redshift galaxies tend to have a higher number of bright, high surface-brightness star-forming clumps, which come to dominate the galaxy morphology and therefore give rise to the population of `clumpy' galaxies seen at high redshift.

\citet{2012MNRAS.427..688L} suggest that the cause of these bright clumps is a combination of high gas fractions at high redshift and the evolving galaxy dynamics. Briefly, the argument is based on the assumption that clumps form in marginally stable discs described by a \citet{1964ApJ...139.1217T} parameter $Q$ where

\begin{equation}
Q = \frac{\kappa \sigma_t(R)}{\pi G \Sigma_0} \approx 1,
\label{eq:q}
\end{equation}

for epicyclic frequency $\kappa \approx \sqrt{2}v/r$ and velocity $v$ at galactocentric radius $r$. \citet{2012MNRAS.421.3488H} argued that star-forming galaxies tend to self-regulate to maintain $Q \sim 1$; instabilities cause the gas to collapse to form stars, while feedback from the star formation stabilises the disc. 

The mass required for collapse on a scale $R$ is the Jeans' mass, $M_J$, given by

\begin{equation}
M_J = \frac{3R}{2G}\sigma.
\label{eq:mj}
\end{equation}

If $M_J$ is related to the maximum mass of clumps in a galaxy, $M_0$ (i.e. the cut-off of the mass function), then \citet{2012MNRAS.427..688L} showed that combining Equations \eqref{eq:q} and \eqref{eq:mj} with an assumed turbulent power spectrum gives 

\begin{equation}
M_0 = \frac{3 \pi^3 G^2}{2} \frac{\Sigma_0^3}{\kappa^4}.
\label{eq:mj2}
\end{equation}

Thus, $M_0$ depends on the disc surface density and epicyclic frequency. To relate this predicted mass to the observable $L_{\rm{H}\alpha}$, we use an empirical relation between the H$\alpha$-derived SFR and mass of local molecular clouds, SFR$\left(M_{\sun}\rmn{yr}^{-1}\right) = 4.6 \pm 2.6 \times 10^{-8} M_{\sun}$ \citep{2010ApJ...724..687L}. This relation is valid for high-density gas, and thus appropriate for star-forming clumps, and is consistent with the independently derived SFRs and gas  masses of star-forming clumps reported in a lensed $z = 2.3$ galaxy by \citet{2011ApJ...742...11S}. Adopting this conversion results in cut-off masses, $M_0$, from our clump luminosity functions of

\[
\frac{M_0}{M_{\odot}} = 
\begin{cases}
4^{+3}_{-2} \times 10^{7}, &z = 0 \\
1.0^{+0.2}_{-0.2} \times 10^{8}, &1 < z < 1.5 \\
9^{+7}_{-7} \times 10^{8}, &1.5 < z < 3 \\
3^{+2}_{-2} \times 10^{9}, &z > 3.
\end{cases}
\]

We can also relate the SFR surface density, $\Sigma_{SFR}$, of clumps to the disc surface density, $\Sigma_0$, using the Kennicutt-Schmidt law. The relationship between the surface density of gas, $\Sigma_{\rm gas}$, and $\Sigma_{\rm SFR}$ found by \citet{1998ApJ...498..541K} is

\begin{equation}
\frac{\Sigma_{\rm SFR}}{M_{\odot}\,\rm{yr}^{-1}\rm{kpc}^{-2}} = A \left( \frac{\Sigma_{\rm g}}{M_{\odot}\rm{pc}^{-2}}\right) ^{n},
\label{eq:ks}
\end{equation}

where $A = \left( 2.5 \pm 0.7 \right) \times 10^{-4}$ and $n = 1.4 \pm 0.15$. The disc surface density $\Sigma_0$ is a combination of gas and stars ($\Sigma_{\rm{g}}$ and $\Sigma_{\ast}$ respectively), which contribute differently to the disc stability. Following \citet{2001MNRAS.323..445R}, we use

\begin{equation}
\Sigma_0 = \Sigma_{\rm{g}} + \left( \frac{2}{1 + f_{\sigma}^2} \right)\Sigma_{\ast},
\label{eq:sigdisc}
\end{equation}

where $f_{\sigma} = \sigma_{\ast}/\sigma_{\rm{g}} \approx 2$ is the ratio of the velocity disperson of the stars to that of the gas \citep{2003AJ....126.2896K}. We define the gas fraction, $f_{\rm gas}$, such that

\begin{equation}
f_{\rm gas} = \frac{M_{\rm gas}}{M_{\rm gas} + M_{\ast}} \approx \frac{\Sigma_{\rm g}}{\Sigma_{\rm g} + \Sigma_{\ast}},
\label{eq:fgas}
\end{equation}

if we measure the gas and stars over the same area. Combining Equations \ref{eq:ks}, \ref{eq:sigdisc} and \ref{eq:fgas}, we have

\begin{equation}
\Sigma_{\rm SFR} = A \left( \Sigma_0 \left( \frac{f_{\rm gas}\left( 1 + f_{\sigma} \right)}{f_{\rm gas}\left( 1 + f_{\sigma} \right) + 2\left( 1 - f_{\rm gas}\right)}\right)\right)^{n}.
\label{eq:sigmasfr}
\end{equation}

Thus, from Equations \ref{eq:mj2} and \ref{eq:sigmasfr}, we find that the cut-off mass, $M_0$, and the clump SFR surface density, $\Sigma_{\rm SFR}$, depend on $\Sigma_0$, $\kappa$ and $f_{\rm gas}$. In order to understand how $M_0$ and $\Sigma_{\rm SFR}$ evolve with redshift, we therefore need to understand the evolution of $\Sigma_0$, $\kappa$ and $f_{\rm gas}$.

We can estimate $\Sigma_0$ and $f_{\rm gas}$ from our data using the $\Sigma_{\rm SFR}$ derived from H$\alpha$ or H$\beta$ and applying the Kennicutt-Schmidt law from Equation \ref{eq:ks} to derive $\Sigma_{\rm gas}$. To estimate $\Sigma_{\ast}$, we use the total $M_{\ast}$ from the SED fitting and apportion the stellar mass according to the fractional flux in each pixel of the reddest available \emph{HST} image. We then combine the $\Sigma_{\rm gas}$ and $\Sigma_{\ast}$ maps to create a map of $\Sigma_0$. There are many uncertainties in estimating of $f_{\rm gas}$ from H$\alpha$, specifically, the assumption of a constant Kennicutt-Schmidt law and in the reliability of the SED-derived stellar masses; nonetheless, we note that our estimate for the Cosmic Eye of $M_{\rm gas} \sim 10^9$\,M$_{\odot}$ is very close to the CO-derived value of $M_{\rm gas} = \left(9.0 \pm 1.6\right) \times 10^8$\,M$_{\odot}$ \citep{2010ApJ...724L.153R}. In our sample, we find a mean $<f_{\rm gas}> = 0.4 \pm 0.1$, consistent with the evolution observed in CO \citep[e.g.][]{2011ApJ...730L..19G,2012ApJ...758L..35L}.

We estimate $\kappa$ by taking the velocity at the half-light radius $r_{1/2}$ along the dynamical axis from the disc fitting, where there is no inclination correction for the radius. This measurement is highly uncertain as the measurement of $r_{1/2}$ is affected by the clumpy morphologies and the irregular source plane PSF, and the measurement of the velocity includes uncertainty in the inclination. We find a median $\kappa \sim 90$\,km\,s$^{-1}$, consistent with the predictions of \citet{2011MNRAS.410.1660D}.

\begin{figure}
\includegraphics[height=84mm, angle=90]{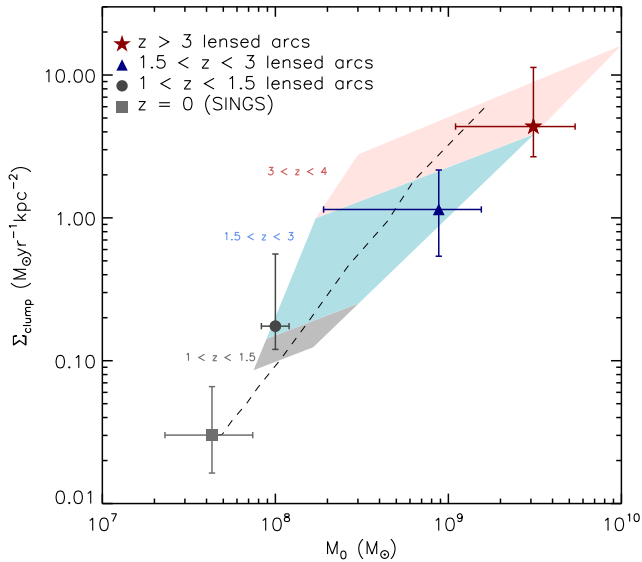}
\caption{Evolution of clump surface brightness $\Sigma_{\rm SFR}$ and the cut-off mass $M_0$ derived from the clumps luminosity functions. The dashed line shows a predicted track based on Equations \ref{eq:mj2} and \ref{eq:sigmasfr} assuming evolution in mass and epicyclic frequency from the simulations of \citet{2011MNRAS.410.1660D} and gas fraction evolution of $f_{\rm gas} \propto \left(1 + z\right)^{2}$ \citep{2011ApJ...730L..19G}, with $f_{\sigma} = 2$. The shaded regions indicate the predicted locations of clumps in each redshift bin, for a range of gas fractions $f_{\rm gas} \propto \left(1 + z\right)^{\left( 2.0 \pm 0.5 \right)}$. The data are consistent with the model, demonstrating that the `clumpy' morphologies of high-$z$ galaxies are driven by evolution in their gas fractions and dynamics.}.
\label{fig:m0sig}
\end{figure}

Using these models for the evolution of $f_{\rm gas}$, $\Sigma_0$ and $\kappa$, we should be able to predict how the star formation surface density of clumps, $\Sigma_{\rm clump}$ and the cut-off mass, $M_0$, evolve. In Figure \ref{fig:m0sig} we present an updated version of the figure in \citet[Figure 9]{2012MNRAS.427..688L} including the IFU data. We find that the evolution in the cut-off of the clump mass function and in the surface brightness of clumps is consistent with the model predictions given by Equations \ref{eq:mj2} and \ref{eq:sigmasfr}, given evolution in the galaxy mass and dynamics from theoretical predictions, and empirical gas fraction evolution. This indicates that galaxies at high redshift appear clumpy because their gas-rich discs fragment on larger scales, leading to star-forming clumps that dominate the morphology of the galaxy. Thus, despite the different appearance of high-redshift clumpy galaxies to local spirals, we find that they do not require a different `mode' of star formation; the clumps can arise naturally as a consequence of the larger gas reservoirs.

\section{Conclusions}
\label{sec:conc}

We have presented integral field spectroscopy around the H$\alpha$ or H$\beta$ emission lines of 12 gravitationally lensed galaxies at $1 < z < 4$, obtained with VLT/SINFONI, Keck/OSIRIS and Gemini/NIFS. We combine these data with 5 galaxies from \citet{2010MNRAS.404.1247J} and investigate the dynamics and star formation properties of 17 high-redshift galaxies. The galaxies all benefit from magnification due to gravitational lensing, increasing the flux by factors of $1.4 - 90\times$ and providing sub-kiloparsec spatial resolution, though we stress that uncertainties are introduced by the anisotropic magnification.

Our combined sample has stellar masses $M_{\ast} \sim 6 \times 10^8 - 6 \times 10^{10}$M$_{\odot}$ and dust extinctions of $A_V \sim 0.4 - 1.1$. The intrinsic star formation rates derived from the H$\alpha$ (or H$\beta$) emission, after correcting for lensing and dust extinction, are SFR$ \sim 0.8 - 40$M$_{\odot}$yr$^{-1}$. The use of gravitational lensing therefore allows us to probe the representative star-forming population at this epoch, with lower $M_{\ast}$ and SFRs than other high-$z$ surveys.

We fit rotating disc models to the sample to derive inclinations and dynamical axes, from which we extract rotation curves and velocity dispersion profiles. All of the galaxies in the combined sample have velocity gradients, and 14 out of 17 are well-fit by an arctan function indicative of rotation. Of the sample, 59\% have velocity profiles indicative of rotation, centrally-peaked velocity dispersion profiles and sufficiently high $v/\sigma$ to be classed as possible rotating discs. Of the remainder, 29\% have disturbed profiles that may be indicative of merging or interacting systems. The remaining two galaxies are undetermined as we do not see a turnover in their rotation curves. The merger fraction is in good agreement with other high-$z$ kinematic surveys \citep[e.g.][]{2009ApJ...706.1364F,2009ApJ...697.2057L,2009ApJ...699..421W,2012A&A...539A..92E}, while, like \citet{2013ApJ...767..104N}, we find that higher spatial resolution causes a higher fraction of the remainder to be classed as rotating discs.

As gravitational lensing allows us to probe intrinsically smaller galaxies, we extend relationships observed between galaxy size, velocity and $v/\sigma$ in unlensed samples to smaller sizes. Our results support previous work finding that larger galaxies generally have a higher contribution to their kinematic support from ordered rotation compared to random motions, but we find a large degree of scatter in these smaller galaxies.

We find that the sample is consistent with the local stellar mass Tully-Fisher relation with no coherent evidence for redshift evolution, in common with the work of \citet{2012ApJ...753...74M}. We further demonstrate that the rotation in our sample, which comprises systematically smaller sizes than unlensed studies, could be dominated by baryons \citep[see also][]{2011ApJ...741..115M}.

We extend the work of \citet{2010MNRAS.404.1247J} and \citet{2012MNRAS.427..688L} by detecting 50 star-forming clumps in our sample and study their luminosities, sizes and velocity dispersions. In common with previous work, we find that the surface brightness evolves with redshift, but we extend this evolution to $z > 3$. The average star formation density in the brightest clumps evolves with redshift as $\log\left(\Sigma_{\rm clump}/{\rm M}_{\odot}\,{\rm yr}^{-1}\,{\rm kpc}^{-2}\right) = \left( 3.5 \pm 0.5 \right) \log\left( 1 + z\right) - \left(1.7 \pm 0.2\right)$. However, we find that the clumps have similar velocity dispersions to unlensed high-$z$ samples while being smaller and less luminous; thus, they introduce a much larger degree of scatter into the $L-\sigma$ and $\sigma-r$ relations observed in other studies. This could be an indication that these clumps are not virialised, and that their velocity dispersions may have additional contributions from star formation feedback or gravitational instability.

We construct luminosity functions of the clumps, and find that they can be fit by a Schechter function in which the cut-off evolves to higher luminosities at higher redshifts. We find that the cut-off evolves as $\log\left(L_0/{\rm erg\,s}^{-1}\right) = \left(2.0 \pm 0.7\right)\log\left( 1 + z\right) + \left(41.0 \pm 0.2\right)$. This supports the picture in which `clumpy' galaxies arise at high-$z$ because gas-rich, turbulent discs fragment on larger scales, resulting in star-forming regions large enough to dominate the morphology of the galaxy. The gas content of high-redshift galaxies remains relatively unexplored due to the small number of observations of molecular gas in normal high-redshift star-forming galaxies. It is to be hoped that the redshift evolution of gas fractions and the universality of the Kennicutt-Schmidt law can be tested as ALMA in full science operations opens up the more normal star-forming population.

\section*{Acknowledgments}

Based in part on observations made with ESO Telescopes at the La Silla Paranal Observatory under programme IDs 083.B-0108, 085.B-0848 and 087.B-0875. Also based in part on observations obtained at the Gemini Observatory under Program ID GN-2010B-Q-61. The Gemini Observatory is operated by the Association of Universities for Research in Astronomy, Inc., under a cooperative agreement with the NSF on behalf of the Gemini partnership: the National Science Foundation (United States), the National Research Council (Canada), CONICYT (Chile), the Australian Research Council (Australia), Minist\'{e}rio da Ci\^{e}ncia, Tecnologia e Inova\c{c}\~{a}o (Brazil) and Ministerio de Ciencia, Tecnolog\'{i}a e Innovaci\'{o}n Productiva (Argentina). RCL acknowledges a studentship from STFC for most of the duration of this project, and funding from the University of Texas at Austin for the latter part. JR is supported by the Marie Curie Career Integration Grant 294074. RGB and IS are supported by STFC and IS further acknowledges a Leverhulme Senior Fellowship. AMS is funded by an STFC Advanced Fellowship, and HE gratefully acknowledges financial support from STScI grants GO-09722, GO-10491, GO-10875 and GO-12166.

\bibliographystyle{mn2e}
\bibliography{bib}

\appendix

\section{Photometry used in SED fitting}
\label{sec:phot}

\begin{table*}
  \caption{\emph{HST} optical photometry used for SED fitting. All fluxes are given in nJy and are corrected for lensing. Upper limits are given at the 3$\sigma$ level.}
  \label{tab:phot1}
  \begin{tabular}{l c c c c c c c c c c}
    \hline
    Name & $F225W$ & $F275W$ & $F336W$ & $F390W$ & $F435W$ & $F475W$ & $F555W$ & $F606W$ & $F625W$ & $F775W$ \\
    \hline
    MACS0744-system3 & $50 \pm 10$ & $80 \pm 20$ & $250 \pm 50$ & $160 \pm 30$ & \ldots & $170 \pm 30$ & $200 \pm 40$ & $230 \pm 40$ & \ldots & $410 \pm 80$ \\
    MACS1149-arcA1.1 & $180 \pm 30$ & $240 \pm 50$ & $380 \pm 70$ & $370 \pm 70$ & $450 \pm 90$ & $430 \pm 80$ & $390 \pm 70$ & $420 \pm 80$ & $450 \pm 80$ & $440 \pm 80$ \\
    MACS0451-system7 & \ldots & \ldots & \ldots & \ldots & \ldots & \ldots & \ldots & $150 \pm 30$ & \ldots & \ldots \\
    A1413-arc2.1a & \ldots & \ldots & \ldots & \ldots & \ldots & \ldots & \ldots & \ldots & \ldots & $300 \pm 60$ \\
    A1689-arc2.1 & \ldots & $0.7 \pm 0.1$ & $2.7 \pm 0.5$ & \ldots & \ldots & $14 \pm 3$ & \ldots & \ldots & $22 \pm 4$ & $29 \pm 6$ \\
    A1689-arc1.2 & \dots & $0.39 \pm 0.07$ & \ldots &  \ldots & \ldots & $5 \pm 1$ & \ldots & \ldots & $11 \pm 2$ & $11 \pm 2$ \\
\hline \\
\end{tabular}
\end{table*}

\begin{table*}
  \caption{\emph{HST} near-infrared photometry used for SED fitting. All fluxes are given in nJy and are corrected for lensing. Upper limits are given at the 3$\sigma$ level.}
  \label{tab:phot2}
  \begin{tabular}{l c c c c c c c}
    \hline
    Name & $F814W$ & $F850LP$ & $F105W$ & $F110W$ & $F125W$ & $F140W$ & $F160W$ \\
     & & & & & & & \\
    \hline
    MACS0744-system3 & $450 \pm 80$ & \ldots & $700 \pm 100$ & $700 \pm 100$ & $480 \pm 90$ & \ldots & \ldots \\
    MACS1149-arcA1.1 & $500 \pm 100$ & $700 \pm 100$ & $810 \pm 150$ & $880 \pm 160$ & $870 \pm 160$ & $920 \pm 170$ & $1000 \pm 200$ \\
    MACS0451-system7 & $160 \pm 30$ & \ldots & \ldots & $280 \pm 50$ & \ldots & $450 \pm 80$ & \ldots \\
    A1413-arc2.1a & \ldots & $360 \pm 70$ & \ldots & \ldots & \ldots & \ldots & \ldots \\
    A1689-arc2.1 & \ldots & $40 \pm 8$ & \ldots & \ldots & $300 \pm 60$ & \ldots & $420 \pm 80$ \\
    A1689-arc1.2 & \ldots & $13 \pm 2$ & \ldots & \ldots & $100 \pm 20$ & \ldots & $120 \pm 20$ \\
\hline \\
\end{tabular}
\end{table*}

\begin{table*}
  \caption{\emph{Spitzer}/IRAC infrared photometry used for SED fitting. All fluxes are given in $\mu$Jy and are corrected for lensing. Upper limits are given at the 3$\sigma$ level.}
  \label{tab:phot3}
  \begin{tabular}{l c c c c}
    \hline
    Name & IRAC & IRAC & IRAC & IRAC \\
     & $3.6\mu$m & $4.5\mu$m & $5.8\mu$m & $8\mu$m \\
    \hline
    MACS0744-system3 & $1900 \pm 300$ & $1400 \pm 300$ & \ldots & \ldots \\
    MACS1149-arcA1.1 & $2000 \pm 400$ & $1500 \pm 300$ & \ldots & \ldots \\
    MACS0451-system7 & $360 \pm 70$ & $380 \pm 70$ & \ldots & \ldots \\
    A1413-arc2.1a & $480 \pm 90$ & $490 \pm 90$ & \ldots & \ldots \\
    A1689-arc2.1 & $< 340$ & $<270$ & $<700$ & $<700$ \\
    A1689-arc1.2 & $140 \pm 30$ & $140 \pm 30$ & $<700$ & $<640$ \\
\hline \\
\end{tabular}
\end{table*}

\label{lastpage}

\end{document}